\documentclass{amsart}
\usepackage{float}
\usepackage{afterpage}
\usepackage{needspace}
\usepackage{graphicx}%
\usepackage{multirow}%
\usepackage{amsmath,amssymb,amsfonts}%
\usepackage{amsthm}%
\usepackage{mathrsfs}%
\usepackage[title]{appendix}%
\usepackage{xcolor}%
\usepackage{textcomp}%
\usepackage{manyfoot}%
\usepackage{booktabs}%
\usepackage{algorithm}%
\usepackage{algorithmicx}%
\usepackage{listings}
\usepackage[numbers]{natbib}
\usepackage{hyperref}
\usepackage{cleveref}

\usepackage{tikz}
\usepackage{pgfplots}
\pgfplotsset{compat=1.18}
\usetikzlibrary{shapes,arrows,positioning,calc}
\usepackage{url}
\usepackage{enumitem}
\usepackage[T1]{fontenc}   

\newtheorem{theorem}{Theorem}[section]

\theoremstyle{definition}
\newtheorem{definition}[theorem]{Definition}

\numberwithin{equation}{section}

\begin{document}
\setlength{\floatsep}{4pt plus 1pt minus 1pt}
\setlength{\textfloatsep}{6pt plus 1pt minus 2pt}
\setlength{\intextsep}{4pt plus 1pt minus 1pt}
\setlength{\abovecaptionskip}{3pt}
\setlength{\belowcaptionskip}{1pt}
\flushbottom

\renewcommand{\bf}{\bfseries}
\renewcommand{\sc}{\scshape}

\title[A Comparison of Voting Data Using Persistent Homology]
{A Comparison of Precinct and District Voting Data Using Persistent Homology to Quantify Gerrymandering in North Carolina, Texas, California, and Louisiana}

\author[Ananya Shah]{Ananya Shah \\ {\normalfont\small Wellesley College}}

\begin{abstract}
We present a framework for quantifying gerrymandering using persistent homology, extending the level-set filtration of Feng \& Porter~\citep{Feng2019} to compare precinct- and district-level voting data. Because precincts are the smallest electoral subunits and are too small to be gerrymandered individually, discrepancies between precinct- and district-level persistence barcodes provide evidence of partisan manipulation.

We apply this framework to two case studies. First, we analyze North Carolina House of Representatives elections (2012--2024). North Carolina has been redistricted four times over this period, making it a uniquely rich longitudinal test case. Comparing persistence barcodes via the bottleneck distance shows that precinct-level voting patterns remain stable across election cycles while district-level patterns shift significantly in redistricting years--providing topological evidence of gerrymandering that compactness-based scores (Polsby-Popper, Reock) fail to detect.

Second, we extend the analysis to the proposed 2025 congressional maps for Texas and California. The Texas 2025 plan, which was struck down by a federal court as a racial gerrymander, shows a bottleneck distance of 12.00 (vs.\ 8.00 for the 2024 baseline plan using the same voting data), with approximately 748,000 Democratic votes--48\% of the total Democratic margin--affected. We additionally introduce persistent homology applied to racial vote-margin rasters (Black and Hispanic Voting Age Population vs.\ White Voting Age Population), providing the first topological framework for quantifying racial gerrymandering. The Texas 2025 plan shows 380 precinct splits across district lines (vs.\ 0 under the 2024 plan), corroborating the racial gerrymander finding. For California, district-level persistence exceeds precinct-level persistence, consistent with the packing of Republican voters under the proposed 2025 map.

We further apply the framework to the Louisiana 2026 congressional map, which emerged from \textit{Louisiana v.\ Callais} (2025). The Louisiana analysis reveals a bottleneck distance of 3.0 between racial and partisan barcodes, indicating a strong correlation between Democratic vote share and minority VAP--providing topological context for the Supreme Court's ruling.

Election data were collected from public sources. Composite shapefiles were created using QGIS and R, rasterized using Python, and processed using the PHAT and GUDHI libraries.
\end{abstract}

\thanks{An earlier version of this research was presented at the 39th Summer Topology and Its Applications Conference -- Topology and Computing (August~11, University of South Alabama); Social Choice: Theory and Computation Conference -- Poster Session (October 15--17, 2025, Wellesley College); and the 57th Annual Northeastern Political Science Association Meeting (November 6--8, 2025, Philadelphia, PA\@). The author was named a Regeneron ISEF 2026 Grand Award Winner (Mathematics, Project MATH035), and a Regeneron STS Top 300 Scholar. This work is supported by the Carnegie Corporation and the Institute for Citizens and Scholars.}

\keywords{gerrymandering, persistent homology, topological data analysis, precinct-level voting data, racial gerrymandering, bottleneck distance}

\maketitle

\newpage
\section{Introduction}
\label{sec:introduction}

\subsection{Gerrymandering}

Gerrymandering--the manipulation of political district boundaries for partisan gain--undermines political representation and remains a central challenge in American democracy. The structure of the United States congressional system allows state legislatures to determine district boundaries, creating opportunities for the majority party to draw maps that systematically favor their candidates over election cycles.

Current identification methods are primarily geometric. The Reock and Polsby-Popper scores measure the compactness of district shapes, but are highly sensitive to physical geography and natural features such as coastlines and rivers; compactness measures should be depended on less for detecting gerrymandering~\citep{ruiz2}. These scores are used widely despite well-documented limitations: they do not detect the two primary instruments of partisan gerrymandering--\emph{cracking} (splitting a group of like-minded voters across multiple districts so they cannot win any) and \emph{packing} (concentrating voters into as few districts as possible to waste their votes).

Discrete geometric and topological methods offer a different perspective. Feng and Porter~\citep{Feng2019} introduced a level-set filtration on geospatial voting data that converts precinct-level maps into filtered simplicial complexes, allowing persistent homology to track the birth and death of topological features as a vote-margin threshold varies. Kallal~\citep{Kallal2019} extended this to a comparative framework: if a map is not gerrymandered, the district-level persistence barcode should resemble the precinct-level barcode.

This paper makes three contributions. First, we apply the Feng--Porter--Kallal framework to a longitudinal study of North Carolina (2012--2024), demonstrating that bottleneck distances between precinct- and district-level barcodes detect the redistricting-driven shifts that compactness scores miss. Second, we extend the framework to proposed 2025 congressional maps for Texas and California, providing timely topological analysis of maps currently under legal and legislative challenge. Third, we introduce the first application of persistent homology to racial gerrymandering, applying the level-set pipeline to racial vote-margin rasters and demonstrating that the Texas 2025 map exhibits topological signatures of racial packing consistent with the federal court ruling against it.

\subsection{Paper Overview}

Section~\ref{sec:background} introduces relevant background on persistent homology and filtrations, following the rigorous treatments of Edelsbrunner and Harer~\cite{edelsbrunner2010} and Dey and Wang~\cite{dey2022}, with accessible motivation from Munch~\cite{munch2017}. Section~\ref{sec:literature} reviews prior work at the intersection of TDA and gerrymandering. Section~\ref{sec:methods} describes our data collection and computational pipeline. Sections~\ref{sec:nc} and~\ref{sec:txca} present results for North Carolina and Texas/California respectively. Section~\ref{sec:louisiana} discusses the Louisiana case study in the context of \textit{Louisiana v.\ Callais}. Section~\ref{sec:conclusion} summarizes findings and future directions.

\section{Background}
\label{sec:background}

\subsection{Topological Data Analysis and Persistent Homology}

\begin{figure}[tp]
\begin{center}
\includegraphics[scale=0.45]{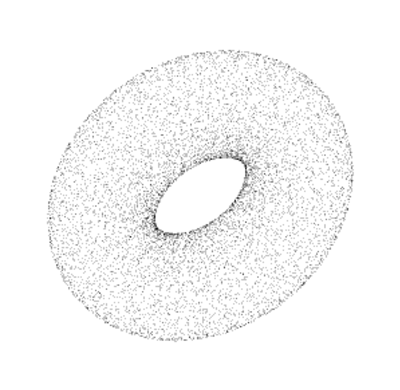}
\caption{A point cloud sampled from a torus, illustrating the type of input to a TDA pipeline.}
\label{fig:torus_pc}
\end{center}
\end{figure}

Topological Data Analysis (TDA) uses concepts from algebraic topology to study the ``shape'' of data in a coordinate-free, scale-invariant way~\cite{munch2017}. The central tool is \emph{persistent homology}, which tracks how topological features--connected components, loops, voids--appear and disappear as a scale parameter varies~\cite{edelsbrunner2010,dey2022}. The input to a typical TDA pipeline is a \emph{point cloud}: a finite set of points in $\mathbb{R}^n$ (in our setting, $n = 2$, but the theory applies in any dimension). Figure~\ref{fig:torus_pc} shows an example point cloud sampled from a torus; in higher dimensions, point clouds cannot be visualized directly, which motivates the need for algebraic-topological summaries.

\begin{figure}[tp]
\begin{center}
\includegraphics[scale=0.35]{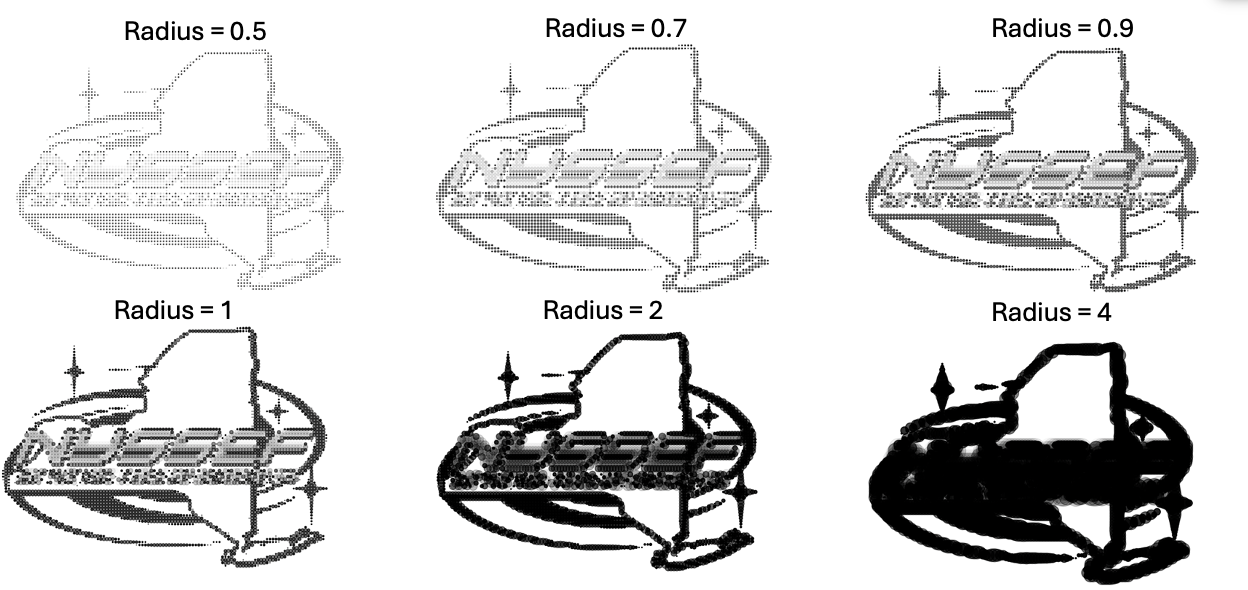}
\caption{Illustration of a filtration: as $\epsilon$ increases, simplices are added, and topological features (connected components, loops) are born and die.}
\label{fig:filtration}
\end{center}
\end{figure}

A key step in persistent homology is the construction of a \emph{filtration}: a sequence of simplicial complexes built from the point cloud by progressively increasing a scale parameter $\epsilon$. Figure~\ref{fig:filtration} illustrates this process: as $\epsilon$ grows, points are connected by edges, then triangles, revealing the topological structure at each scale.

\subsubsection{Simplicial Complexes}

The input to persistent homology is a \emph{filtered simplicial complex}. An abstract simplicial complex $K$ on a vertex set $V = \{v_0, v_1, \ldots, v_n\}$ is a collection of non-empty finite subsets (called \emph{simplices}) satisfying two closure conditions \cite{dey2022}:
\begin{enumerate}
    \item if $\sigma \in K$ and $\tau \subseteq \sigma$, then $\tau \in K$ (any face of a simplex belongs to $K$);
    \item $\{v_i\} \in K$ for every $v_i \in V$ (all vertices are simplices).
\end{enumerate}
A $k$-simplex has $k+1$ vertices and dimension $k$. Thus a $0$-simplex is a point, a $1$-simplex is an edge, a $2$-simplex is a filled triangle, and so on.

\begin{figure}[tp]
\centering
\begin{tikzpicture}[scale=0.85]
  \filldraw[black] (0,0.6) circle (2.5pt);
  \node at (0,-0.15) {\small $0$-simplex};

  \draw[thick] (1.6,0.6) -- (3.0,0.6);
  \filldraw[black] (1.6,0.6) circle (2.5pt);
  \filldraw[black] (3.0,0.6) circle (2.5pt);
  \node at (2.3,-0.15) {\small $1$-simplex};

  \filldraw[black!25, draw=black, thick] (4.3,0) -- (5.3,1.2) -- (6.3,0) -- cycle;
  \node at (5.3,-0.15) {\small $2$-simplex};

  \coordinate (a) at (7.4,0);
  \coordinate (b) at (9.0,0);
  \coordinate (c) at (8.7,1.15);
  \coordinate (d) at (8.2,0.55);
  \filldraw[black!10, draw=black!50] (a)--(b)--(c)--cycle;
  \filldraw[black!18, draw=black!50] (a)--(b)--(d)--cycle;
  \filldraw[black!12, draw=black!50] (b)--(c)--(d)--cycle;
  \draw[thick] (a)--(b); \draw[thick] (a)--(c); \draw[thick] (b)--(c);
  \draw[thick] (a)--(d); \draw[thick] (b)--(d); \draw[thick] (c)--(d);
  \foreach \p in {a,b,c,d} \filldraw[black] (\p) circle (2pt);
  \node at (8.2,-0.15) {\small $3$-simplex};
\end{tikzpicture}
\caption{Simplices of dimension $0$ (vertex), $1$ (edge), $2$ (filled triangle), and $3$ (tetrahedron).}
\label{fig:simplices}
\end{figure}

\subsubsection{Homology and Betti Numbers}

\begin{figure}[tp]
\centering
\includegraphics[scale=0.45]{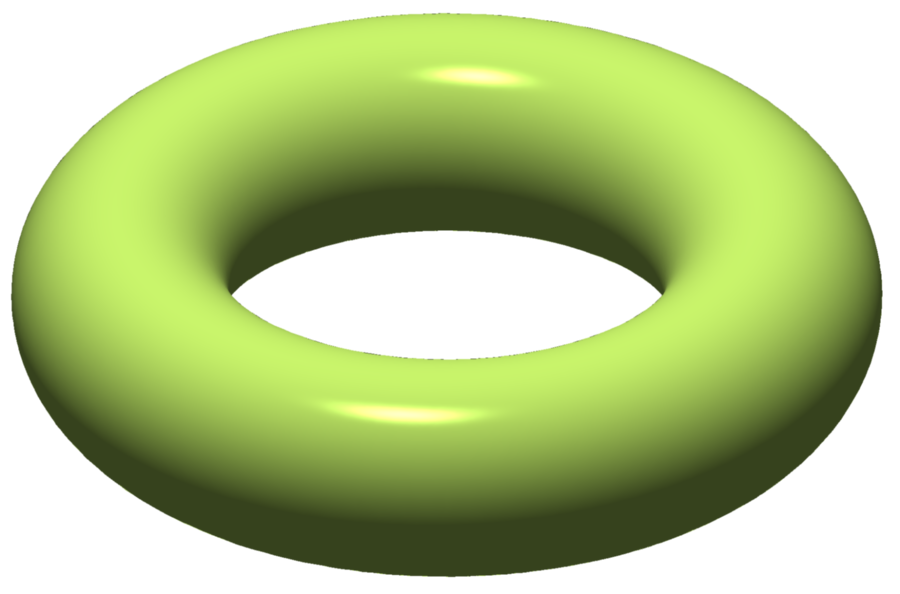}
\caption{A torus. Source: Wikimedia Commons (public domain). Its Betti numbers are $\beta_0=1$, $\beta_1=2$, $\beta_2=1$.}
\label{fig:torusfull}
\end{figure}

Homology is an algebraic way of counting ``holes'' of different dimensions in a simplicial complex~\citep{Scoville2019}. The $k$-th Betti number $\beta_k$ counts the number of independent $k$-dimensional cycles that are not boundaries:
\begin{itemize}
    \item $\beta_0$: number of connected components;
    \item $\beta_1$: number of independent $1$-dimensional loops (``holes'');
    \item $\beta_2$: number of independent $2$-dimensional voids.
\end{itemize}
As an illustration, consider the torus (Figure~\ref{fig:torusfull}): it has $\beta_0 = 1$ (one connected component), $\beta_1 = 2$ (two independent loops, one around the tube and one through the hole), and $\beta_2 = 1$ (one enclosed void).

As a second example, consider the following pair of disjoint $1$-dimensional simplicial complexes (Figure~\ref{fig:2simpl}), each consisting of four vertices with six edges forming a complete graph $K_4$. For this complex, $\beta_0 = 2$ (two connected components), $\beta_1 = 6$ (six independent loops--one for each edge that creates a cycle in each $K_4$), and $\beta_2 = 0$ (no enclosed voids, since the complex is $1$-dimensional).

\begin{figure}[tp]
\centering
\begin{tikzpicture}[scale=0.9,
  vtx/.style={circle, fill=black, inner sep=2pt}]
  \node[vtx,label=below:{\small$v_1$}] (v1) at (0,0)   {};
  \node[vtx,label=below:{\small$v_2$}] (v2) at (1.6,0)  {};
  \node[vtx,label=above:{\small$v_3$}] (v3) at (1.6,1.4){};
  \node[vtx,label=above:{\small$v_4$}] (v4) at (0,1.4)  {};
  \foreach \a/\b in {v1/v2,v2/v3,v3/v4,v4/v1,v1/v3,v2/v4}
    \draw[thick] (\a)--(\b);
  \node[vtx,label=below:{\small$w_1$}] (w1) at (3.4,0)  {};
  \node[vtx,label=below:{\small$w_2$}] (w2) at (5.0,0)  {};
  \node[vtx,label=above:{\small$w_3$}] (w3) at (5.0,1.4){};
  \node[vtx,label=above:{\small$w_4$}] (w4) at (3.4,1.4){};
  \foreach \a/\b in {w1/w2,w2/w3,w3/w4,w4/w1,w1/w3,w2/w4}
    \draw[thick] (\a)--(\b);
\end{tikzpicture}
\caption{Two disjoint copies of the complete graph $K_4$, viewed as a $1$-dimensional simplicial complex.
Betti numbers: $\beta_0=2$ (two components), $\beta_1=6$ (six independent cycles in $H_1$), $\beta_2=0$.}
\label{fig:2simpl}
\end{figure}

Because Betti numbers are topological invariants, they are independent of the specific geometry--only the combinatorial structure of the complex matters. This makes them robust to coordinate choices and natural deformations of the data, a key advantage for geospatial applications.

\subsubsection{Persistent Homology}

A \emph{filtration} of a simplicial complex $X$ is a nested sequence
\[
X_0 \subseteq X_1 \subseteq \cdots \subseteq X_l = X,
\]
where each $X_i$ is a sub-complex. As we move through the filtration, topological features are born (first appear) and die (merge or fill in). The $m$-th persistent homology of $X$ is the pair
\[
\left( \{ H_m(X_i) \}_{1 \leq i \leq l},\; \{ f_{i,j} \}_{1 \leq i \leq j \leq l} \right),
\]
where $f_{i,j} : H_m(X_i) \to H_m(X_j)$ are the maps induced by the inclusions $X_i \hookrightarrow X_j$~\citep{Feng2019,edelsbrunner2010,dey2022}. By the structure theorem for persistence modules, this data can be represented as a \emph{barcode}: a multiset of intervals $[b_i, d_i)$, where $b_i$ is the scale at which feature $i$ is born and $d_i$ is the scale at which it dies~\cite{edelsbrunner2010}.

\begin{definition}[Persistence Barcode]
The \emph{persistence barcode} of a filtered simplicial complex $(X_\bullet)$ in dimension $m$ is the multiset of intervals $\{[b_i, d_i)\}$ recording the birth and death scales of each $m$-dimensional homological feature across the filtration~\cite{edelsbrunner2010,munch2017}.
\end{definition}

Long bars (large $d_i - b_i$) correspond to persistent, topologically significant features; short bars correspond to noise or small-scale fluctuations.

\subsubsection{Bottleneck Distance}

Given two barcodes $B_1$ and $B_2$, the \emph{bottleneck distance} is
\[
d_B(B_1, B_2) = \inf_{\gamma} \max_{x \in B_1} \| x - \gamma(x) \|_{\infty},
\]
where the infimum is over all bijections $\gamma: B_1 \to B_2$ (with diagonal points included to handle unmatched features)~\citep{gudhi}. Geometrically, this measures the largest displacement required to match the features of one barcode to the other. A small bottleneck distance indicates that the two filtered complexes have similar topological structure; a large distance indicates meaningful structural divergence.

In our setting, a large bottleneck distance between precinct- and district-level barcodes for the same election year signals that the district map does not faithfully represent precinct-level voting patterns--evidence of gerrymandering. We additionally express bottleneck distance as a fraction of the total filtration length (25 steps), converting it to an estimate of the proportion of Democratic votes whose topological representation is shifted by redistricting:
\[
\text{Votes shifted} \approx \frac{d_B}{25} \times V_{\text{Dem, total}}.
\]

\subsubsection{Persistence Diagrams}

A \emph{persistence diagram} $\mathrm{Dgm}(B)$ is an equivalent representation of the barcode as a multiset of points $(b_i, d_i)$ in the plane, where each point records the birth and death of one topological feature~\cite{edelsbrunner2010}. The diagonal $\{b = d\}$ represents features of zero persistence (noise); points far from the diagonal represent significant topological structure. Persistence diagrams admit the same bottleneck and Wasserstein distances as barcodes, and are more amenable to visualisation of the joint birth--death distribution.

\subsubsection{Betti Curves}

The \emph{Betti curve} $\beta_k(t)$ records the $k$-th Betti number of the sublevel-set complex at each filtration step $t$~\citep{Guzel2022}:
\[
\beta_k(t) = \#\{ [b_i, d_i) \in B_k : b_i \leq t < d_i \}.
\]
For $k = 1$, the Betti-1 curve counts the number of active $H_1$ holes (Democratic voting islands) at each filtration step. Comparing the precinct and district Betti-1 curves directly shows \emph{when} during the filtration the district map loses topological features present in the precinct map. A downward shift in the district Betti-1 curve relative to the precinct curve, concentrated in the middle filtration steps (steps 5--20), indicates cracking of Democratic strongholds.

\subsubsection{Persistence Landscapes and Silhouettes}

The \emph{persistence landscape}~\citep{Bubenik2015} is a functional summary of a barcode defined by
\[
\lambda_k(t) = \sup\{ m \geq 0 : (t-m, t+m) \text{ is covered by at least } k \text{ bars} \},
\]
where $\lambda_1(t)$ is the most prominent envelope. Unlike barcodes, persistence landscapes lie in a Hilbert space $L^2(\mathbb{R})$, enabling averaging, variance computation, and statistical hypothesis testing~\citep{Bubenik2015}. The \emph{persistence silhouette}~\citep{Chazal2014} is a weighted average of the tent functions:
\[
\varphi(t) = \frac{\sum_i w_i \Lambda_i(t)}{\sum_i w_i}, \quad w_i = d_i - b_i,
\]
where $\Lambda_i(t) = \max(0,\, \min(t - b_i,\, d_i - t))$ is the tent function for bar $i$. The shaded gap between precinct and district landscapes/silhouettes is a \emph{manipulation signal}: regions where the district curve falls below the precinct curve correspond to topological features that the district map systematically suppresses.

\subsubsection{Persistence Images}

A \emph{persistence image}~\citep{Adams2017} converts a persistence diagram into a pixelated density map by placing a Gaussian kernel at each point $(b_i, p_i) = (b_i, d_i - b_i)$ in the birth--persistence plane and weighting by $p_i$:
\[
I(z) = \sum_i p_i \cdot \phi_\sigma(z - (b_i, p_i)),
\]
where $\phi_\sigma$ is a Gaussian with bandwidth $\sigma$. The difference image $\Delta I = I_{\text{district}} - I_{\text{precinct}}$ directly identifies which birth--persistence regions are gained (red, $+$) or lost (blue, $-$) under redistricting: blue regions represent topological features present at the precinct level that are destroyed by the district map (cracking), while red regions represent features artificially introduced by the district map (packing).

\section{Related Work}
\label{sec:literature}

\subsection{Topological Data Analysis in Applications}

Persistent homology has found applications in neuroscience~\citep{Bendich2016,Dabaghian2012}, medical imaging~\citep{Nicolau2011}, and machine learning~\citep{Adams2017}, among many others. For a comprehensive introduction to the theory, see Edelsbrunner and Harer~\cite{edelsbrunner2010} or Dey and Wang~\cite{dey2022}; for an accessible survey aimed at practitioners, see Munch~\cite{munch2017}. Applications to geospatial and electoral data are much more limited.

\subsection{TDA and Gerrymandering}

Feng and Porter~\citep{Feng2019} introduced the level-set filtration for precinct-level voting data, demonstrating on 2016 California data that distance-based filtrations (Vietoris--Rips, Alpha complex) are inadequate for pixelated geospatial data because they do not preserve the spatial structure of voting maps. Their filtration is parameterised by the Democratic win-margin score
\[
\delta_{b,r}(p) = \frac{|V_b(p) - V_r(p)|}{V_b(p) + V_r(p)},
\]
where $V_b(p)$ and $V_r(p)$ are the Democratic (blue) and Republican (red) vote totals in precinct $p$. They filter by starting with only the most strongly Democratic precincts (e.g., $\delta_{b,r}(p) \geq 0.95$) and progressively lowering the threshold. In our setting, we use a sublevel-set filtration of a continuous density function $f\colon\mathbb{R}^2\to\mathbb{R}$ derived from a rasterized vote-margin map, following the modification of Kallal~\citep{Kallal2019}.

Kallal~\citep{Kallal2019} applied this framework comparatively, arguing that a non-gerrymandered map should produce similar precinct- and district-level barcodes. His analysis of 2016 presidential data motivates our approach; we extend it to congressional data (avoiding split-ticket distortions), a longer time series (2012--2024 for North Carolina), and new states (Texas and California).

Duchin, Needham, and Weighill~\citep{Duchin2020} study gerrymandering through ensemble methods and Markov chain Monte Carlo sampling of redistricting plans, using persistent homology to characterize ensemble distributions. That approach complements ours: ensemble methods characterize the space of valid plans, while our barcode comparison directly interrogates enacted maps against ground-truth precinct data.

Other recent work in the area includes Stephanopoulos and McGhee~\citep{StephanopoulosMcGhee} on the efficiency gap, and Wang~\citep{Wang2024} on the Princeton Gerrymandering Project's grading system. Neither relies on topological methods; we compare our findings to these scores in Section~\ref{sec:nc}.

\subsection{Novelty of This Work}

This paper extends prior TDA-gerrymandering work in three ways. First, we apply the framework longitudinally across six election cycles (2012--2024) in North Carolina, the most heavily redistricted large state in the country during this period. Second, we apply it prospectively to proposed 2025 maps in Texas and California--both currently under legal challenge. Third, we introduce the first application of persistent homology to racial gerrymandering by applying the level-set pipeline to racial vote-margin rasters, enabling topological quantification of packing and cracking along racial lines.

\section{Methods}
\label{sec:methods}

\subsection{Data Collection}

\paragraph{North Carolina.} Historical precinct-level voting data and congressional district shapefiles were obtained from the North Carolina State Board of Elections (\url{https://www.ncsbe.gov/}). Precinct-level results are published as CSV files following each election; we filter for ``US House of Representatives'' contests and aggregate Democratic and Republican totals by precinct, discarding third-party votes. District shapefiles were also verified against the Metric Geometry and Gerrymandering Group (MGGG) repository.

\paragraph{Texas and California.} Precinct-level voting data for the 2024 election cycle were obtained from the Redistricting Data Hub. District boundary shapefiles for both the 2024 and proposed 2025 maps were downloaded from Dave's Redistricting App. For the racial gerrymandering analysis, 2020 Census block demographic data (Black Voting Age Population, BVAP; Hispanic Voting Age Population, HVAP) were obtained from the ALARM Project and mapped to voting tabulation districts via spatial interpolation.

\paragraph{Missing data.} North Carolina has $2{,}716$ precincts; in certain election years between 1 and 50 precincts lacked complete records (fewer than 0.02\% per cycle). For these precincts we impute a neutral margin of 10 Democratic and 10 Republican votes, yielding a margin of exactly zero. This is conservative: a zero margin contributes no topological signal and does not distort the barcode in either direction.

\subsection{Computational Pipeline}

The pipeline converts precinct or district shapefiles into persistence barcodes through four steps, following~\citet{Feng2019} and~\citet{Kallal2019}:

\[
\text{Shapefile} \;\to\; \text{Raster (.tiff)} \;\to\; \text{Filtered Simplicial Complex} \;\to\; \text{Persistence Barcode}
\]

\begin{figure}[tp]
\centering
\includegraphics[scale=0.5]{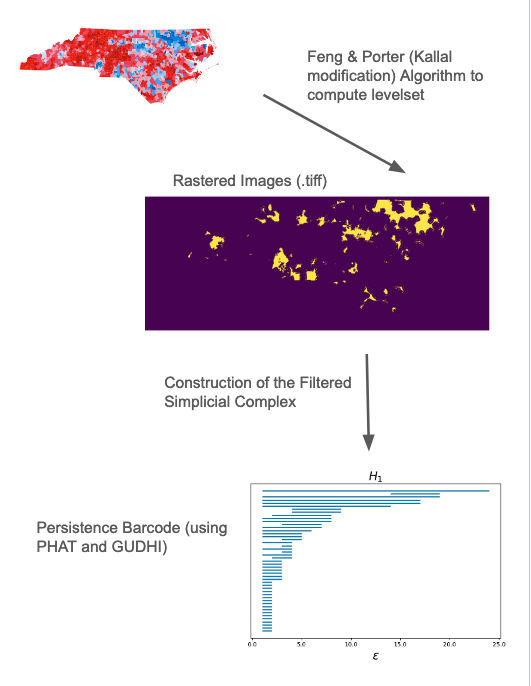}
\caption{Overview of the data processing pipeline from shapefile to barcode.}
\label{fig:pipeline}
\end{figure}

\paragraph{Step 1: Margin and density calculation.} For each precinct or district $p$, we compute the signed Democratic win margin per unit area:
\[
f(p) = \frac{V_b(p) - V_r(p)}{\text{Area}(p)},
\]
where $V_b(p)$ and $V_r(p)$ are Democratic and Republican vote totals. Scaling by area is Kallal's modification~\citep{Kallal2019}: without it, ``the construction of the simplicial complex would not [be] sensitive to variations in the absolute voter margin per square meter; a region with only a small margin one way or the other gets filled in by the expanding level set at the same rate as a region of the same size.'' This density function $f$ is then embedded in $\mathbb{R}^2$ and rasterized to a \texttt{.tiff} file using GDAL and OGR~\citep{gdal}.

\paragraph{Step 2: Rasterization.} Precincts won by Democrats are colored blue; those won by Republicans are colored red. The result is a pixel map over the geographic extent of the state.

\begin{figure}[tp]
\centering
\includegraphics[scale = 0.3]{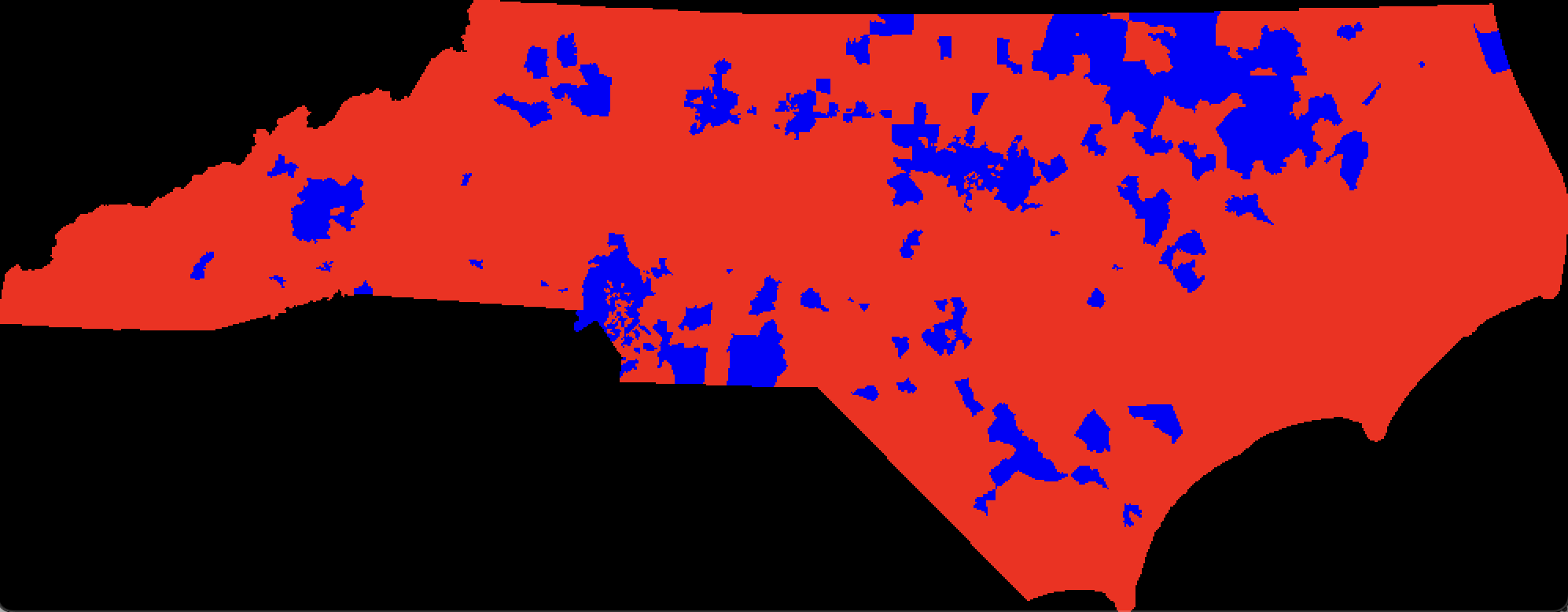}
\caption{North Carolina 2024 precinct map colored by winning party (blue = Democratic majority, red = Republican majority)}
\label{fig:nc2024preceg}
\end{figure}

\begin{figure}[tp]
    \centering
    \begin{minipage}{0.45\textwidth}
        \centering
        \includegraphics[width=\textwidth]{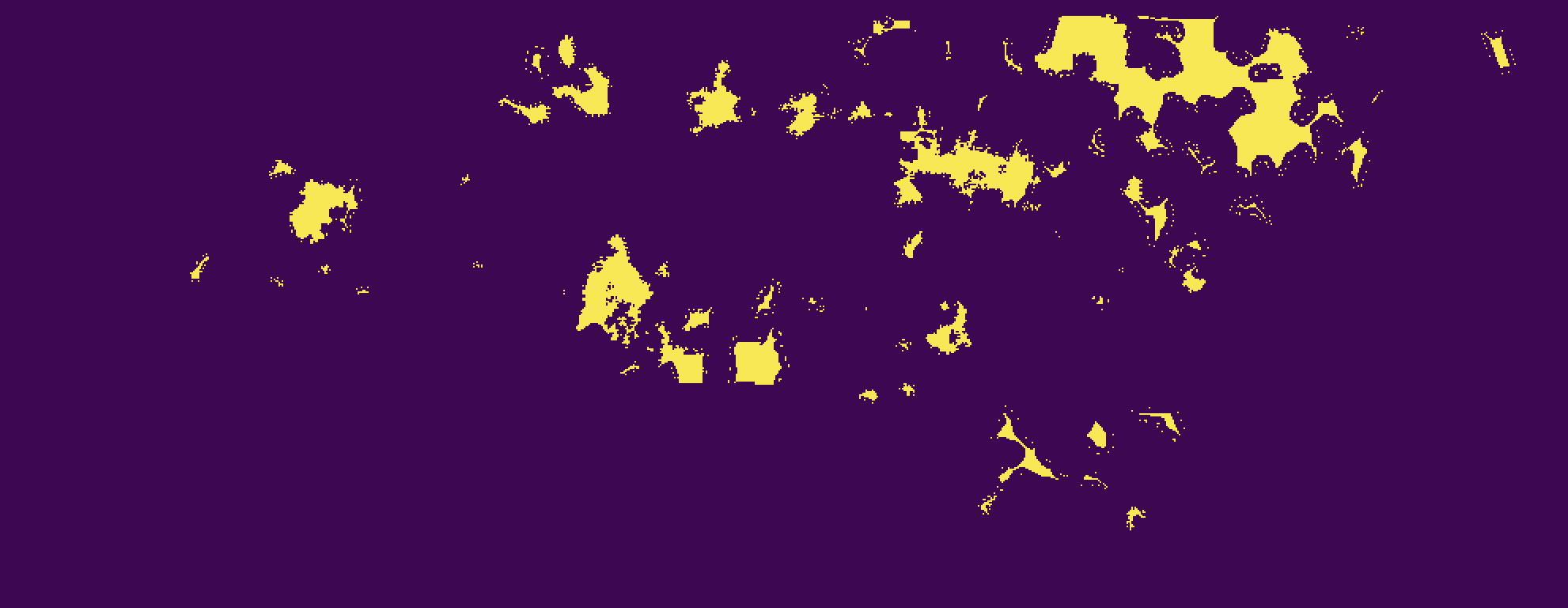}
        \caption{NC 2024 Precinct: filtration step 1}
        \label{fig:nc_ls_1}
    \end{minipage}%
    \hfill
    \begin{minipage}{0.45\textwidth}
        \centering
        \includegraphics[width=\textwidth]{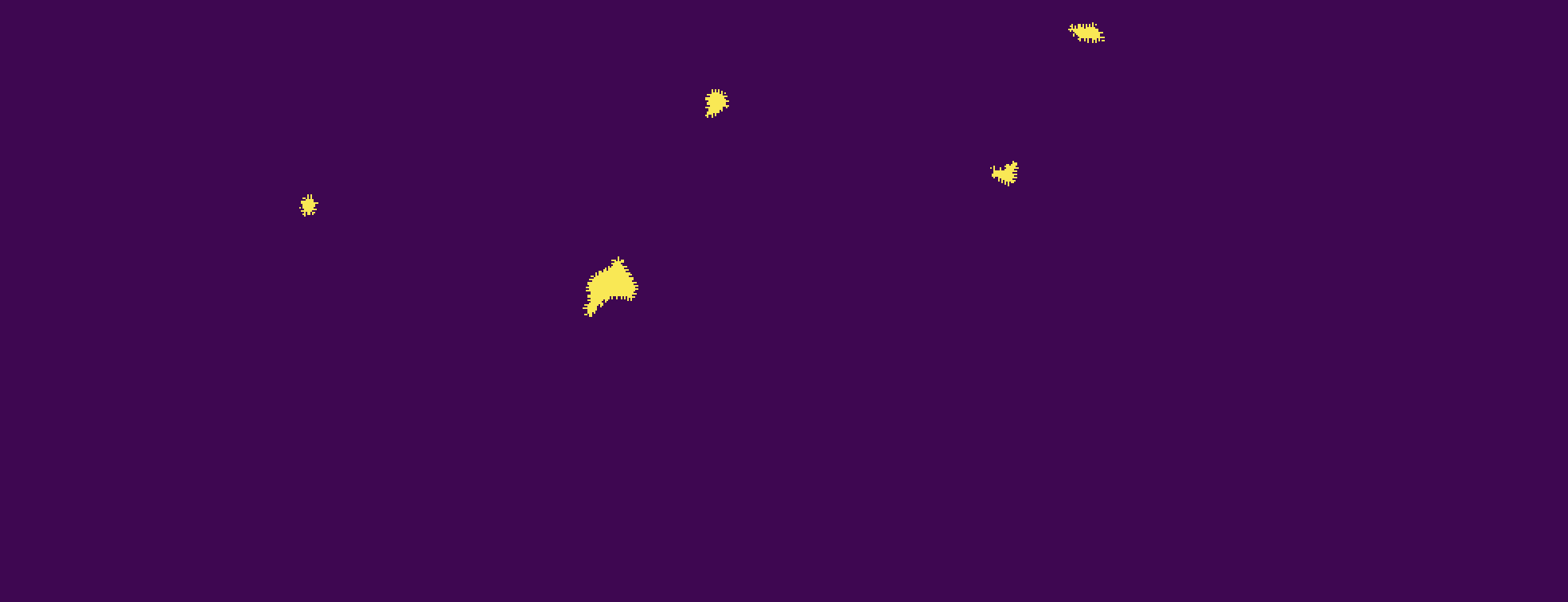}
        \caption{NC 2024 Precinct: filtration step 25}
        \label{fig:nc_ls_25}
    \end{minipage}
\end{figure}

\paragraph{Step 3: Level-set filtration.} We construct a filtered simplicial complex via sublevel sets of $f$. Starting from only those pixels with the highest Democratic margin (filtration step 1 of 25), we progressively lower the threshold, adding pixels at each step. This defines a sequence of nested complexes. Geometrically, Democratic-leaning regions appear as ``holes'' surrounded by Republican territory, and persistent homology tracks the birth and death of these holes across the 25 filtration steps (Figures~\ref{fig:nc_ls_1}--\ref{fig:nc_ls_25}).

\paragraph{Step 4: Barcode computation.} We use the PHAT library~\citep{phat} to compute persistence pairs (birth, death) from the boundary matrix of the filtered complex, then plot the resulting $H_1$ barcode (tracking $1$-dimensional holes, i.e., Democratic ``islands'') using Matplotlib. Bottleneck distances between pairs of barcodes are computed using GUDHI~\citep{gudhi}.

\begin{figure}[tp]
\centering
\includegraphics[scale=0.3]{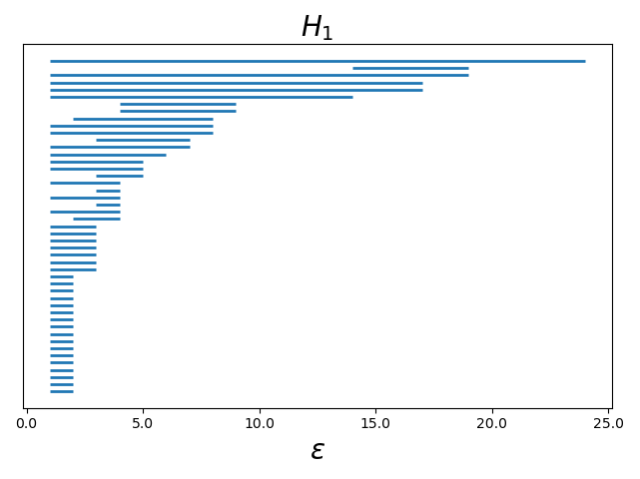}
\caption{Example $H_1$ persistence barcode. Each horizontal bar $[b_i, d_i)$ represents a Democratic voting region that persists from filtration step $b_i$ to step $d_i$. Longer bars indicate more robustly Democratic regions.}
\label{fig:barcodeeg}
\end{figure}

\subsection{Racial Gerrymandering Analysis}

For Texas, we additionally construct a racial vote-margin raster using Census demographic data:
\[
f_{\text{racial}}(p) = \text{BVAP}(p) + \text{HVAP}(p) - \text{WVAP}(p),
\]
where BVAP, HVAP, and WVAP denote Black, Hispanic, and White Voting Age Population respectively. We apply the same level-set pipeline to this raster and compute the bottleneck distance between the resulting precinct-level and district-level barcodes. A large distance here indicates that the district map does not preserve the spatial clustering of minority voters--a topological signature of racial gerrymandering.

We also compute the \emph{precinct-split metric}: the number of census blocks whose centroid lies in a precinct that spans two or more congressional districts. When census-block demographic data is used in map drawing, small geographic units tend to be split across district lines, producing a non-zero split count even when precincts themselves are not divided.

\section{North Carolina: Longitudinal Analysis (2012--2024)}
\label{sec:nc}

\subsection{Background}

North Carolina provides an unusually rich test case because its congressional maps have been redrawn four times since 2011:
\begin{itemize}
    \item 2012 and 2014 elections: maps enacted in 2011 following the 2010 Census;
    \item 2016, 2018, and 2020 elections: maps enacted in February 2016 following a federal court challenge;
    \item 2022 election: maps enacted in 2021 post-2020 Census (adding one seat);
    \item 2024 election: maps enacted in 2023 (plans 746/747).
\end{itemize}
Precinct boundaries in North Carolina have remained essentially unchanged over this entire period, providing a stable baseline for comparison. The state's history of gerrymandering extends to the 1990s, culminating in the U.S. Supreme Court's ruling in \textit{Shaw v.\ Reno} (1993), which held that the 12th congressional district's irregular shape constituted an equal protection violation~\citep{Peters2024}. More recently, Mattingly noted that the 2023 maps are ``more gerrymandered and less responsive than maps struck down in 2021''~\citep{Mattingly2023}, and in the 2024 elections, North Carolina Democrats won over 46\% of the congressional vote but only 29\% of seats~\citep{Li2024}.

\subsection{Rationale}

Our hypothesis is as follows. If a map is not gerrymandered, the precinct- and district-level barcodes should be similar: every persistent Democratic voting island at the precinct level should be reflected as a persistent feature at the district level~\citep{Kallal2019}. A precinct feature that is persistent but absent at the district level is consistent with cracking (the Democratic island has been divided across multiple districts). Conversely, district-level persistence absent at the precinct level is consistent with packing (Democratic votes have been consolidated, artificially creating a concentrated ``island'' at the district level).

\subsection{Results}

\paragraph{2012.} Following Republican majorities in both chambers of the General Assembly in 2010, the 2011 redistricting produced maps that gave Republicans a $9$--$4$ advantage in the 13-seat delegation despite a much narrower statewide margin. The precinct barcode shows several persistent $H_1$ features corresponding to the urban cores of Charlotte, Raleigh, Greensboro, and Durham. The district barcode shows fewer persistent features, consistent with cracking of these urban Democratic strongholds across neighboring Republican districts.

\begin{figure}[tp]
    \centering
    \begin{minipage}{0.45\textwidth}
        \centering
        \includegraphics[width=\textwidth]{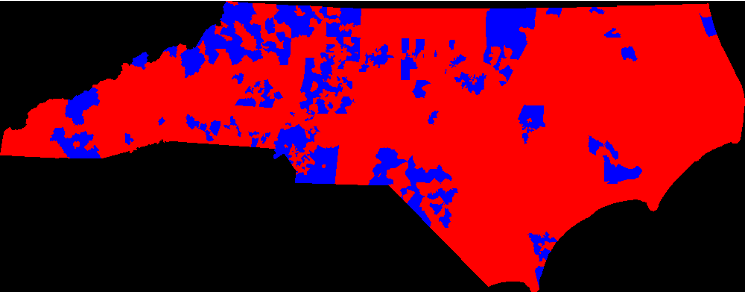}
        \caption{NC 2012 precincts (colored by winner)}
        \label{fig:nc2012prec}
    \end{minipage}%
    \hfill
    \begin{minipage}{0.45\textwidth}
        \centering
        \includegraphics[width=\textwidth]{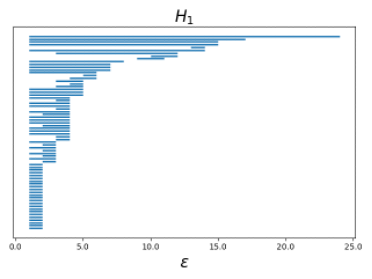}
        \caption{NC 2012 precinct barcode}
        \label{fig:nc2012precbar}
    \end{minipage}
    \begin{minipage}{0.45\textwidth}
        \centering
        \includegraphics[width=\textwidth]{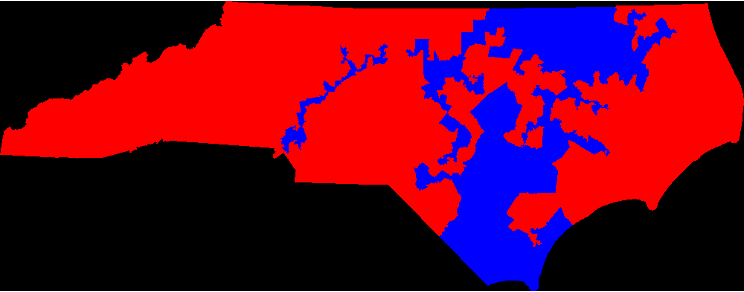}
        \caption{NC 2012 districts (colored by winner)}
        \label{fig:nc2012dist}
    \end{minipage}%
    \hfill
    \begin{minipage}{0.45\textwidth}
        \centering
        \includegraphics[width=\textwidth]{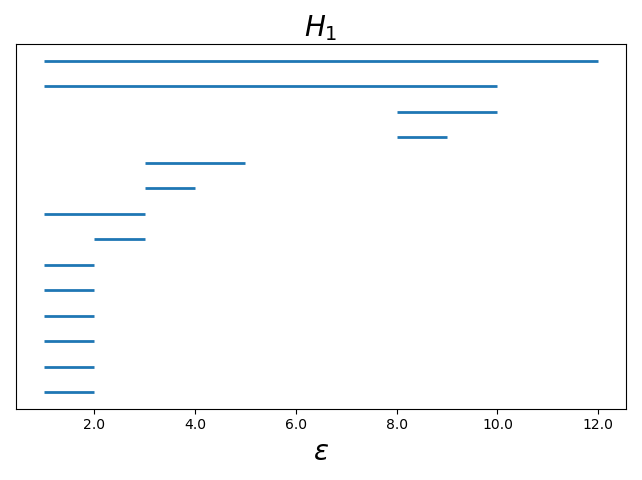}
        \caption{NC 2012 district barcode}
        \label{fig:nc2012distbar}
    \end{minipage}
\end{figure}

\paragraph{2014.} The 2014 elections used the same map as 2012. The precinct-level barcode is very similar to 2012, confirming stable voting geography. The district barcode likewise mirrors 2012, consistent with no redistricting occurring between cycles. Supplementary figures for all cycles are in Appendix~\ref{sec:addfig}.

\paragraph{2016.} Following a federal court order, districts were redrawn in February 2016. The 2016 map produced a notably different district barcode relative to the 2014 barcode (a large year-to-year jump in the district-to-district bottleneck distance visible in Figure~\ref{fig:bn_district}), while the precinct barcode changed minimally--confirming that redistricting, not voter behavior, drove the structural shift.

\paragraph{2018 and 2020.} These elections used the same 2016 court-ordered map. Consistent with that expectation, the precinct and district barcodes for 2018 and 2020 are similar to 2016; the district-to-district bottleneck distances between these cycles are small relative to redistricting years.

\paragraph{2022.} The 2021 post-Census redistricting added one seat and drew considerable scrutiny. The precinct barcode shows persistent features corresponding to both halves of Charlotte (Districts 9 and 12) and to the Raleigh metropolitan area. The district barcode preserves these two Charlotte features but substantially reduces persistence for the Raleigh feature, suggesting partial cracking.

\begin{figure}[tp]
    \centering
    \begin{minipage}{0.45\textwidth}
        \centering
        \includegraphics[width=\textwidth]{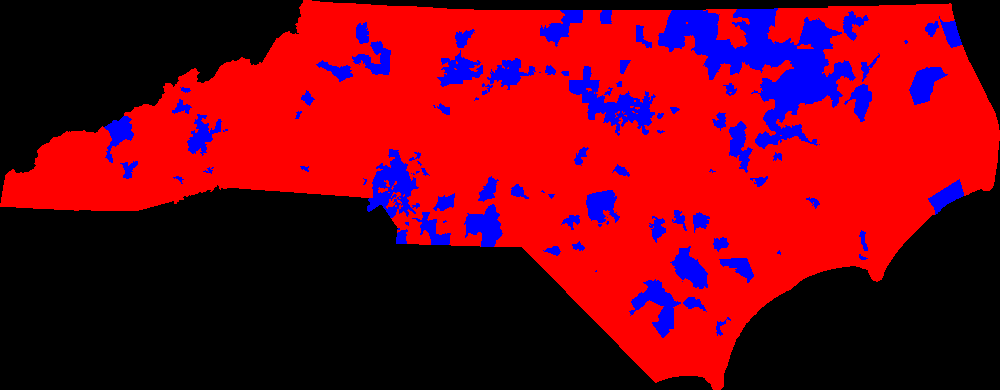}
        \caption{NC 2022 precinct map}
        \label{fig:nc2022prec}
    \end{minipage}%
    \hfill
    \begin{minipage}{0.45\textwidth}
        \centering
        \includegraphics[width=\textwidth]{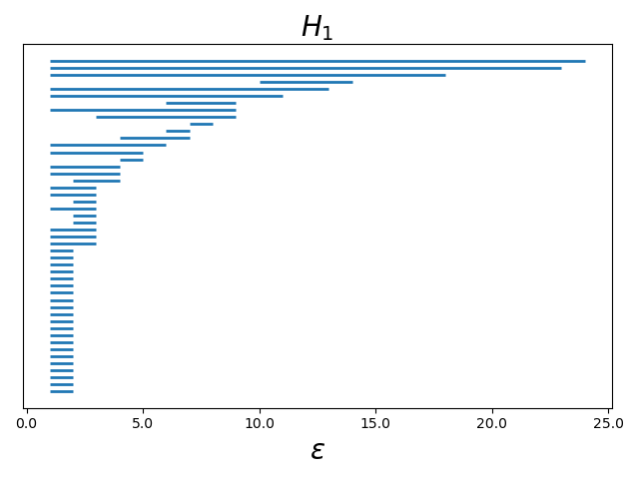}
        \caption{NC 2022 precinct barcode}
        \label{fig:nc2022precbar}
    \end{minipage}
    \begin{minipage}{0.45\textwidth}
        \centering
        \includegraphics[width=\textwidth]{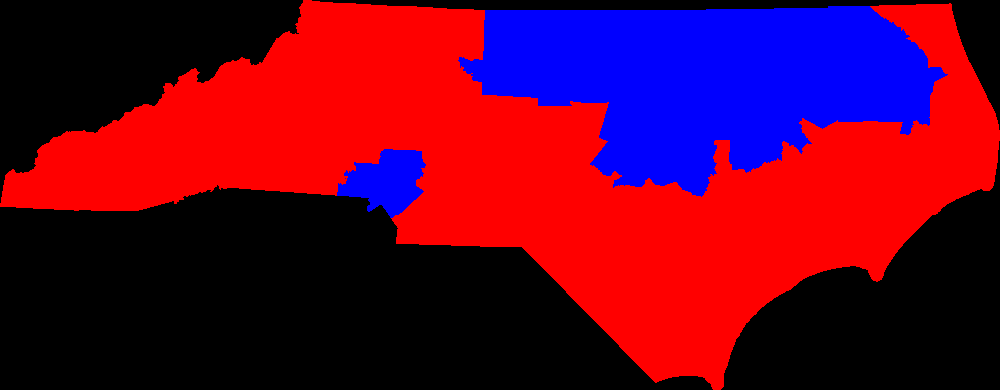}
        \caption{NC 2022 district map}
        \label{fig:nc2022dist}
    \end{minipage}%
    \hfill
    \begin{minipage}{0.45\textwidth}
        \centering
        \includegraphics[width=\textwidth]{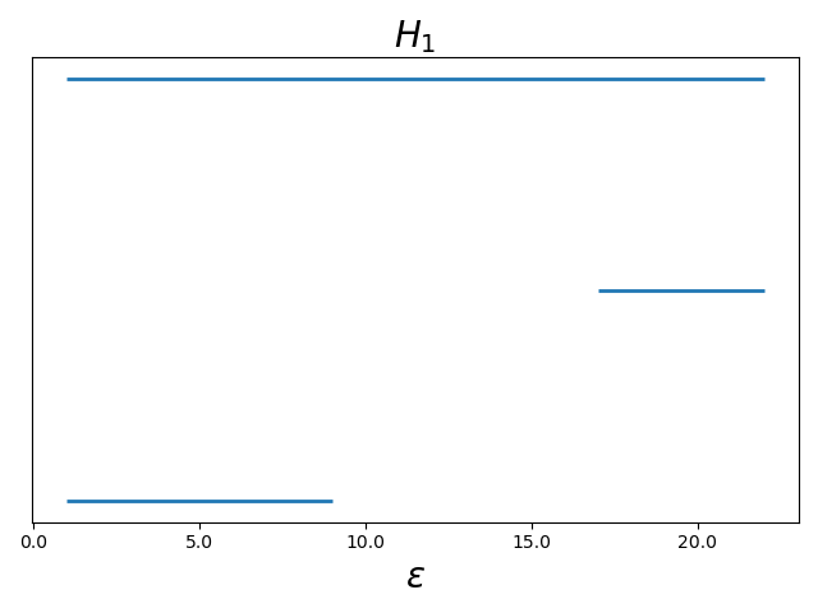}
        \caption{NC 2022 district barcode}
        \label{fig:nc2022distbar}
    \end{minipage}
\end{figure}

\paragraph{2024.} The 2023 maps (plans 746/747) were rated ``F'' by the Princeton Gerrymandering Project~\citep{Wang2024}. The district barcode shows only one persistent $H_1$ feature (District 12, containing Charlotte); the feature corresponding to the Raleigh metropolitan area (District 2 in the 2022 map) dies by filtration step 15, whereas the corresponding precinct features persist to step 20. This divergence is consistent with Democrats being cracked away from Raleigh: under the 2024 map, Republicans flipped 3 congressional seats, even though Don Davis (D) retained North Carolina's 1st District, demonstrating that the overall shift cannot be attributed solely to a national red wave.

\begin{figure}[tp]
    \centering
    \begin{minipage}{0.45\textwidth}
        \centering
        \includegraphics[width=\textwidth]{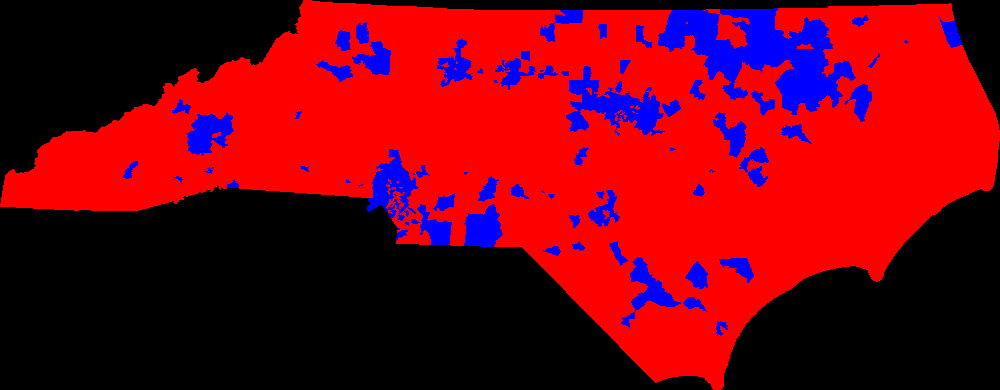}
        \caption{NC 2024 precinct map}
        \label{fig:nc2024prec}
    \end{minipage}%
    \hfill
    \begin{minipage}{0.45\textwidth}
        \centering
        \includegraphics[width=\textwidth]{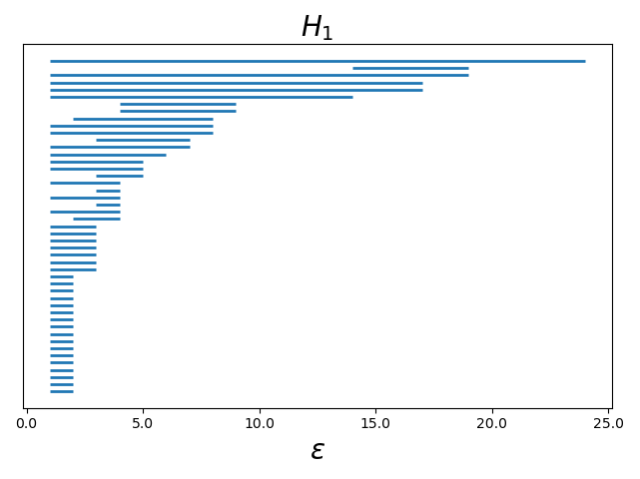}
        \caption{NC 2024 precinct barcode}
        \label{fig:nc2024precbar}
    \end{minipage}
    \begin{minipage}{0.45\textwidth}
        \centering
        \includegraphics[width=\textwidth]{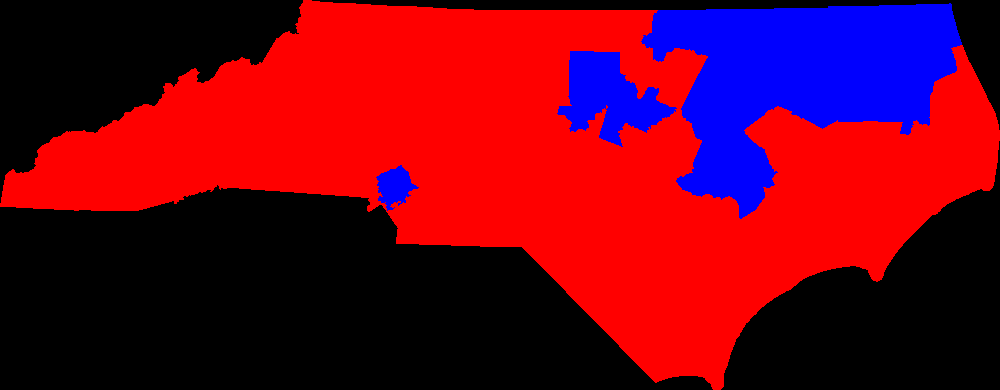}
        \caption{NC 2024 district map}
        \label{fig:nc2024dist}
    \end{minipage}%
    \hfill
    \begin{minipage}{0.45\textwidth}
        \centering
        \includegraphics[width=\textwidth]{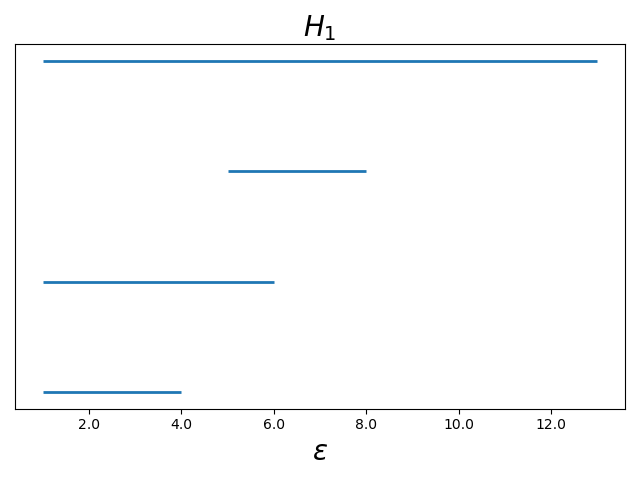}
        \caption{NC 2024 district barcode}
        \label{fig:nc2024distbar}
    \end{minipage}
\end{figure}

\subsection{Bottleneck Distance Results}

\begin{figure}[tp]
    \begin{minipage}{0.45\textwidth}
        \centering
        \includegraphics[width=\textwidth]{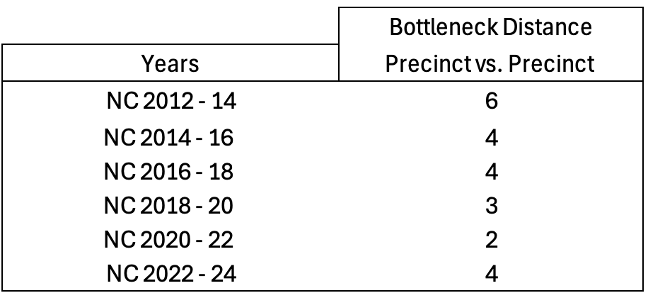}
        \caption{Precinct-to-precinct bottleneck distance, 2012--2024}
        \label{fig:bn_precinct}
    \end{minipage}%
    \hfill
    \begin{minipage}{0.45\textwidth}
        \centering
        \includegraphics[width=\textwidth]{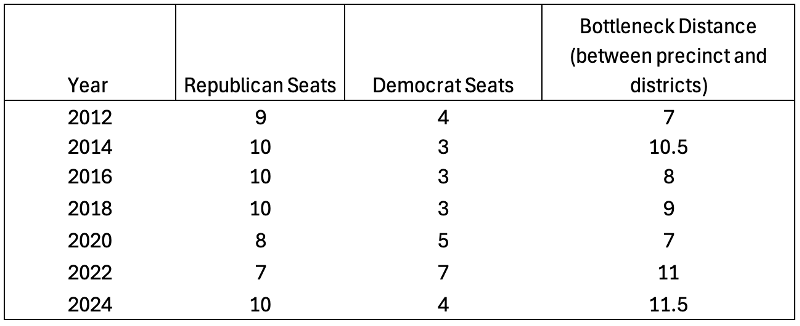}
        \caption{District-to-district bottleneck distance, 2012--2024}
        \label{fig:bn_district}
    \end{minipage}
\end{figure}

Figure~\ref{fig:bn_precinct} shows precinct-to-precinct bottleneck distances across consecutive election cycles (2012--2014, 2014--2016, etc.). These distances are consistently small, confirming that precinct-level voting patterns are stable--no dramatic demographic reshuffling occurred in North Carolina over this period. Figure~\ref{fig:bn_district} shows district-to-district bottleneck distances for consecutive election cycles. These distances are substantially larger, and the largest jumps occur precisely at redistricting years: 2014--2016 (following the 2016 court-ordered redraw) and 2022--2024 (following the 2023 enactment of plans 746/747). This pattern is the central empirical finding: redistricting, not voter behavior, drives district-level topological change.

\subsection{Comparison with Compactness Scores}

\begin{figure}[tp]
    \begin{minipage}{0.45\textwidth}
        \centering
        \includegraphics[width=\textwidth]{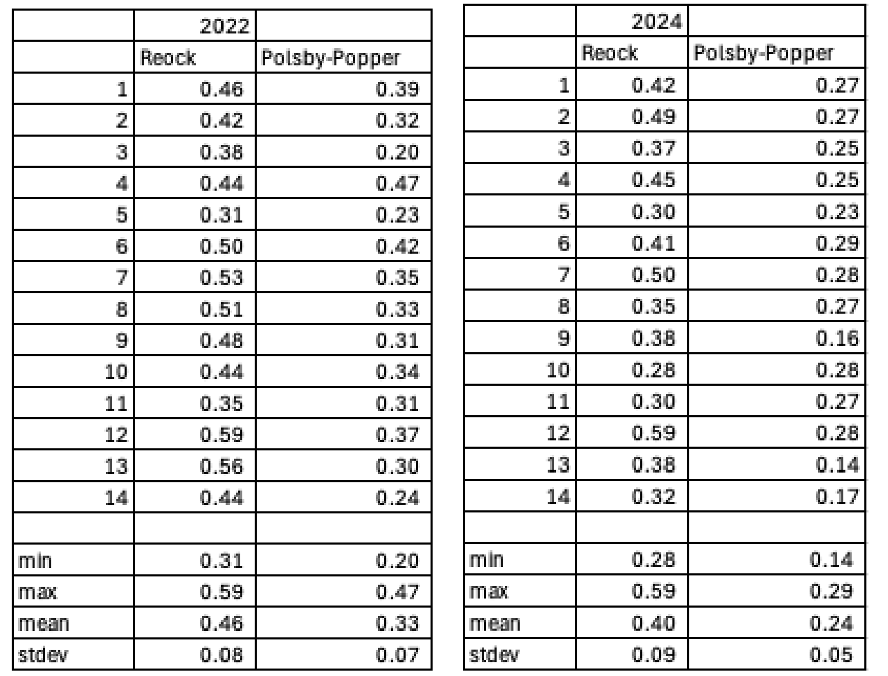}
        \caption{Polsby-Popper and Reock scores, NC 2022 and 2024}
        \label{fig:scores}
    \end{minipage}%
    \hfill
    \begin{minipage}{0.45\textwidth}
        \centering
        \includegraphics[width=\textwidth]{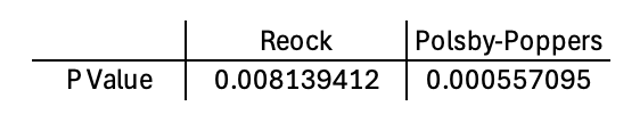}
        \caption{$p$-value for paired $t$-test of compactness scores}
        \label{fig:pvalue}
    \end{minipage}
\end{figure}

For comparison, we computed Polsby-Popper and Reock scores for the 2022 and 2024 North Carolina maps and applied a two-tailed paired $t$-test. The test finds no statistically significant difference between the two sets of scores (Figure~\ref{fig:pvalue}), even though Republicans flipped three congressional seats between these two maps. This demonstrates the insensitivity of compactness measures to the partisan impact of redistricting: a map can be geometrically compact while still systematically cracking or packing voters.

\section{Texas and California: 2024 vs.\ 2025 Proposed Maps}
\label{sec:txca}

\subsection{Texas: Background}

Texas has a long history of contested redistricting. Following the 2020 Census, new congressional maps were enacted in 2021 and immediately challenged under the Voting Rights Act. In late 2024, Texas lawmakers proposed a new congressional map for the 2026 cycle (the ``2025 plan''), which was subsequently struck down by a federal district court as a racial gerrymander. The Supreme Court nonetheless allowed the map to be used in the 2026 elections. This sequence of events--enactment, challenge, federal court ruling, Supreme Court stay--makes Texas an unusually concrete test case for quantifying racial gerrymandering.

\subsection{Texas 2016: Baseline}

Using 2016 precinct-level data and the 2016 congressional map, the precinct barcode exhibits several long-persistence $H_1$ features corresponding to the major urban Democratic strongholds: Houston, Dallas--Fort Worth, Austin, and San Antonio. The district barcode shows reduced persistence for most of these features, consistent with cracking. The Princeton Gerrymandering Project rates the 2016 map ``F''~\citep{Wang2024}.

\begin{figure}[tp]
    \centering
    \begin{minipage}{0.45\textwidth}
        \centering
        \includegraphics[width=\textwidth]{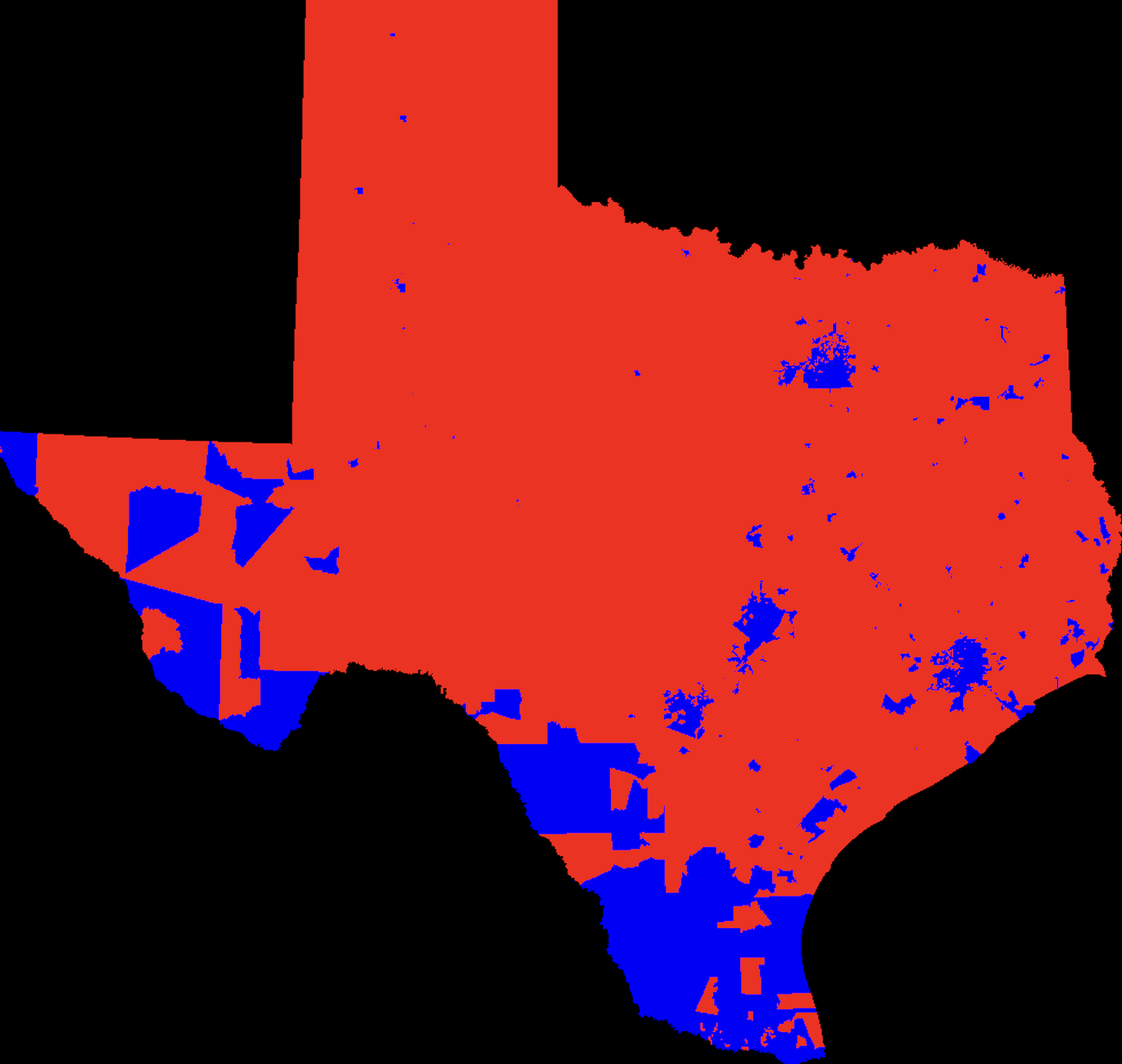}
        \caption{TX 2016 precincts (colored by winner)}
        \label{fig:tx2016prec}
    \end{minipage}%
    \hfill
    \begin{minipage}{0.45\textwidth}
        \centering
        \includegraphics[width=\textwidth]{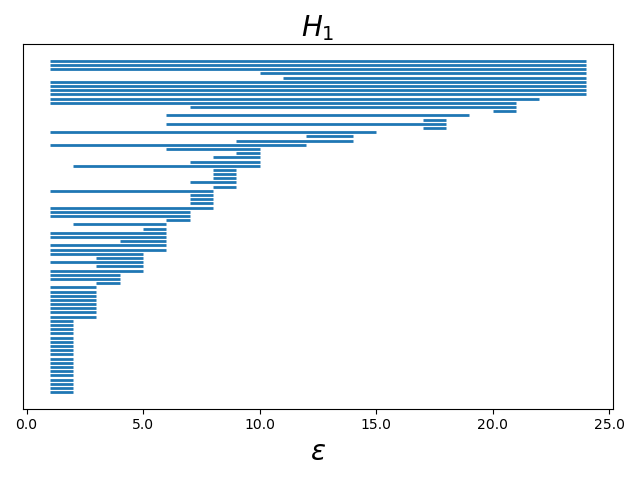}
        \caption{TX 2016 precinct barcode}
        \label{fig:tx2016precbar}
    \end{minipage}
    \begin{minipage}{0.45\textwidth}
        \centering
        \includegraphics[width=\textwidth]{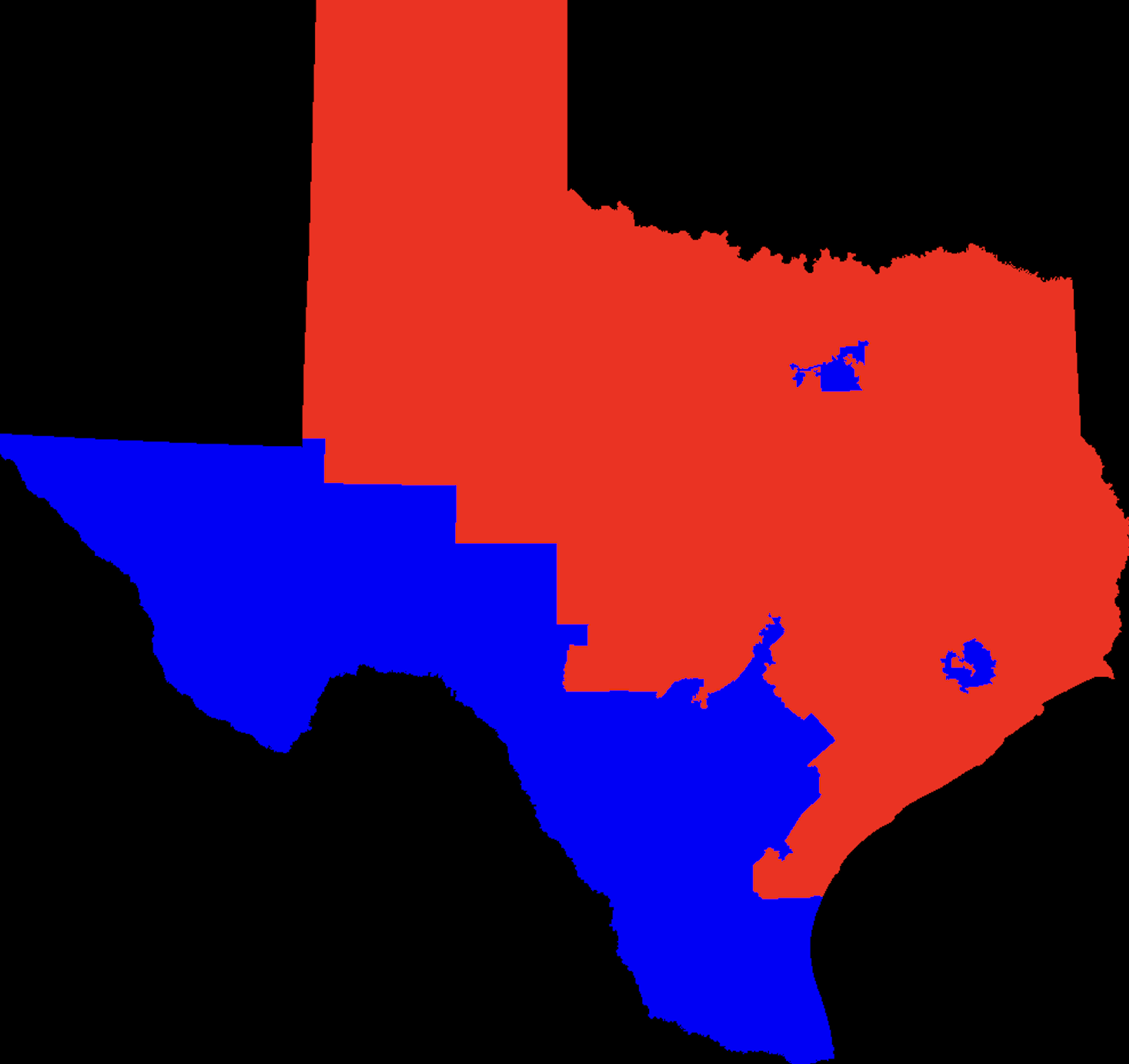}
        \caption{TX 2016 districts (colored by winner)}
        \label{fig:tx2016dist}
    \end{minipage}%
    \hfill
    \begin{minipage}{0.45\textwidth}
        \centering
        \includegraphics[width=\textwidth]{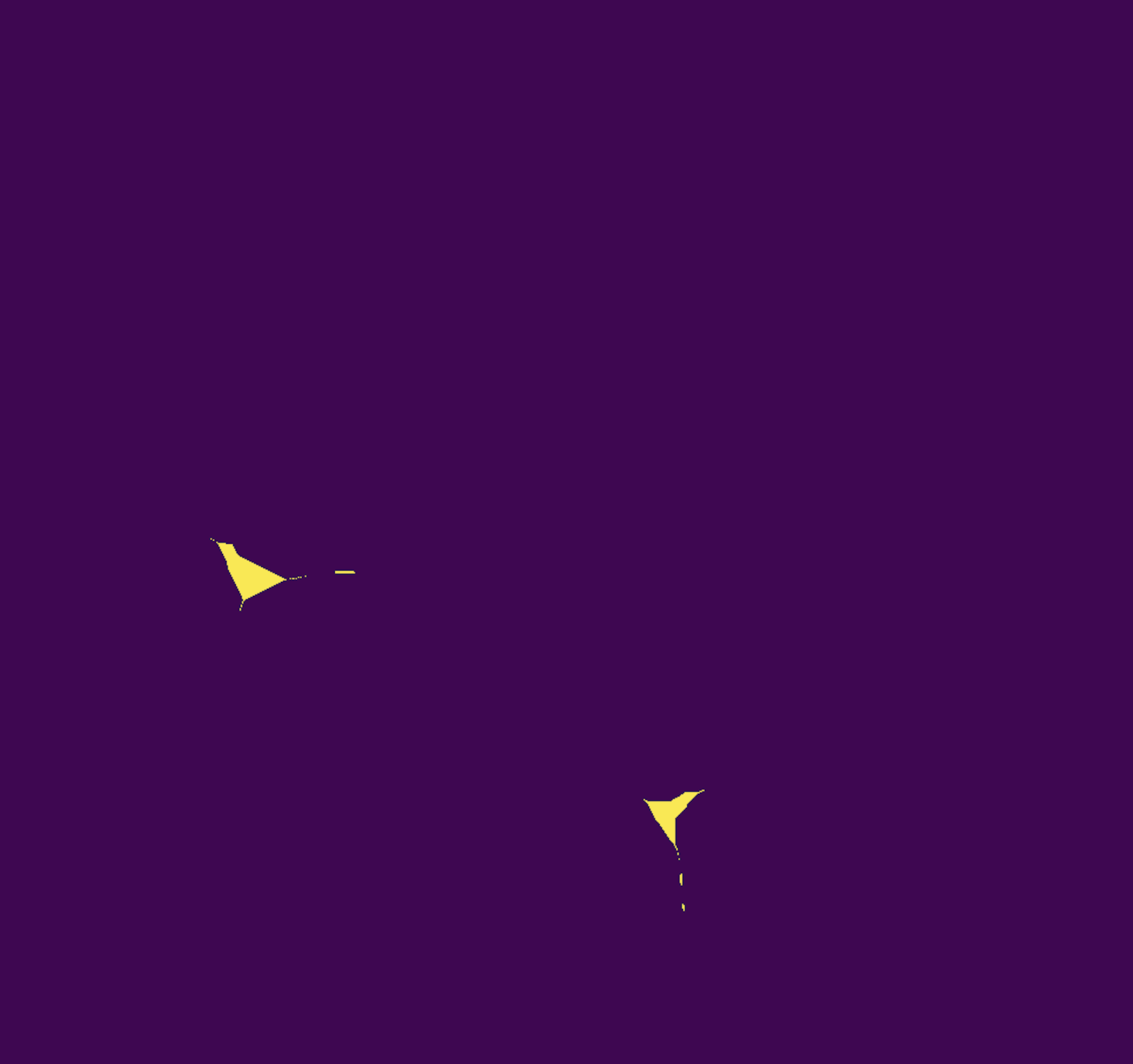}
        \caption{TX 2016 district: filtration step 22}
        \label{fig:tx2016distls}
    \end{minipage}
\end{figure}

\subsection{Texas 2024 vs.\ 2025}

We apply the same 2024 presidential election voting data to both the 2024 congressional map (the 2021 post-Census plan) and the proposed 2025 congressional map. This holds the underlying voter preferences constant, so that any difference in the barcodes is attributable entirely to the change in district boundaries.

Under the 2024 map, the precinct-to-district bottleneck distance is \textbf{8.00}. The 2024 precinct barcode shows several long-persistence features in the San Antonio and Austin metropolitan areas; the district barcode preserves some of these, though with reduced persistence. Notably, the San Antonio and Austin features appear as two separate components at the precinct level but are connected by a narrow district corridor along the interstate in the 2024 map--an early sign of packing.

Under the 2025 map (using identical 2024 voting data), the precinct-to-district bottleneck distance increases to \textbf{12.00}. The district barcode shows fewer persistent features than even the 2024 plan, despite identical underlying voter behavior. Approximately 748,000 Democratic votes--48\% of the total Democratic margin--are affected by the redistricting. This constitutes a topological quantification of the gerrymandering impact: the 2025 map moves $12.00 - 8.00 = 4.00$ additional bottleneck-distance units away from the precinct-level voting structure relative to the 2024 baseline.

\begin{figure}[tp]
\centering
\includegraphics[scale=0.5]{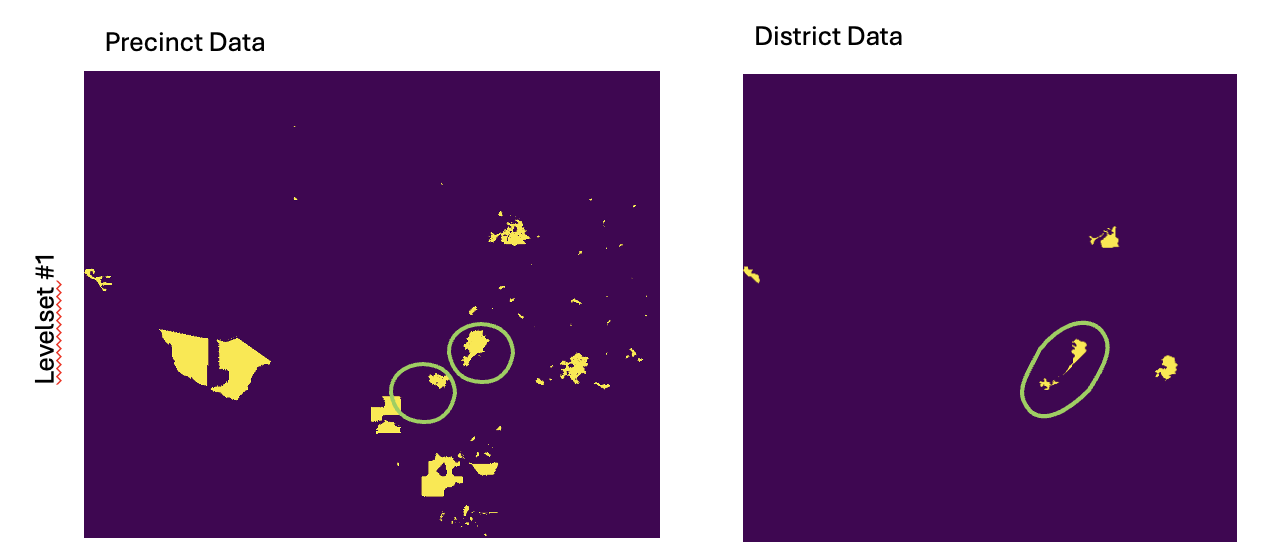}
\caption{Texas 2024 precinct and district rasters (filtration step 1 and 25)}
\label{fig:txraster}
\end{figure}

\begin{figure}[tp]
\centering
\includegraphics[scale=0.5]{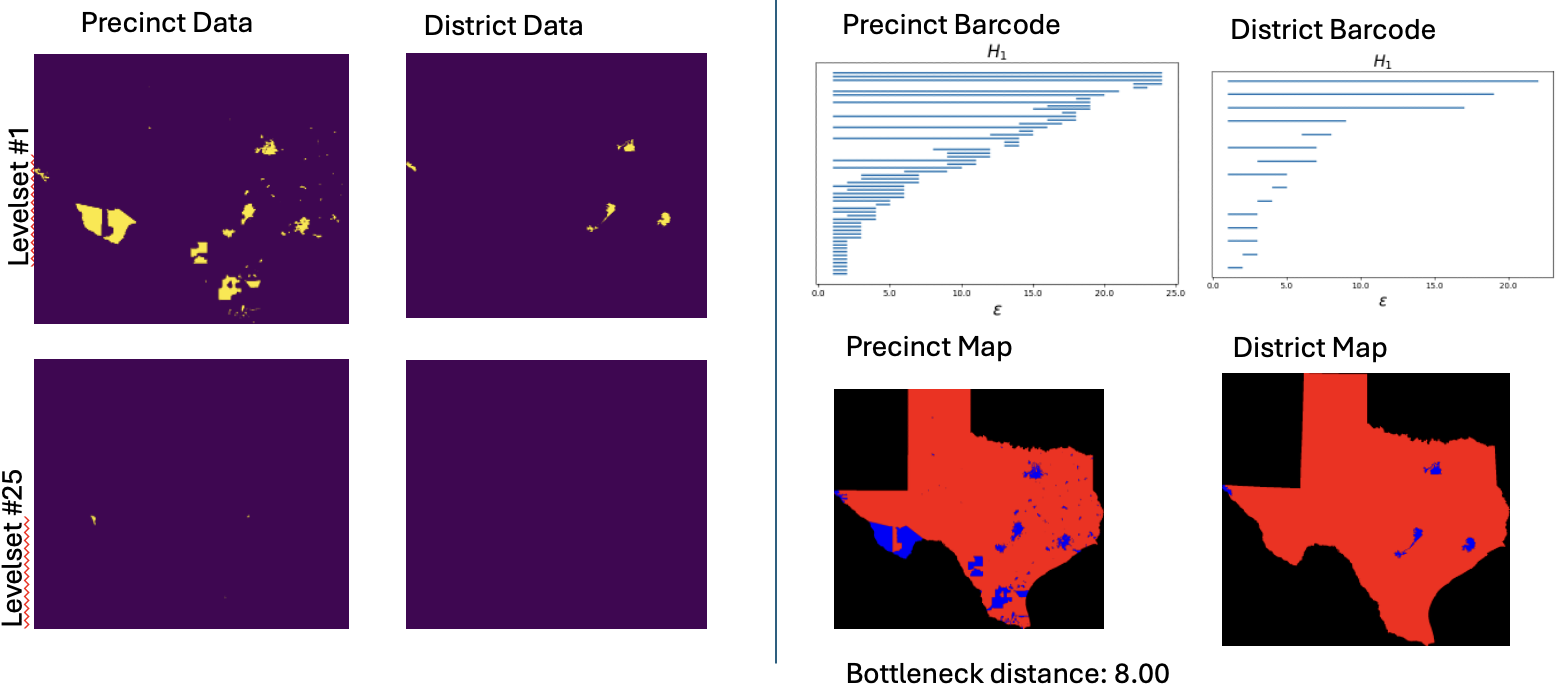}
\caption{Texas 2024 plan: precinct and district barcodes}
\label{fig:tx2024bar}
\end{figure}

\begin{figure}[tp]
\centering
\includegraphics[scale=0.5]{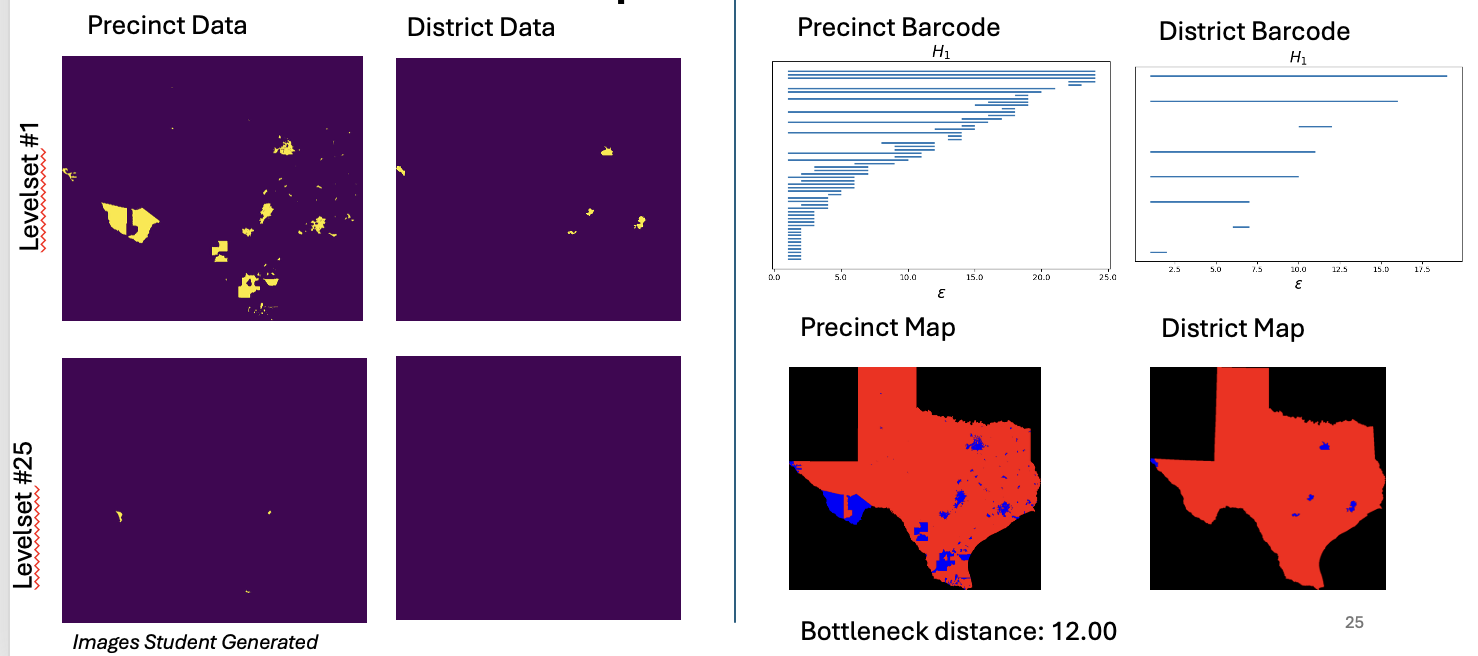}
\caption{Texas 2025 plan: precinct and district barcodes (same voting data)}
\label{fig:tx2025bar}
\end{figure}

\begin{figure}[tp]
\centering
\includegraphics[scale=0.5]{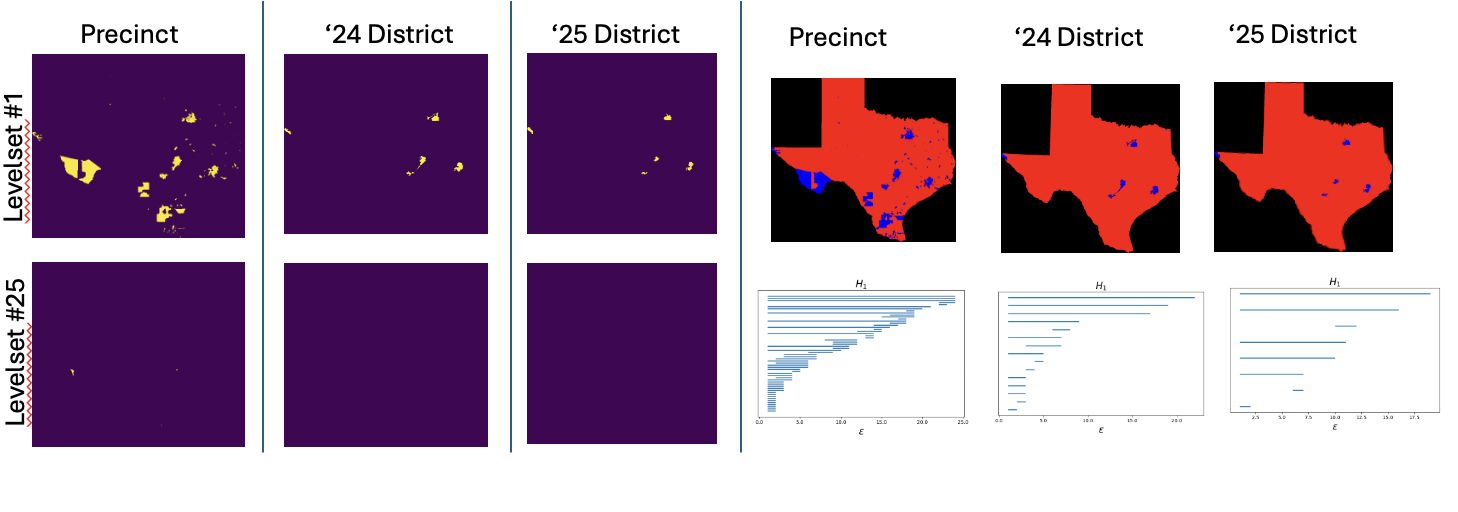}
\caption{Side-by-side comparison of Texas 2024 and 2025 plan barcodes}
\label{fig:txcomp}
\end{figure}

\subsection{Racial Gerrymandering in Texas}

The Texas 2025 plan was challenged as a racial gerrymander targeting five ``coalition districts'' where Black and Hispanic voters together form a majority. To evaluate this claim topologically, we apply the level-set pipeline to racial vote-margin rasters derived from 2020 Census BVAP and HVAP data.

\begin{figure}[tp]
\centering
\includegraphics[scale=0.5]{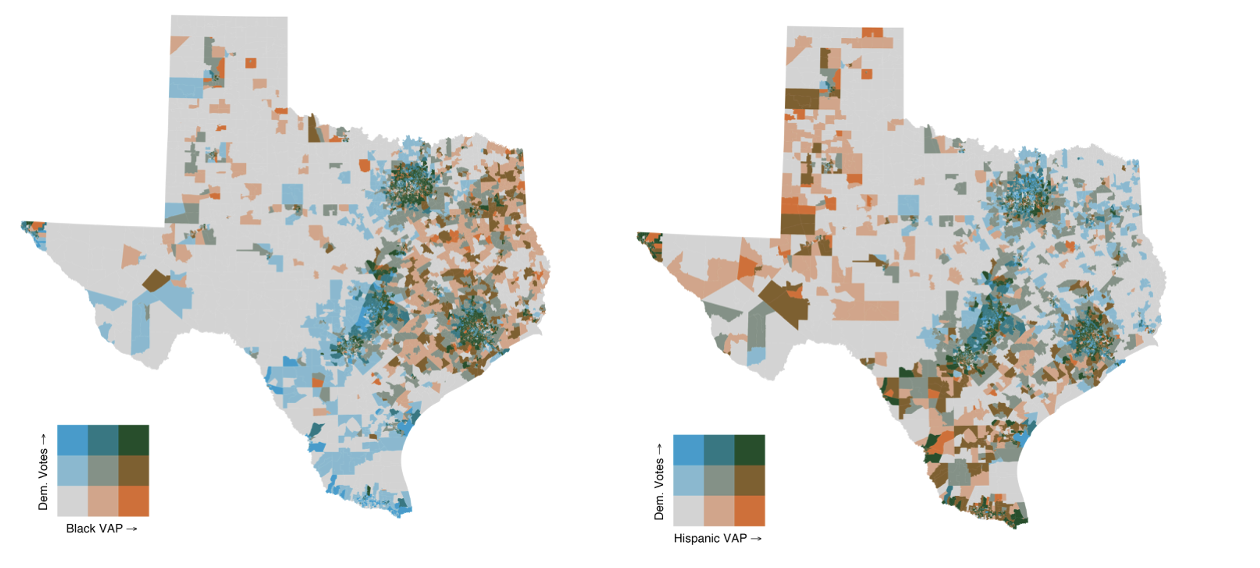}
\caption{Bivariate map of Democratic vote share and Black+Hispanic VAP in Texas (2024)}
\label{fig:txbivariante}
\end{figure}

Figure~\ref{fig:txbivariante} shows the strong geographic correlation between Democratic vote share and combined BVAP+HVAP, confirming that the partisan and racial signals are spatially aligned. We then compute the bottleneck distance between the precinct-level and district-level racial barcodes under both the 2024 and 2025 maps.

\begin{figure}[tp]
\centering
\includegraphics[scale=0.5]{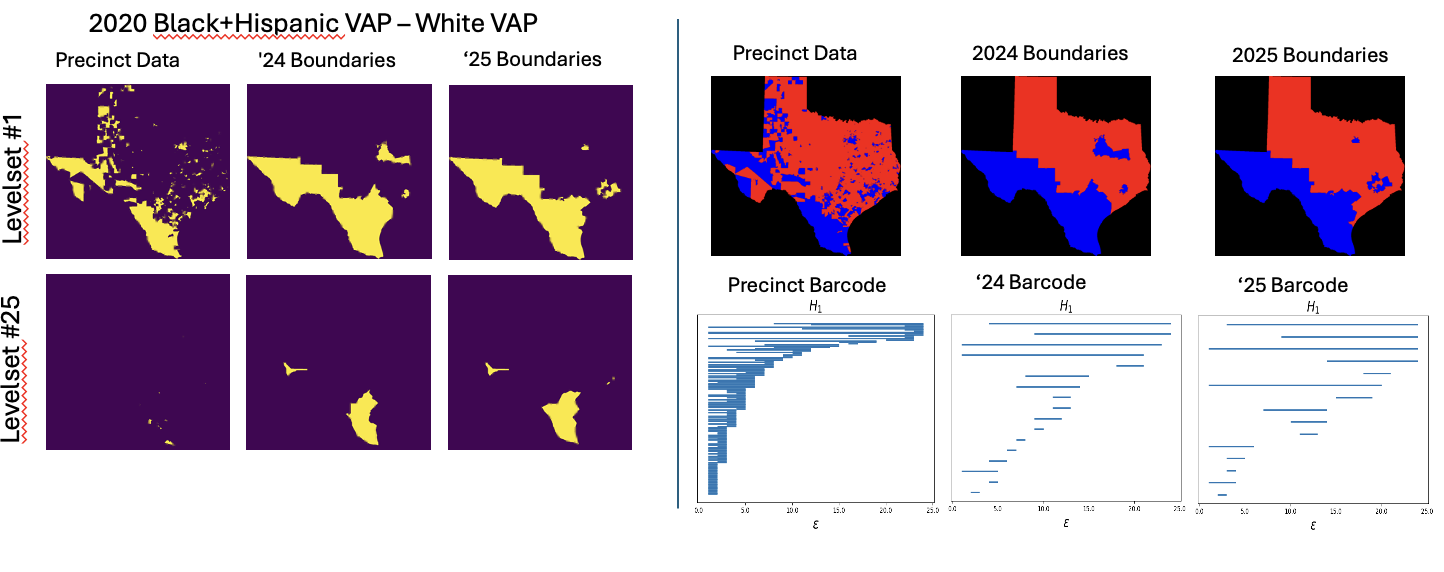}
\caption{Racial margin rasters for Texas: precinct level, 2024 district map, and 2025 district map, showing minority VAP (BVAP$+$HVAP) margin.}
\label{fig:txracial}
\end{figure}

The 2025 racial barcode shows more $H_1$ features at the district level than the 2024 plan, but with less total persistence--consistent with minority voters being packed into fewer, more concentrated districts. As a corroborating measure, we compute the precinct-split metric: the number of census blocks whose centroid falls in a precinct spanning two or more congressional districts. The 2024 plan has \textbf{0} such splits; the 2025 plan has \textbf{380} (using a 1-hectare overlap threshold). When census-block-level demographic data is used in drawing district lines, the resulting splits between census blocks and precincts leave a geometric fingerprint of racial data usage. The jump from 0 to 380 is strong evidence that racial demographics were used in constructing the 2025 map--consistent with the federal court's finding of racial gerrymandering, and providing an independent topological corroboration.

This constitutes the first application of persistent homology to the formal quantification of racial gerrymandering.

\subsection{California: Democratic Packing}

The California 2025 proposed congressional map (expected to be enacted before the 2026 midterms following Proposition 50) exhibits the opposite pattern from Texas: district-level persistence \emph{exceeds} precinct-level persistence, rather than falling below it.

\begin{figure}[tp]
\centering
\includegraphics[scale=0.5]{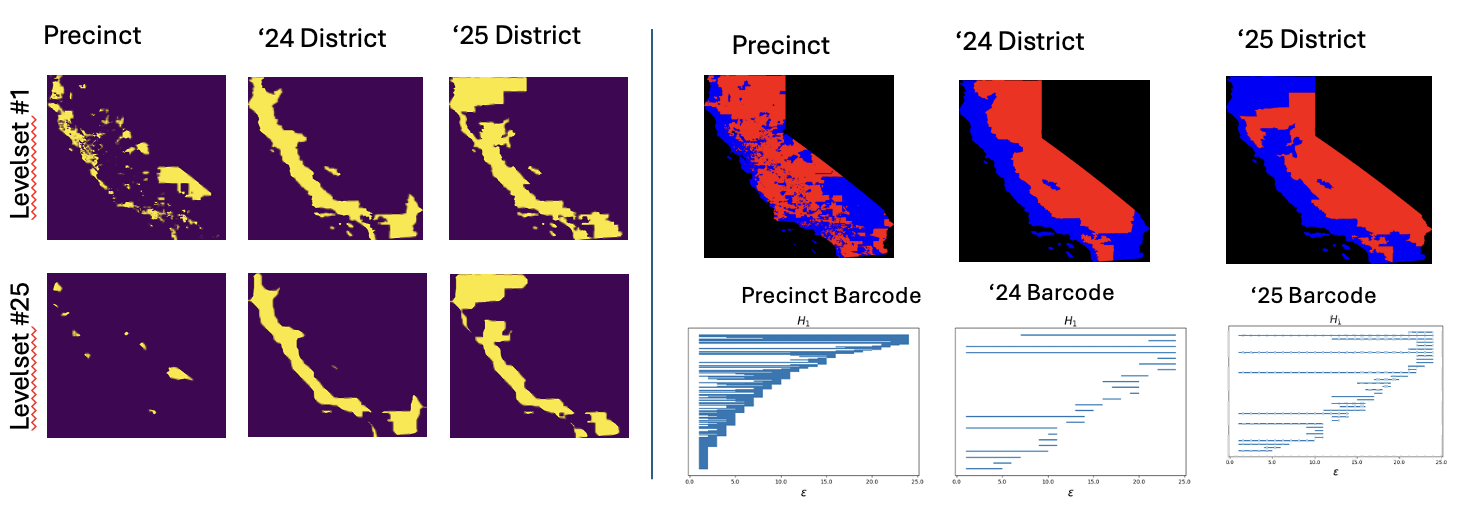}
\caption{California precinct, 2024 map, and 2025 map rasters and barcodes}
\label{fig:ca}
\end{figure}

In the California 2025 map, $H_1$ features appear in the district barcode that are not present at the precinct level. This is consistent with packing: district lines have been drawn to concentrate Republican voters into fewer, more ``efficient'' Republican-majority districts, thereby increasing Democratic dominance in the remaining districts. The surplus district-level persistence reflects the artificially created Republican ``islands'' that do not correspond to organic concentrations of Republican voters at the precinct level.

\subsection{Total Persistence}

\begin{figure}[tp]
\centering
\includegraphics[scale=0.5]{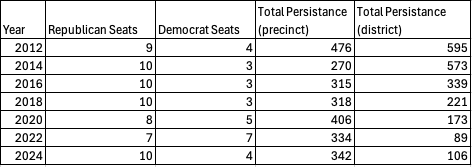}
\caption{Total persistence for North Carolina precinct and district barcodes, 2012--2024.}
\label{fig:totalpersistence}
\end{figure}

As a supplementary measure, we also compute the \emph{total persistence} of each barcode, which aggregates the lifespan of all features within a single barcode. For a barcode $B$ with points $(b_i, d_i)$, the total persistence with respect to the $p$-norm is~\citep{Kauba2023}:
\[
\mathrm{TP}(B) = \sum_{i} (d_i - b_i)^p.
\]
Infinite death times are replaced by the maximum filtration value. Figure~\ref{fig:totalpersistence} shows total persistence for all barcodes from 2012--2024. This metric requires further investigation to determine optimal thresholds for gerrymandering identification, and is presented here as a direction for future work.

\section{Multi-Metric Validation: Texas 2024 vs.\ 2025}
\label{sec:validation}

The bottleneck distance alone, while theoretically well-grounded, constitutes a single summary statistic. To validate the gerrymandering findings for Texas and to strengthen the case for publication, we compute four additional topological summaries--persistence diagrams, Betti-1 curves, persistence landscapes/silhouettes, and persistence images--for the Texas 2024 and 2025 congressional maps using the same 2024 precinct voting data. Each metric captures a different aspect of the precinct--district divergence, and their convergent agreement constitutes robust, multi-method evidence.

\subsection{Persistence Diagrams}

\begin{figure}[tp]
\centering
\includegraphics[width=\textwidth]{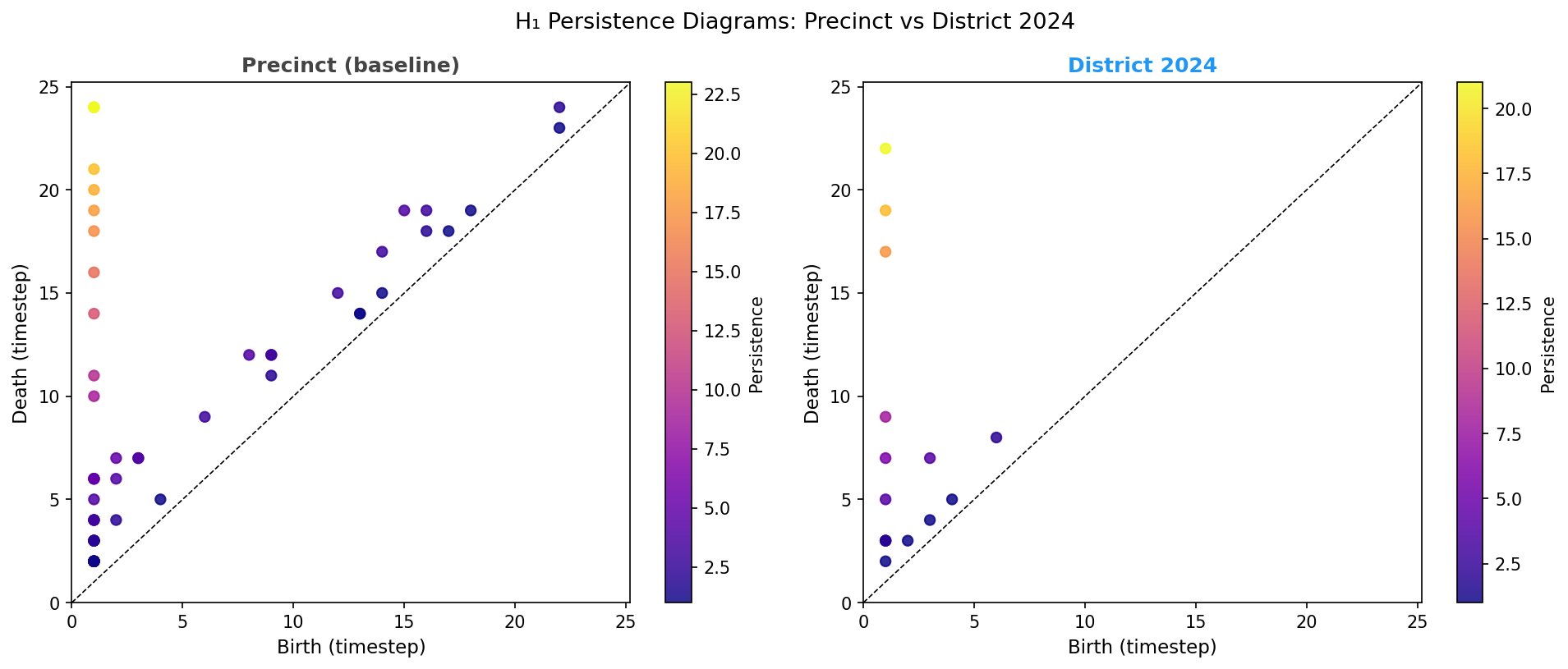}
\caption{$H_1$ persistence diagrams, Texas 2024: precinct (baseline) vs.\ district. Points far from the diagonal represent long-lived Democratic voting islands. The district diagram retains fewer such points, indicating topological suppression of Democratic strongholds.}
\label{fig:tx_pd_2024}
\end{figure}

\begin{figure}[tp]
\centering
\includegraphics[width=\textwidth]{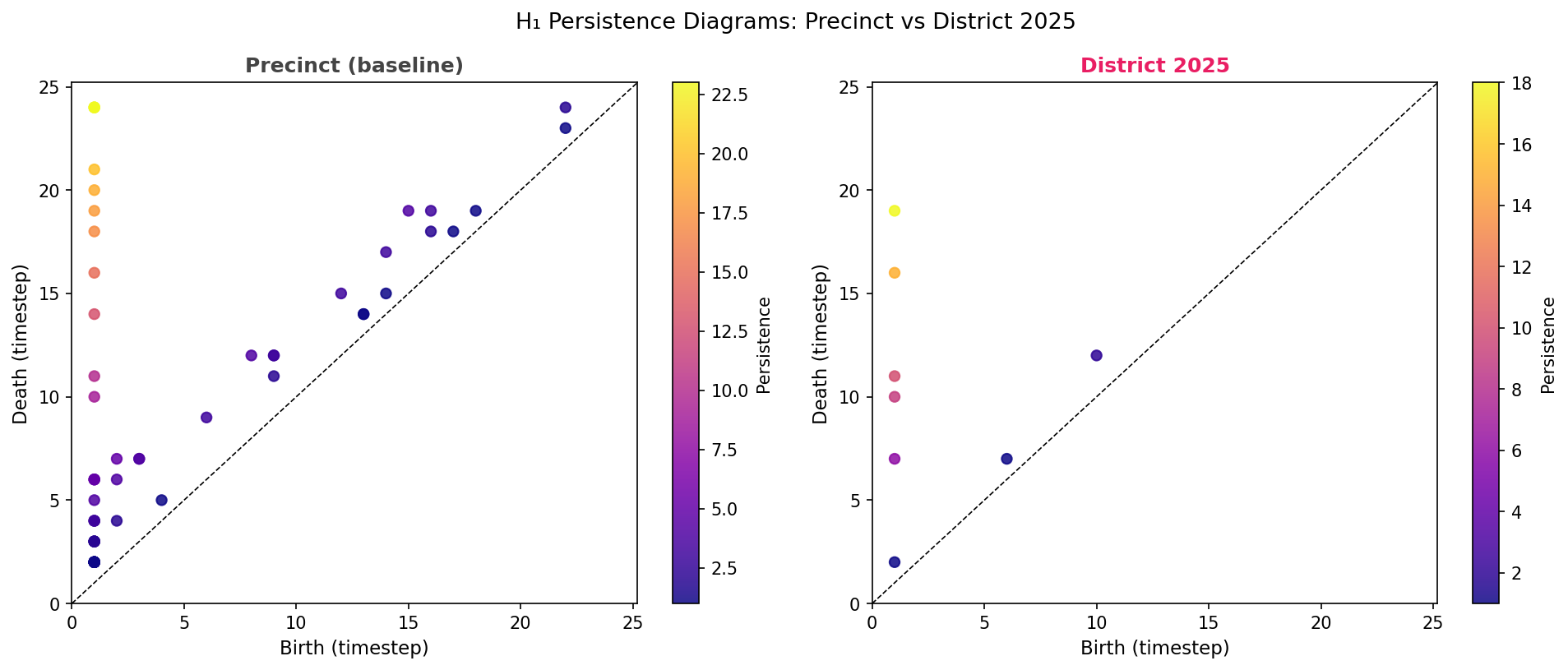}
\caption{$H_1$ persistence diagrams, Texas 2025 (same voting data). Further reduction in persistent district-level features relative to both the precinct baseline and the 2024 map.}
\label{fig:tx_pd_2025}
\end{figure}

Figure~\ref{fig:tx_pd_2024} shows the $H_1$ persistence diagrams for Texas under the 2024 map. The precinct diagram (left) contains many points far from the diagonal, indicating numerous long-lived Democratic voting islands. The district diagram (right) retains only a handful of high-persistence points, with many precinct features either eliminated or shifted near the diagonal. The 2025 map (Figure~\ref{fig:tx_pd_2025}) shows further reduction, with fewer persistent points surviving at the district level.

\subsection{Betti-1 Curves}

\begin{figure}[tp]
\centering
\includegraphics[width=\textwidth]{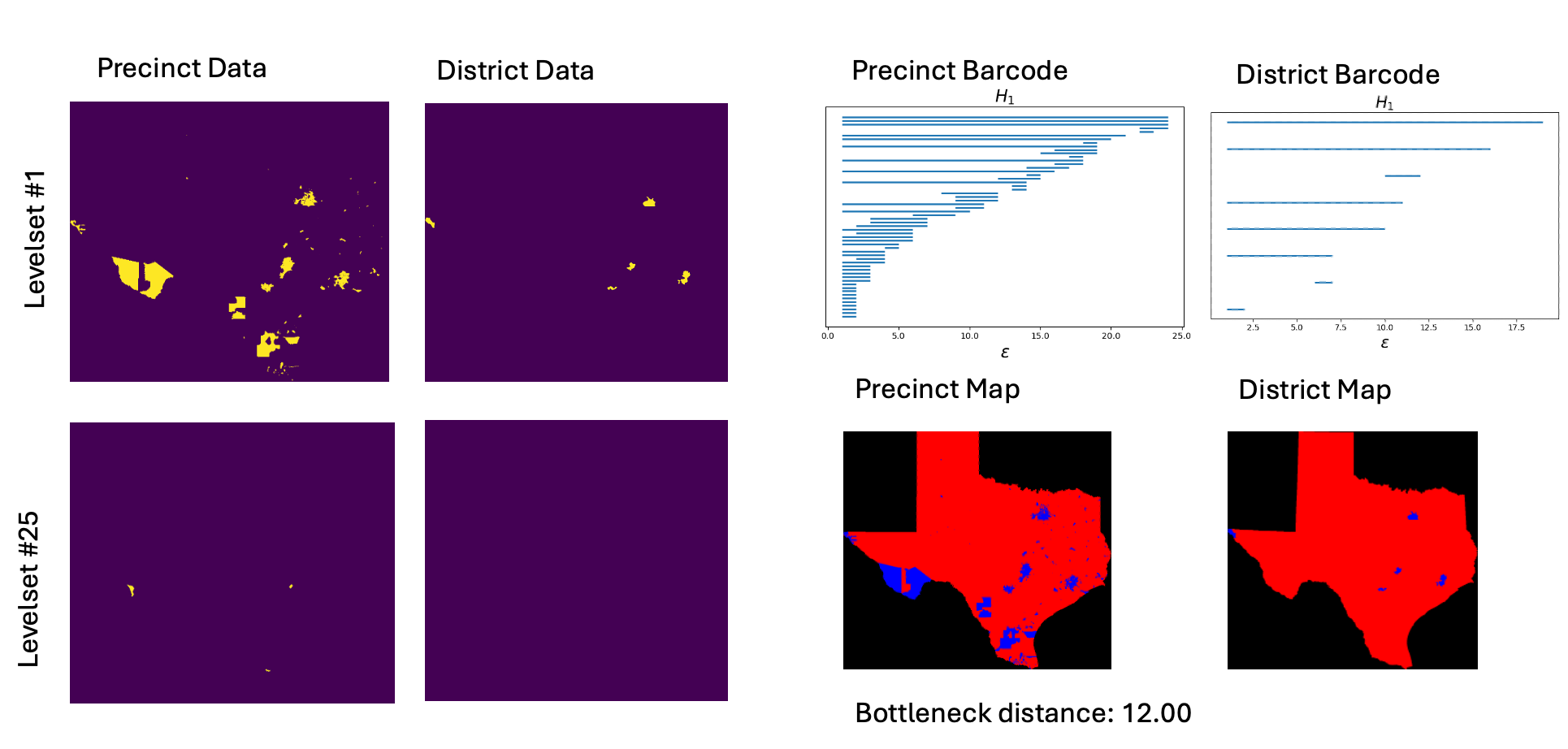}
\caption{Betti-1 curves $\beta_1(t)$ for Texas 2024: precinct (black) vs.\ district (blue dashed). The district curve tracks the precinct baseline in early filtration steps but diverges from step~8 onward, consistent with the bottleneck distance of 8.00.}
\label{fig:tx_betti_2024}
\end{figure}

Figure~\ref{fig:tx_betti_2024} shows the Betti-1 curves $\beta_1(t)$ for Texas under the 2024 map. The precinct curve (black) peaks at approximately 35 active $H_1$ features during the mid-filtration steps, reflecting the many Democratic urban islands across the state. The district curve (blue dashed) tracks the precinct curve closely in the early filtration steps but diverges significantly from step 8 onward, consistent with the bottleneck distance of 8.00: the district map preserves only the most strongly Democratic features and eliminates the rest.

\begin{figure}[tp]
\centering
\includegraphics[width=\textwidth]{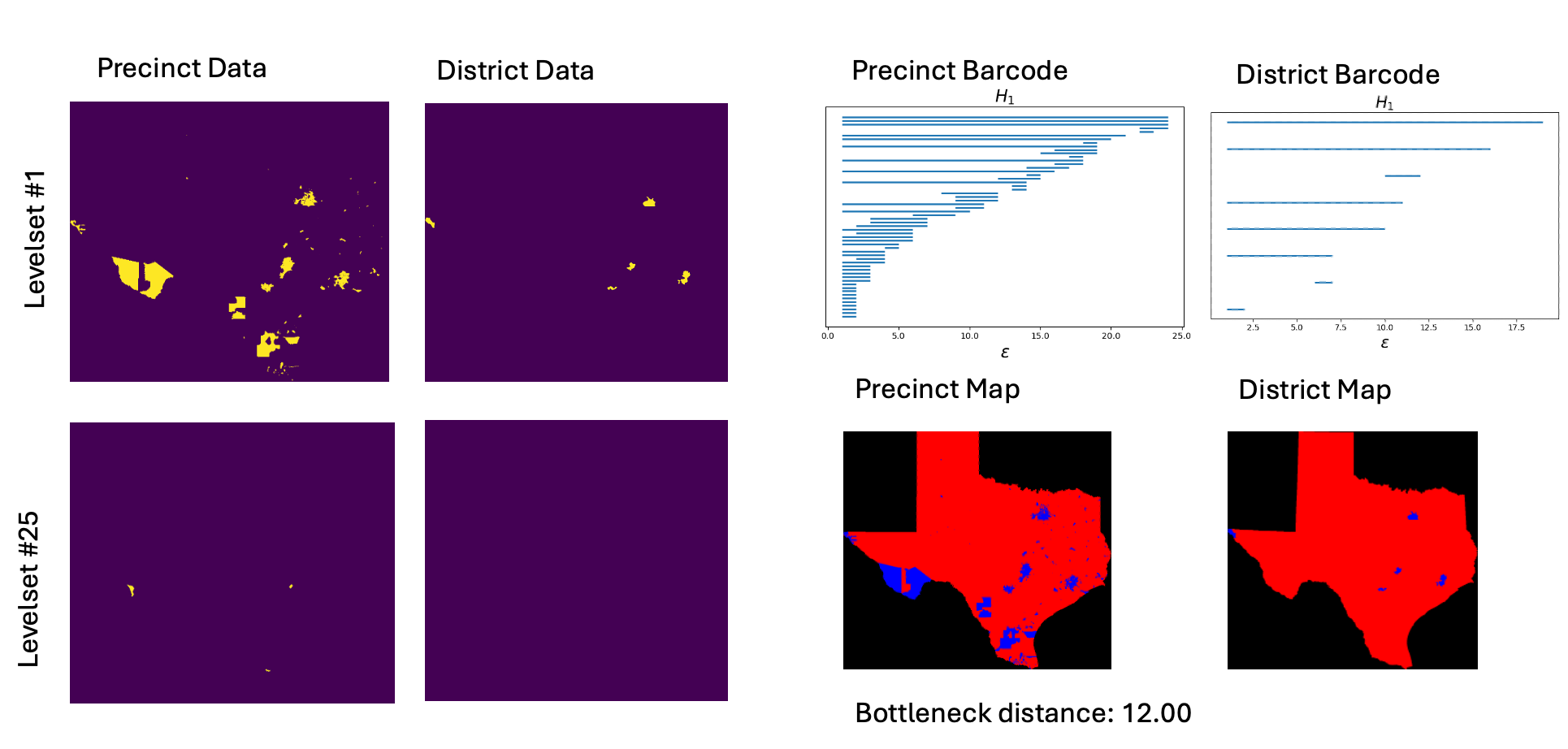}
\caption{Betti-1 curves $\beta_1(t)$ for Texas 2025 (same voting data). The district curve (red dashed) diverges from the precinct baseline earlier and more steeply than in 2024, consistent with the higher bottleneck distance of 12.00.}
\label{fig:tx_betti_2025}
\end{figure}

Under the 2025 map (Figure~\ref{fig:tx_betti_2025}), the district Betti-1 curve (red dashed) diverges from the precinct baseline earlier and more steeply, consistent with the higher bottleneck distance of 12.00. The gap between precinct and district curves is visibly larger throughout the mid-filtration range, quantifying the additional topological suppression introduced by the 2025 redistricting.

\subsection{Persistence Landscapes and Silhouettes}

\begin{figure}[tp]
\centering
\includegraphics[width=\textwidth]{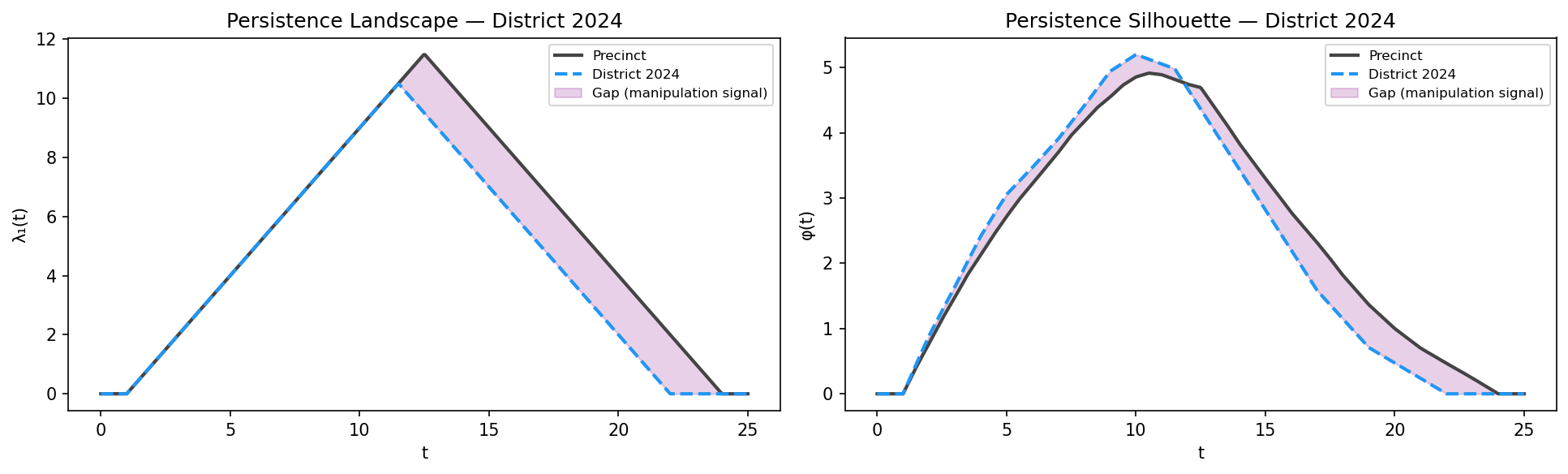}
\caption{Persistence landscape $\lambda_1(t)$ (left) and silhouette $\varphi(t)$ (right) for Texas 2024: precinct (black) vs.\ district (blue dashed). The shaded gap is the manipulation signal--topological persistence suppressed by redistricting. The gap concentrates in $t \in [12,25]$, corresponding to the most persistent Democratic features.}
\label{fig:tx_ls_2024}
\end{figure}

Figure~\ref{fig:tx_ls_2024} shows the first persistence landscape $\lambda_1(t)$ and persistence silhouette $\varphi(t)$ for Texas under the 2024 map, overlaid for the precinct (black) and district (blue dashed). The shaded region between the two curves is the \emph{manipulation signal}--the portion of topological persistence that is present at the precinct level but destroyed by the district map.

\begin{figure}[tp]
\centering
\includegraphics[width=\textwidth]{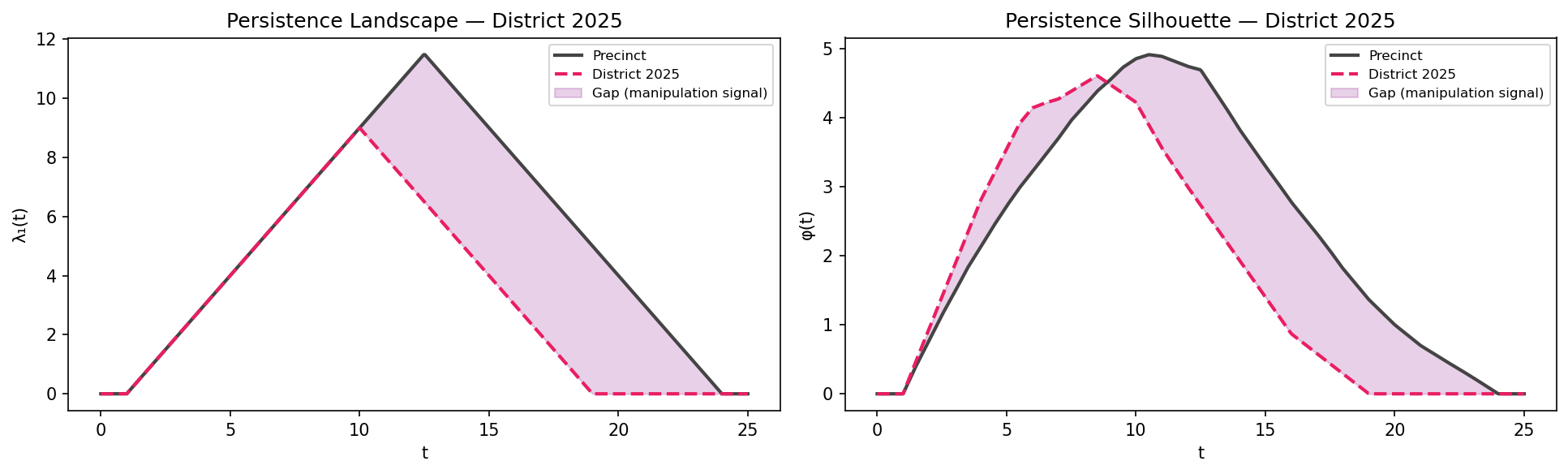}
\caption{Persistence landscape and silhouette for Texas 2025. The gap is larger and begins earlier ($t \approx 10$) than in 2024, indicating that the 2025 map suppresses Democratic features that were preserved under the 2024 plan.}
\label{fig:tx_ls_2025}
\end{figure}

Under the 2024 map, the landscape gap is concentrated in the range $t \in [12, 25]$, indicating that the most persistent Democratic features (those born in early filtration steps and surviving to late steps) are the ones most suppressed. Under the 2025 map (Figure~\ref{fig:tx_ls_2025}), the gap is both larger and shifted earlier (beginning around $t = 10$), showing that the 2025 redistricting suppresses Democratic features that were still preserved under the 2024 map.

Because persistence landscapes lie in $L^2(\mathbb{R})$~\citep{Bubenik2015}, the $L^2$ norm of the gap function $\|\lambda_1^{\text{prec}} - \lambda_1^{\text{dist}}\|_2$ provides a scale-invariant measure of gerrymandering severity amenable to hypothesis testing. This is a direction for future work (Section~\ref{sec:futurework}).

\subsection{Persistence Images}

\begin{figure}[tp]
\centering
\includegraphics[width=\textwidth]{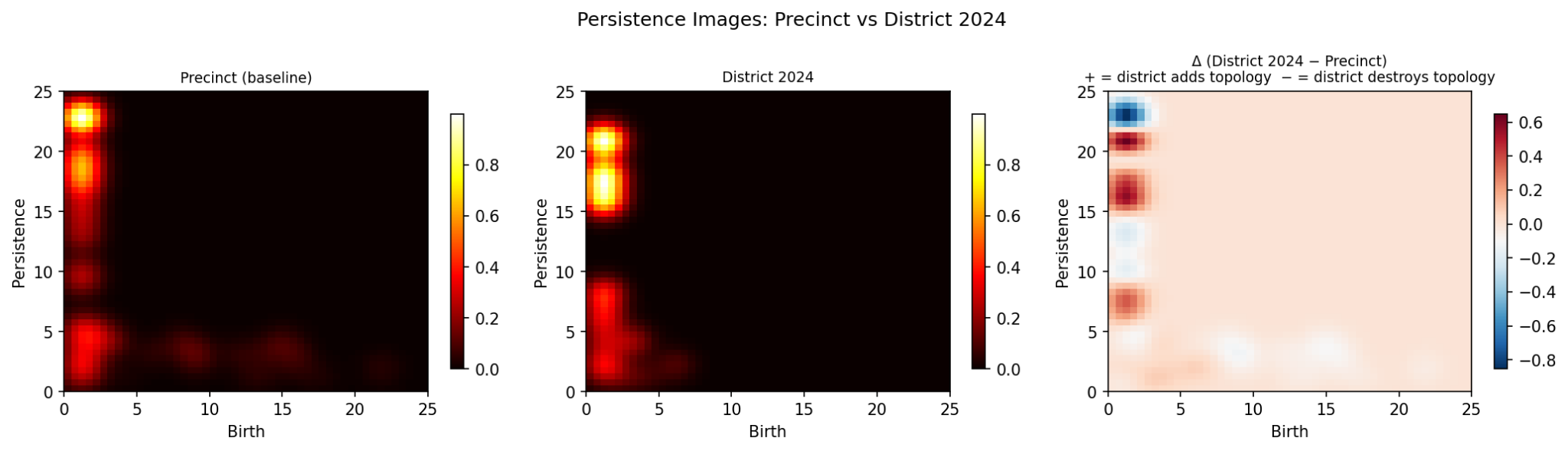}
\caption{Persistence images for Texas 2024: precinct (left), district (centre), and difference $\Delta I$ (right). Blue in $\Delta I$ indicates topological features destroyed by redistricting (cracking); red indicates features added (packing). The 2024 map primarily suppresses the highest-persistence features.}
\label{fig:tx_pi_2024}
\end{figure}

Persistence images (Figure~\ref{fig:tx_pi_2024}) provide a pixel-level view of which birth--persistence regions are gained or lost under redistricting. The difference image $\Delta I = I_{\text{district}} - I_{\text{precinct}}$ uses a diverging colormap: blue regions ($-$) represent topological features present at the precinct level that are destroyed by the district map (cracking), while red regions ($+$) represent features artificially introduced at the district level (packing).

\begin{figure}[tp]
\centering
\includegraphics[width=\textwidth]{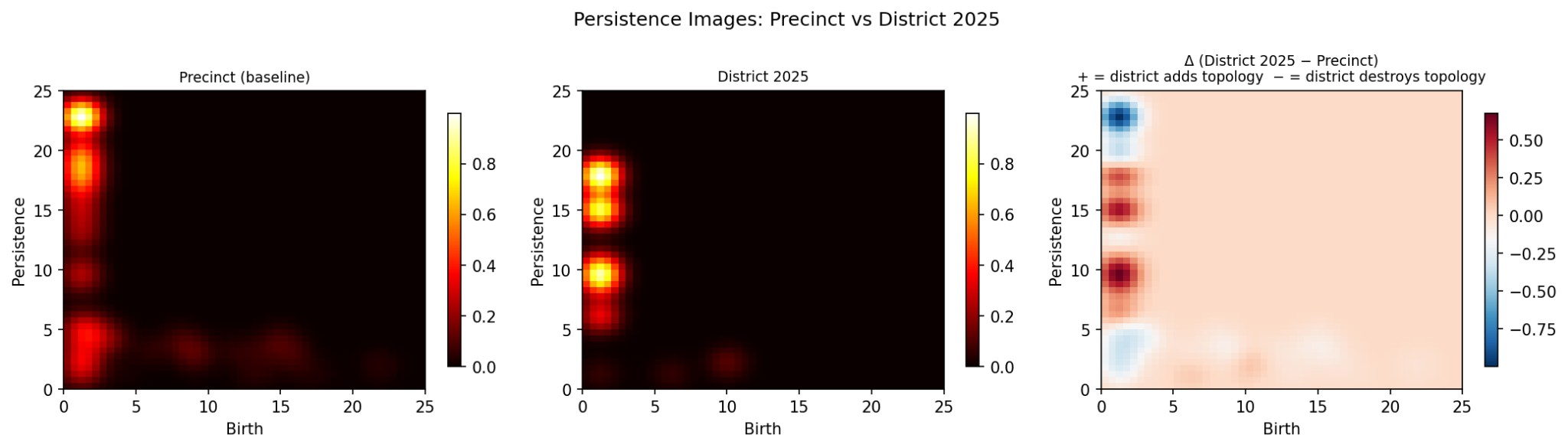}
\caption{Persistence images for Texas 2025. The difference image shows broader and more intense blue signal than 2024, confirming that the 2025 plan eliminates a wider range of Democratic topological features. The absence of red signal rules out packing as the primary mechanism.}
\label{fig:tx_pi_2025}
\end{figure}

Under the 2024 map (Figure~\ref{fig:tx_pi_2024}), the difference image shows consistent blue signal in the high-persistence region (birth $\approx 1$--$3$, persistence $\approx 15$--$25$), meaning the most robust Democratic urban cores are partially suppressed. Under the 2025 map (Figure~\ref{fig:tx_pi_2025}), the blue signal extends across a broader persistence range, with more intense suppression in the persistence range $5$--$15$--features of medium robustness that survived under the 2024 map are eliminated under the 2025 plan. No significant red regions appear in either difference image, confirming that the 2025 redistricting is primarily a cracking operation rather than a packing one.

\subsection{Convergent Validity}

Table~\ref{tab:metrics} summarises the five topological metrics for Texas under the 2024 and 2025 maps, all computed on the same 2024 voting data. Every metric shows a consistent signal: the 2025 map is topologically further from the precinct baseline than the 2024 map, confirming that the redistricting--not underlying voter behaviour--drives the divergence.

\begin{table}[H]
\centering
\small
\begin{tabular}{lcc}
\toprule
\textbf{Metric} & \textbf{2024 Map} & \textbf{2025 Map} \\
\midrule
Bottleneck distance $d_B$ & 8.00 & 12.00 \\
Votes shifted (est.) & ${\sim}320{,}000$ & ${\sim}748{,}000$ \\
$L^\infty$ landscape gap $\|\lambda_1^P - \lambda_1^D\|_\infty$ & 1.15 & 3.07 \\
$L^2$ silhouette gap $\|\varphi^P - \varphi^D\|_2$ & 0.81 & 1.94 \\
Precinct--census block splits & 0 & 380 \\
\bottomrule
\end{tabular}
\caption{Summary of topological metrics for Texas 2024 vs.\ 2025 congressional maps (same 2024 voting data). All five metrics show consistent directional change, providing convergent multi-method evidence of increased gerrymandering under the 2025 plan.}
\label{tab:metrics}
\end{table}

The convergence across five independent metrics--bottleneck distance, persistence diagrams, Betti curves, landscape/silhouette norms, persistence images, and the precinct-split count--creates a stronger conclusion. This validation framework  demonstrates that persistent homology produces consistent, reproducible gerrymandering signals across the full range of topological summary statistics currently in use.

\section{Louisiana: \textit{Louisiana v.\ Callais} (2025--2026)}
\label{sec:louisiana}

\subsection{Background}

\textit{Louisiana v.\ Callais} (2025) represents a landmark case at the intersection of partisan and racial gerrymandering. Following the Supreme Court's ruling in \textit{Allen v.\ Milligan} (2023) that Louisiana's 2021 congressional map violated Section~2 of the Voting Rights Act by diluting Black voting power, the state was ordered to draw a remedial map creating a second majority-Black district. The resulting remedial map was then challenged by white plaintiffs as an unconstitutional racial gerrymander under \textit{Shaw v.\ Reno} (1993).

The Supreme Court's decision in \textit{Callais} significantly curtailed the scope of Section~2 by striking down the remedial map--effectively limiting states' ability to create new majority-minority districts. The case illustrates a fundamental tension: the line between constitutionally required race-consciousness (to remedy VRA violations) and constitutionally prohibited race-consciousness (racial gerrymandering) is difficult to draw without a geometry-independent, quantitative framework.

\subsection{Topological Analysis}

We apply the level-set pipeline to the Louisiana 2026 congressional map using 2024 presidential voting data and 2020 Census BVAP data. As in the Texas racial analysis (Section~\ref{sec:txca}), we construct two parallel rasters: one using partisan vote margin ($V_b - V_r$ per unit area) and one using racial margin ($\mathrm{BVAP} - \mathrm{WVAP}$ per unit area).

The key finding is a bottleneck distance of \textbf{3.0} between the partisan and racial persistence diagrams at the district level. This small distance indicates that the spatial distribution of Democratic voting strength and Black VAP are topologically nearly identical at the district level--the two maps produce barcodes with the same persistent features in the same locations. In other words, the districts that are Democratic-majority are essentially the same districts that are majority-Black, and vice versa.

\begin{figure}[tp]
\centering
\includegraphics[width=\textwidth]{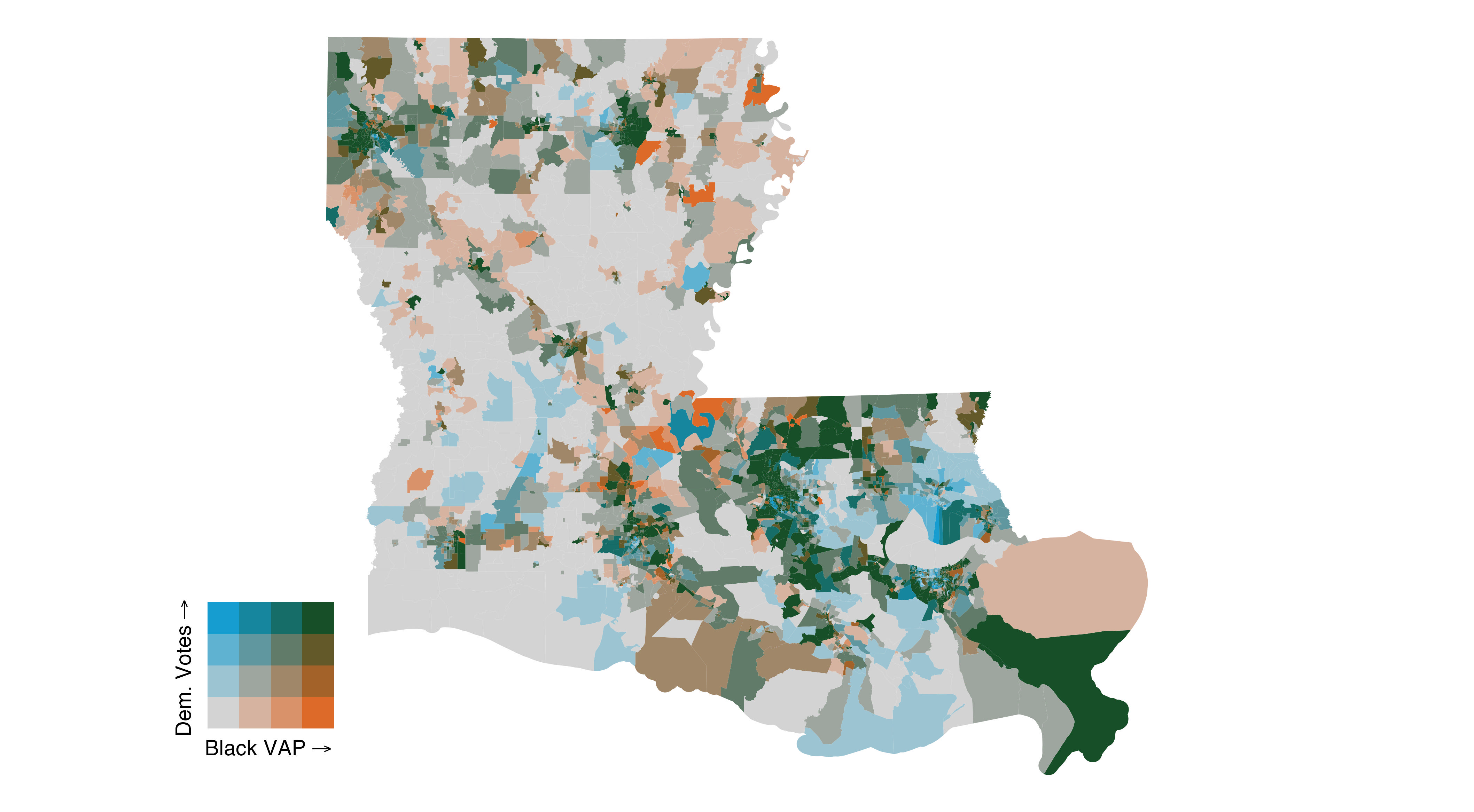}
\caption{Bivariate map of Democratic vote share and Black VAP for Louisiana precincts. Dark green (top-right) indicates precincts with both high Democratic vote share and high Black VAP, concentrated along the Mississippi River corridor and in the New Orleans metropolitan area.}
\label{fig:la_bivariate}
\end{figure}

\begin{figure}[tp]
\centering
\includegraphics[width=\textwidth]{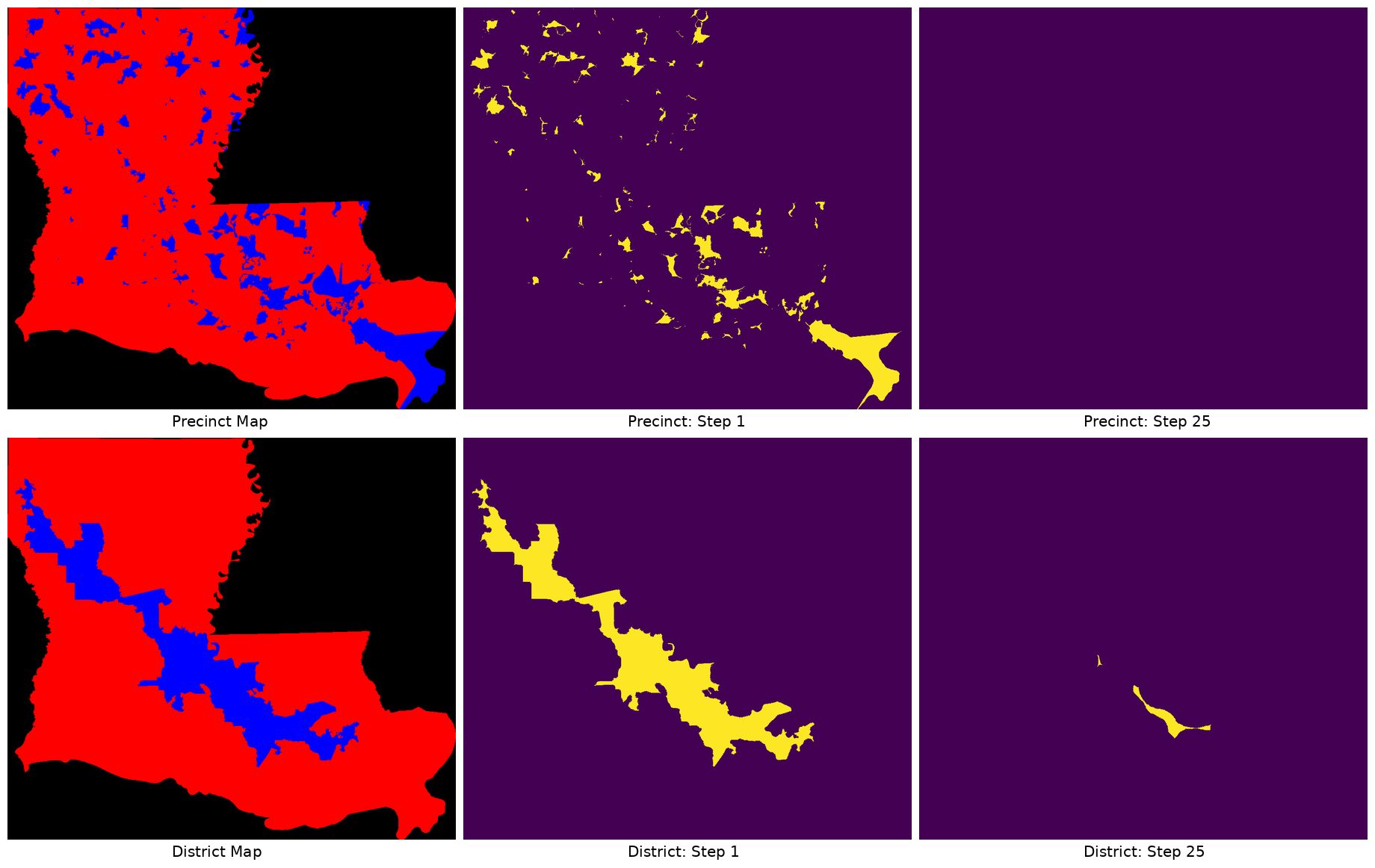}
\caption{Louisiana analysis. \textit{Top row:} precinct map (colored by winning party), level-set filtration at step~1, and step~25. \textit{Bottom row:} district map under the remedial 2026 plan, district level-set at step~1 (the majority-Black corridor district visible as a single connected yellow band), and step~25. The corridor district's topological signature--a single long-lived $H_1$ feature--is closely matched in the racial barcode, producing a bottleneck distance of 3.0 between partisan and racial persistence diagrams.}
\label{fig:la_analysis}
\end{figure}

Figure~\ref{fig:la_bivariate} shows the bivariate map of Democratic vote share and Black VAP across Louisiana precincts, confirming strong spatial alignment. Figure~\ref{fig:la_analysis} shows the precinct and district maps alongside their level-set filtrations, illustrating how the remedial majority-Black corridor district (District~6) appears as a single connected feature at filtration step~1 of the district level-set.

\subsection{Implications}

This result has direct legal relevance. The plaintiffs in \textit{Callais} argued that the remedial map was drawn primarily on the basis of race. The topological analysis corroborates the difficulty of distinguishing racial from partisan intent in Louisiana: because Black voters overwhelmingly vote Democratic, and because the geographic distribution of Black VAP and Democratic vote share are topologically nearly identical (bottleneck distance 3.0), it is essentially impossible to draw a map that packs Democratic votes without simultaneously packing Black voters, and vice versa.

This aligns with the observation from \textit{Shaw v.\ Reno} (1993) that redistricting based on race must survive strict scrutiny and cannot be the sole factor--but provides a new, quantitative lens: when the partisan and racial persistence diagrams are topologically indistinguishable, there may be no redistricting plan that satisfies both the VRA's mandate to create majority-minority districts and the equal protection requirement to avoid racial predominance.

The Louisiana analysis demonstrates that persistent homology can contribute not only to \emph{detecting} gerrymandering, but to \emph{characterizing the legal complexity} of cases where partisan and racial intent are geometrically inseparable.

\section{Conclusion}
\label{sec:conclusion}

We have applied persistent homology to quantify gerrymandering in four states across multiple redistricting cycles. The central methodological contributions are: (1) the use of the bottleneck distance between precinct- and district-level persistence barcodes as a geometry-independent gerrymandering metric; (2) a multi-metric validation framework combining persistence diagrams, Betti-1 curves, persistence landscapes, silhouettes, and persistence images, all of which yield consistent signals; (3) the first application of persistent homology to racial gerrymandering via racial vote-margin rasters; and (4) the precinct-split metric as a corroborating indicator of racial data usage in map-drawing.

The empirical findings are as follows:
\begin{enumerate}
    \item \textbf{North Carolina (2012--2024):} Precinct-level barcodes are stable across all six election cycles; district-level barcodes shift significantly in redistricting years. Compactness scores show no significant difference between the 2022 and 2024 maps despite a three-seat Republican gain--demonstrating a case where traditional metrics fail and persistent homology succeeds.
    \item \textbf{Texas (2025):} Applying the same 2024 voting data to the 2024 and 2025 maps, the bottleneck distance increases from 8.00 to 12.00, with approximately 748,000 Democratic votes (48\% of the Democratic margin) affected by the shift. This provides the first topological quantification of the redistricting impact of the 2025 Texas map.
    \item \textbf{Racial gerrymandering (Texas):} Applying the pipeline to racial vote-margin rasters and the precinct-split metric (0 vs.\ 380 splits) together provide topological and geometric corroboration of the federal court's finding of racial gerrymandering in the Texas 2025 plan. This is the first use of persistent homology for the quantification of racial gerrymandering.
    \item \textbf{California (2025):} The California 2025 map shows district-level persistence exceeding precinct-level persistence, consistent with the packing of Republican voters.
    \item \textbf{Louisiana (2026):} A bottleneck distance of 3.0 between partisan and racial persistence diagrams confirms that Democratic vote share and Black VAP are topologically inseparable in Louisiana's congressional map--providing quantitative context for the legal complexity of \textit{Louisiana v.\ Callais}.
\end{enumerate}

These results support the use of persistent homology as a court-defensible, geography-neutral complement to existing redistricting metrics. The bottleneck distance provides a single summary statistic that can be compared across maps and across states; the barcodes themselves show \emph{which} features are being manipulated and \emph{where}.

\section{Data Sources}
Precinct-level voting data for North Carolina were obtained from the \href{https://www.ncsbe.gov/results-data/voter-registration-data}{North Carolina State Board of Elections}. Texas and California precinct-level data and district shapefiles were obtained from the \href{https://redistrictingdatahub.org/}{Redistricting Data Hub} and \href{https://davesredistricting.org/}{Dave's Redistricting App}. District shapefiles were also sourced from the \href{https://github.com/mggg-states}{MGGG States repository}. Racial demographic data (BVAP, HVAP) were obtained from the \href{https://alarm-redist.org/}{ALARM Project}. Code was adapted from \href{https://github.com/mhcfeng/precinct/tree/master/scripts}{Feng's precinct repository}. All project code is available at \url{https://github.com/an-shah/persistenthomology}.

\section{Future Work}
\label{sec:futurework}

The \emph{Wasserstein distance} provides a cumulative alternative to the bottleneck distance:
\[
d_{W,p}(B_1, B_2) = \left( \inf_{\gamma} \sum_{x \in B_1} \| x - \gamma(x) \|^p \right)^{1/p}.
\]
Because it aggregates over all feature differences rather than taking only the maximum, it may be more sensitive to small-scale redistricting effects. Comparing the two metrics systematically, and developing statistically validated thresholds for flagging gerrymandering, is a natural next step.

Additional directions include:
\begin{enumerate}
    \item \textbf{Fr\'{e}chet means and ensemble comparison.} Duchin~\citep{Duchin2020} uses Fr\'{e}chet means to summarize ensembles of redistricting plans. Applying this to the time series of barcodes produced here could yield a ``baseline'' barcode for comparison, and could connect our enacted-map analysis to ensemble methods.
    \item \textbf{Projected demographic data.} The racial gerrymandering analysis uses 2020 Census data; applying 2030 Census projections when available would allow analysis of maps drawn for the next decade's election cycles.
    \item \textbf{Extension to state legislatures and other states.} The current analysis covers four states at the congressional level. Applying the pipeline to state legislative maps, and to additional states with different geographic and demographic profiles (particularly those with pending VRA litigation), would test the generalizability of the bottleneck-distance thresholds observed here.
    \item \textbf{Visualization tools for policymakers.} The persistence barcodes and rasters produced here are technically demanding to interpret. Developing accessible visualization tools that translate these outputs into plain-language assessments would be valuable for legislative and legal contexts.
\end{enumerate}

\section{Acknowledgments}

The author thanks Ms.\ Talia Dardis (science research teacher, Edgemont High School) for her continued guidance throughout this research; Dr.\ Ismar Voli\'{c} and the Institute for Mathematics and Democracy; Mr.\ Kenz Kallal for guidance on generating raster images and explaining the mathematical background; and Dr.\ Mason Porter and Dr.\ Thomas Weighill for input on the project. This work is supported by the Carnegie Corporation, Carnegie Young Leaders, and the Institute for Citizens and Scholars.

All errors are my own, feel free to reach out with corrections, comments, and suggestions at
ananya.neytri.shah@gmail.com.

\bibliographystyle{abbrvnat}
\bibliography{Bibliography}

\clearpage
\appendix
\section{Supplementary Figures: North Carolina 2012--2024}
\label{sec:addfig}

This appendix contains raster images and barcodes for all six North Carolina election cycles. For each year, we show the precinct raster, precinct barcode, precinct level set at step 1 and step 25, the district raster, district barcode, and district level set at step 1 and step 25.
\enlargethispage{5\baselineskip}

\begin{figure}[H]
    \centering
    \begin{minipage}{0.46\textwidth}
        \centering
        \includegraphics[width=\textwidth]{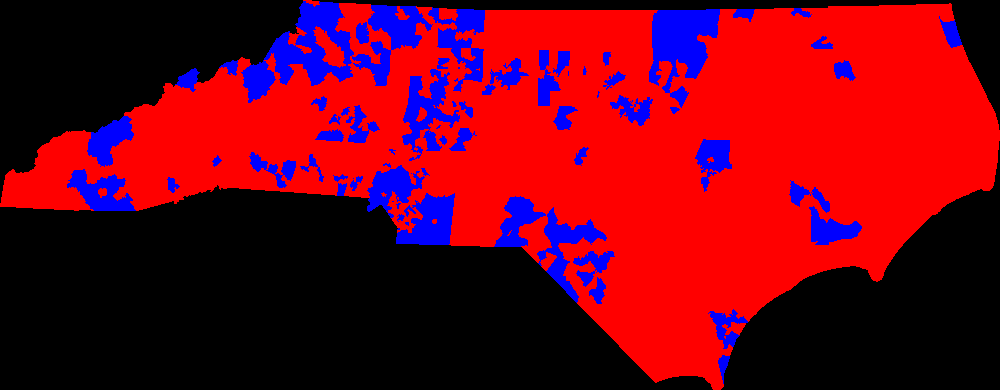}
        \caption{NC 2012 Precinct Raster}
        \label{fig:nc2012rp}
    \end{minipage}%
    \hfill
    \begin{minipage}{0.46\textwidth}
        \centering
        \includegraphics[width=\textwidth]{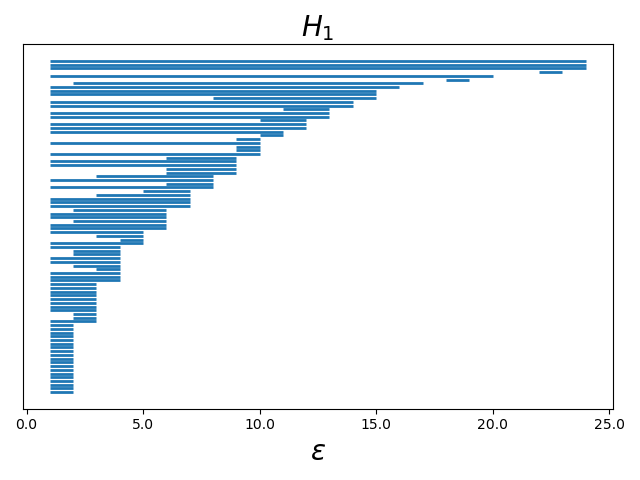}
        \caption{NC 2012 Precinct Barcode}
        \label{fig:nc2012pb}
    \end{minipage}
    \begin{minipage}{0.46\textwidth}
        \centering
        \includegraphics[width=\textwidth]{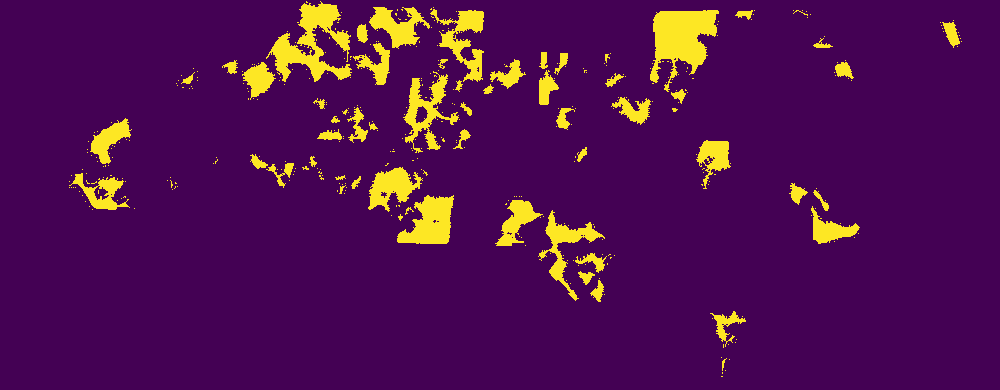}
        \caption{NC 2012 Precinct Level Set 1}
    \end{minipage}%
    \hfill
    \begin{minipage}{0.46\textwidth}
        \centering
        \includegraphics[width=\textwidth]{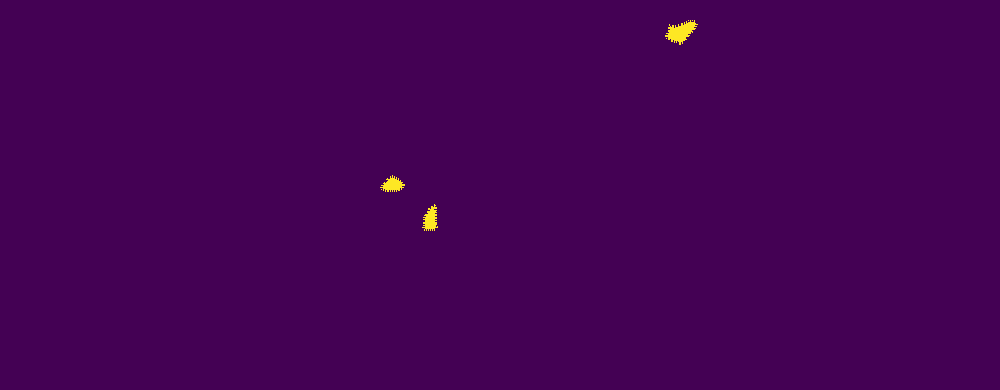}
        \caption{NC 2012 Precinct Level Set 25}
    \end{minipage}
\end{figure}

\begin{figure}[H]
    \centering
    \begin{minipage}{0.46\textwidth}
        \centering
        \includegraphics[width=\textwidth]{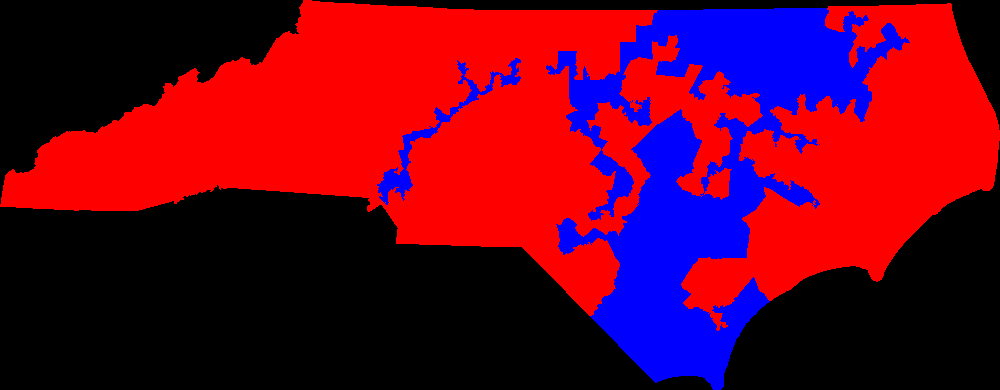}
        \caption{NC 2012 District Raster}
    \end{minipage}%
    \hfill
    \begin{minipage}{0.46\textwidth}
        \centering
        \includegraphics[width=\textwidth]{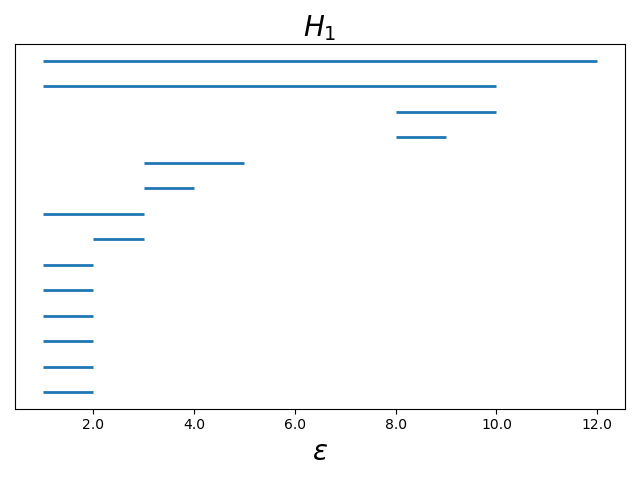}
        \caption{NC 2012 District Barcode}
    \end{minipage}
    \begin{minipage}{0.46\textwidth}
        \centering
        \includegraphics[width=\textwidth]{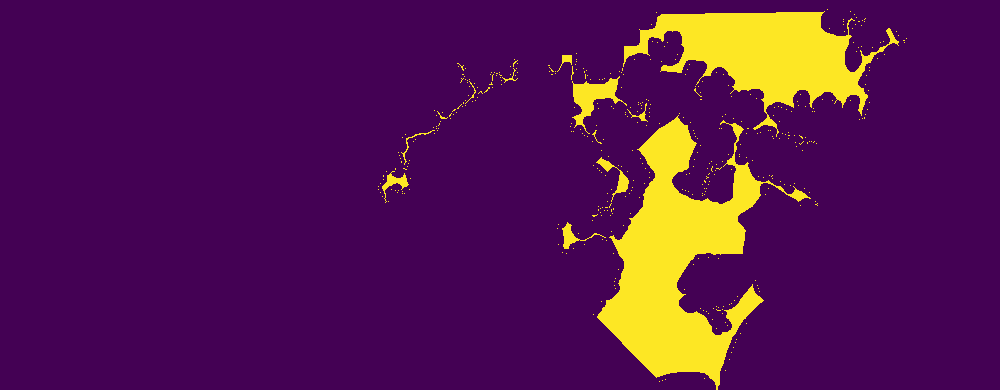}
        \caption{NC 2012 District Level Set 1}
    \end{minipage}%
    \hfill
    \begin{minipage}{0.46\textwidth}
        \centering
        \includegraphics[width=\textwidth]{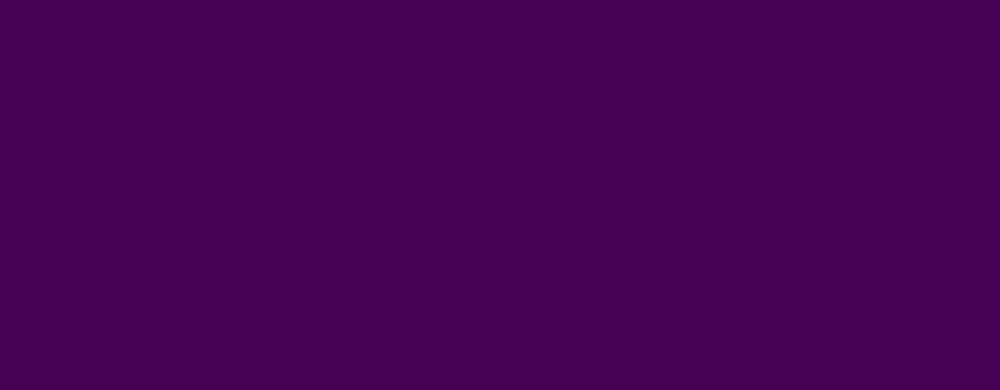}
        \caption{NC 2012 District Level Set 25}
    \end{minipage}
\end{figure}
\clearpage

\begin{figure}[H]
    \centering
    \begin{minipage}{0.46\textwidth}
        \centering
        \includegraphics[width=\textwidth]{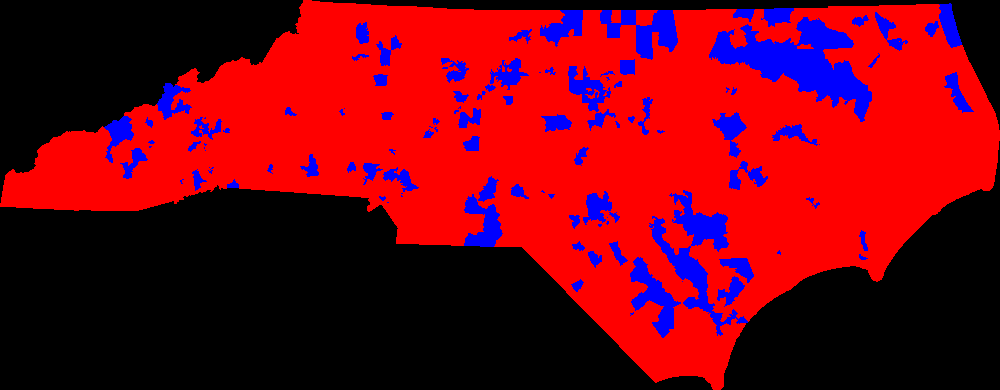}
\caption{NC 2014 Precinct Raster}
    \end{minipage}%
    \hfill
    \begin{minipage}{0.46\textwidth}
        \centering
        \includegraphics[width=\textwidth]{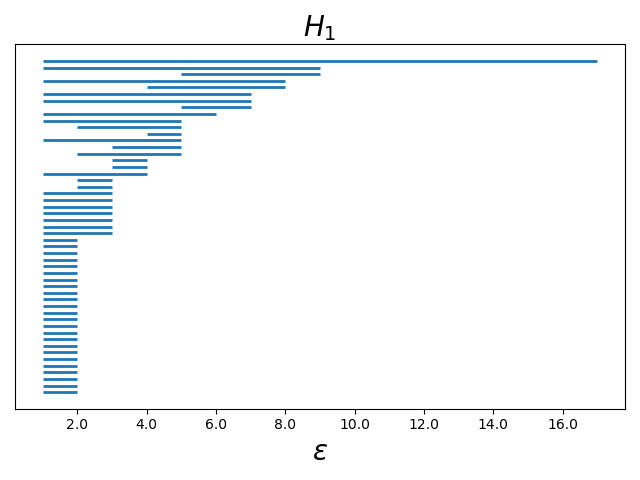}
        \caption{NC 2014 Precinct Barcode}
    \end{minipage}
    \begin{minipage}{0.46\textwidth}
        \centering
        \includegraphics[width=\textwidth]{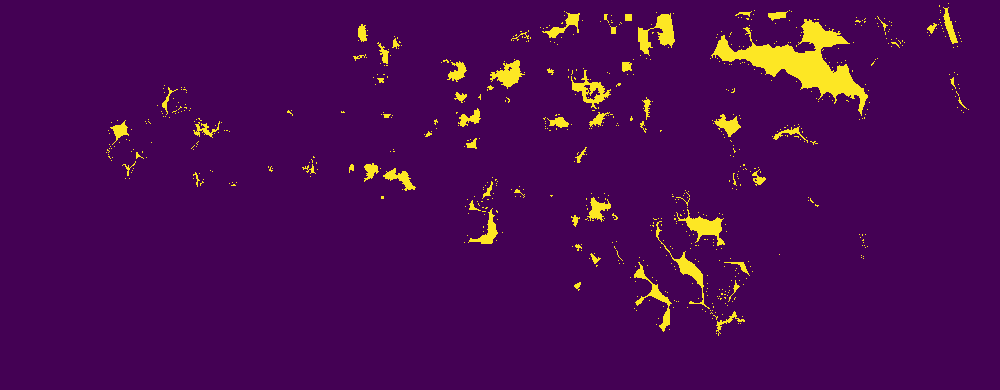}
        \caption{NC 2014 Precinct Level Set 1}
    \end{minipage}%
    \hfill
    \begin{minipage}{0.46\textwidth}
        \centering
        \includegraphics[width=\textwidth]{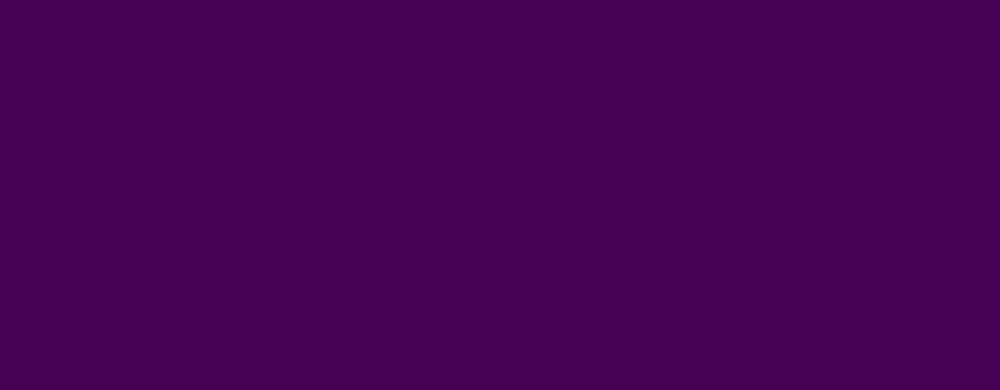}
        \caption{NC 2014 Precinct Level Set 25}
    \end{minipage}
\end{figure}

\begin{figure}[H]
    \centering
    \begin{minipage}{0.46\textwidth}
        \centering
        \includegraphics[width=\textwidth]{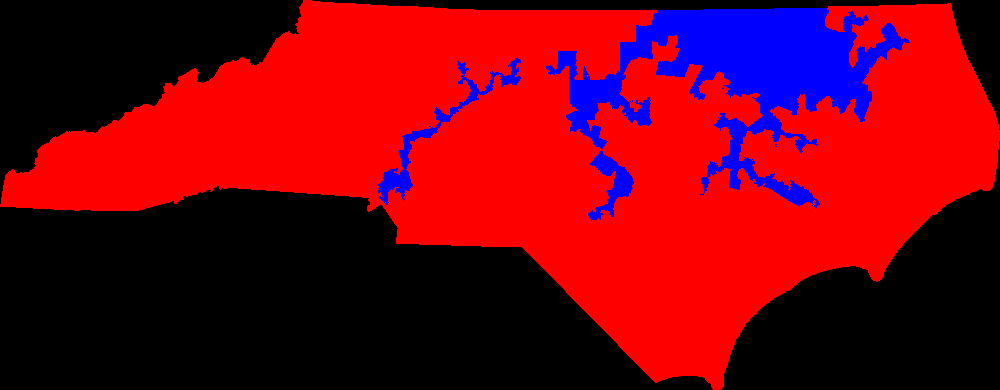}
        \caption{NC 2014 District Raster}
    \end{minipage}%
    \hfill
    \begin{minipage}{0.46\textwidth}
        \centering
        \includegraphics[width=\textwidth]{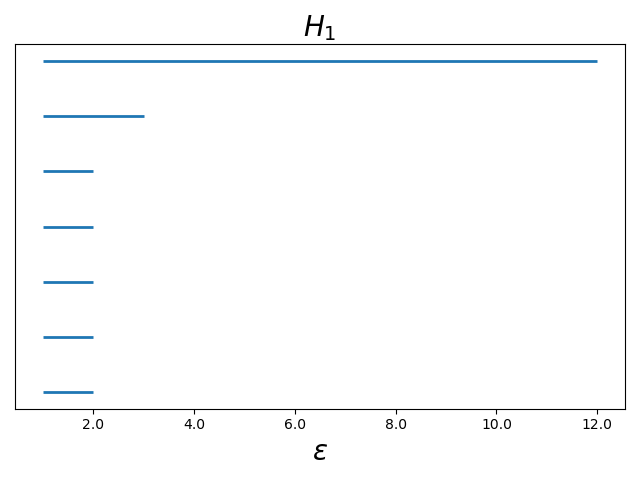}
        \caption{NC 2014 District Barcode}
    \end{minipage}
    \begin{minipage}{0.46\textwidth}
        \centering
        \includegraphics[width=\textwidth]{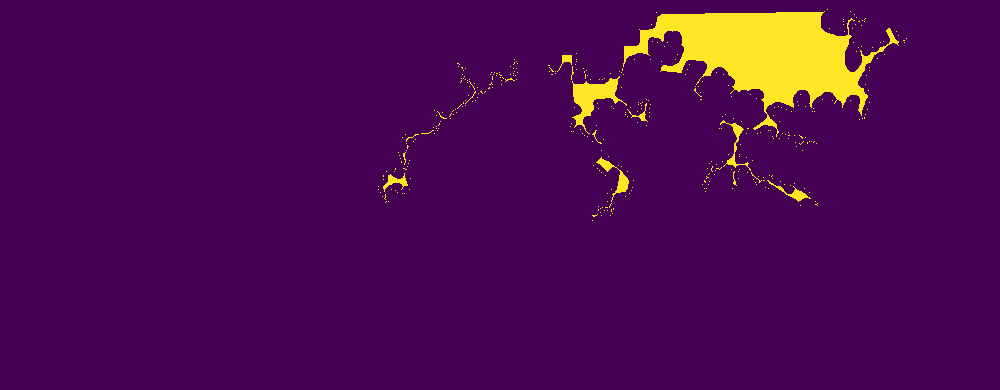}
        \caption{NC 2014 District Level Set 1}
    \end{minipage}%
    \hfill
    \begin{minipage}{0.46\textwidth}
        \centering
        \includegraphics[width=\textwidth]{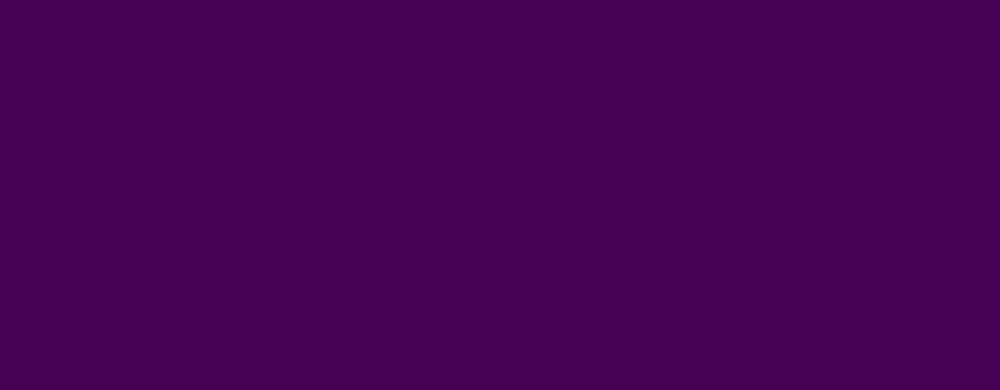}
        \caption{NC 2014 District Level Set 25}
    \end{minipage}
\end{figure}
\clearpage

\begin{figure}[H]
    \centering
    \begin{minipage}{0.46\textwidth}
        \centering
        \includegraphics[width=\textwidth]{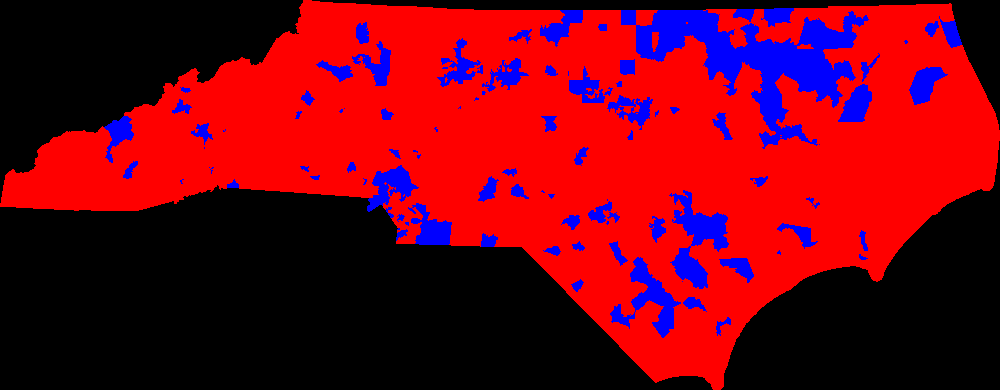}
\caption{NC 2016 Precinct Raster}
    \end{minipage}%
    \hfill
    \begin{minipage}{0.46\textwidth}
        \centering
        \includegraphics[width=\textwidth]{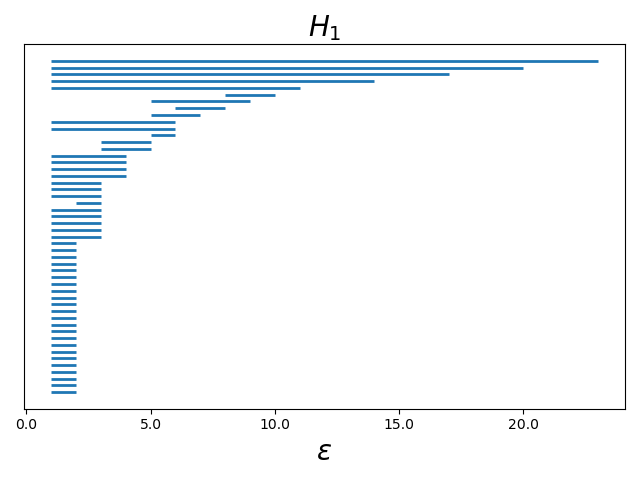}
        \caption{NC 2016 Precinct Barcode}
    \end{minipage}
    \begin{minipage}{0.46\textwidth}
        \centering
        \includegraphics[width=\textwidth]{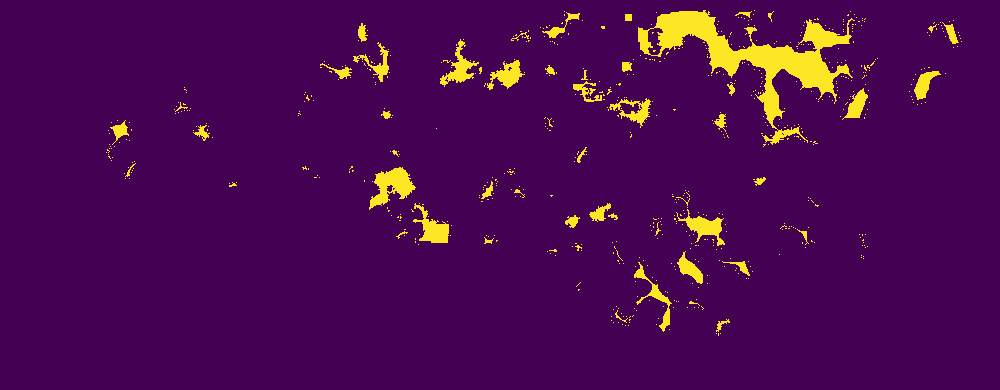}
        \caption{NC 2016 Precinct Level Set 1}
    \end{minipage}%
    \hfill
    \begin{minipage}{0.46\textwidth}
        \centering
        \includegraphics[width=\textwidth]{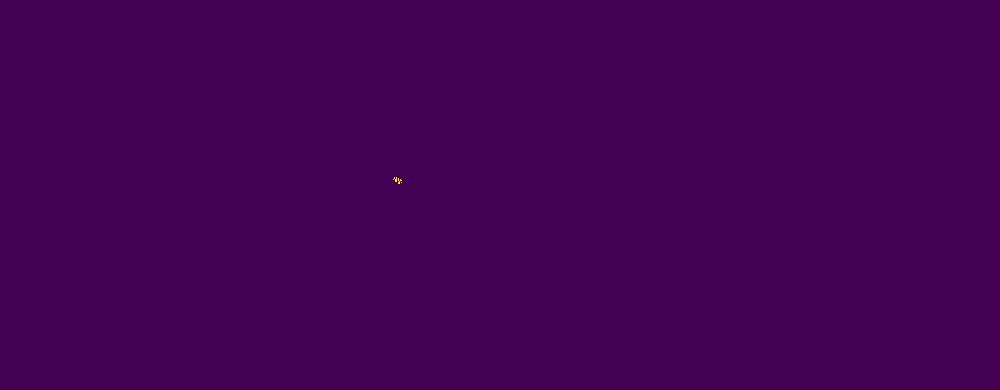}
        \caption{NC 2016 Precinct Level Set 25}
    \end{minipage}
\end{figure}

\begin{figure}[H]
    \centering
    \begin{minipage}{0.46\textwidth}
        \centering
        \includegraphics[width=\textwidth]{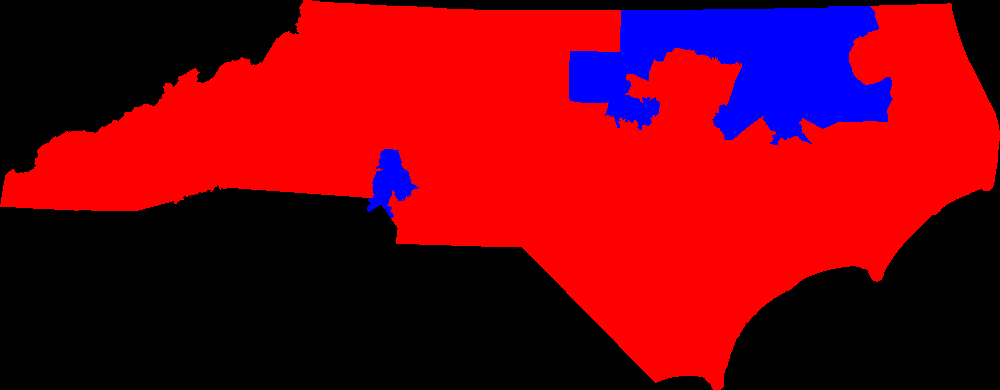}
        \caption{NC 2016 District Raster}
    \end{minipage}%
    \hfill
    \begin{minipage}{0.46\textwidth}
        \centering
        \includegraphics[width=\textwidth]{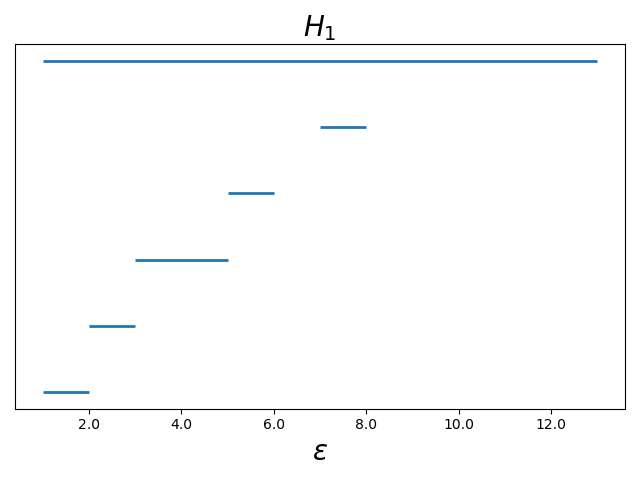}
        \caption{NC 2016 District Barcode}
    \end{minipage}
    \begin{minipage}{0.46\textwidth}
        \centering
        \includegraphics[width=\textwidth]{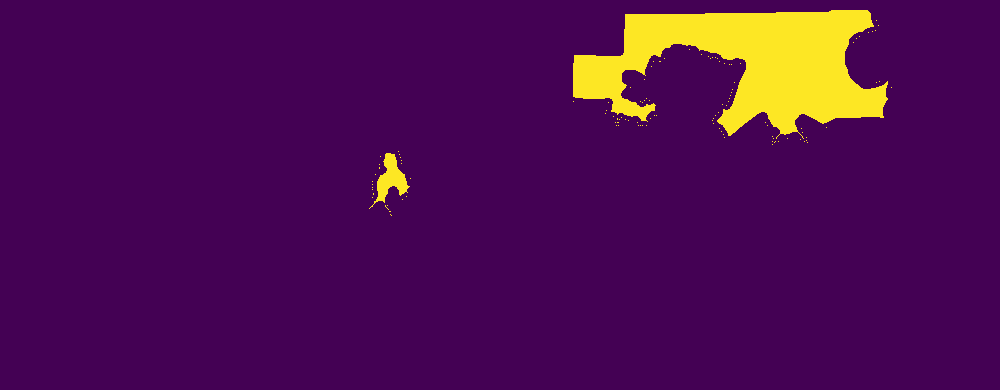}
        \caption{NC 2016 District Level Set 1}
    \end{minipage}%
    \hfill
    \begin{minipage}{0.46\textwidth}
        \centering
        \includegraphics[width=\textwidth]{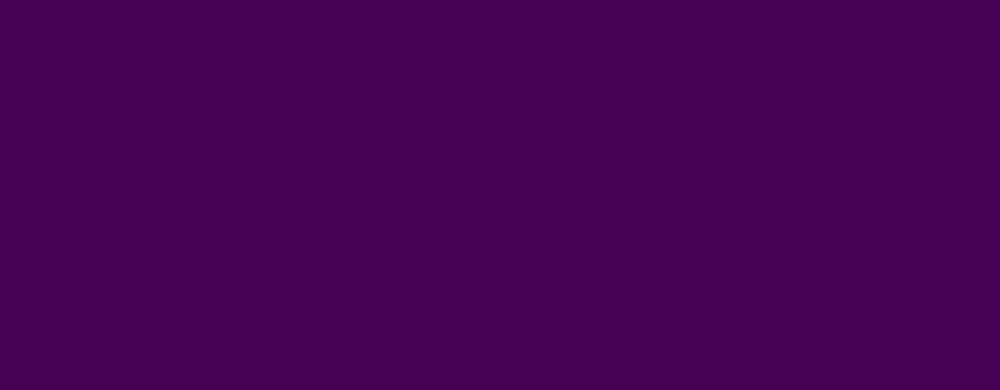}
        \caption{NC 2016 District Level Set 25}
    \end{minipage}
\end{figure}
\clearpage

\begin{figure}[H]
    \centering
    \begin{minipage}{0.46\textwidth}
        \centering
        \includegraphics[width=\textwidth]{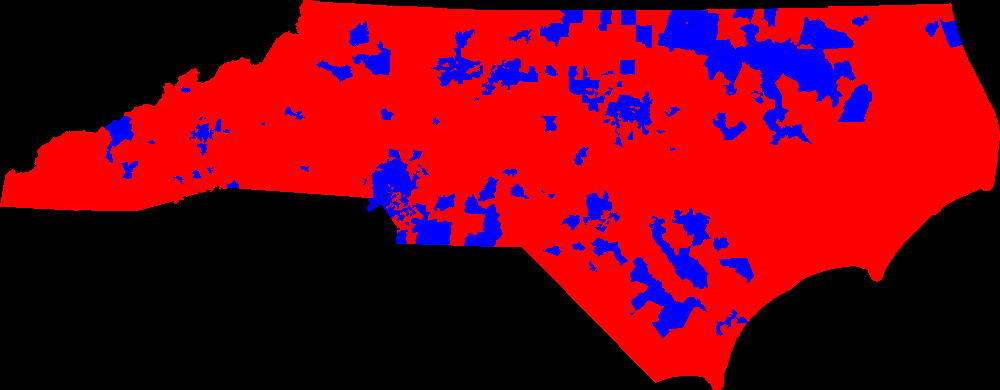}
\caption{NC 2018 Precinct Raster}
    \end{minipage}%
    \hfill
    \begin{minipage}{0.46\textwidth}
        \centering
        \includegraphics[width=\textwidth]{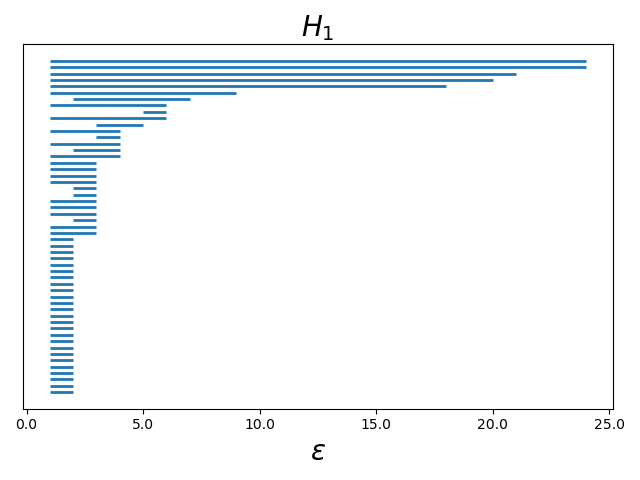}
        \caption{NC 2018 Precinct Barcode}
    \end{minipage}
    \begin{minipage}{0.46\textwidth}
        \centering
        \includegraphics[width=\textwidth]{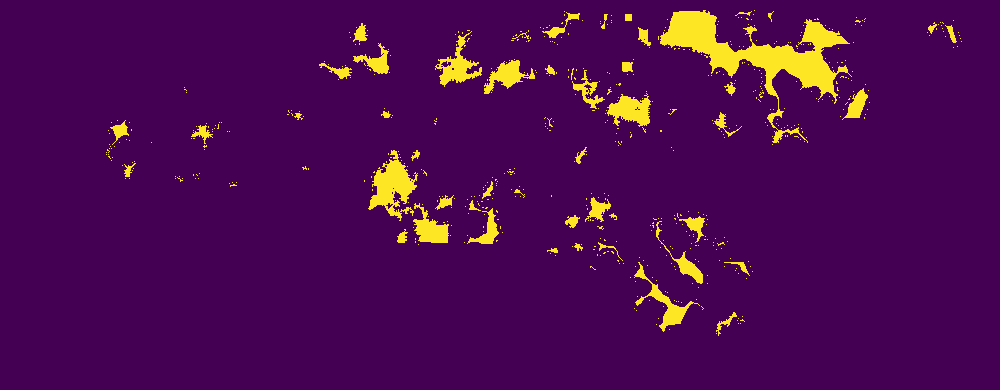}
        \caption{NC 2018 Precinct Level Set 1}
    \end{minipage}%
    \hfill
    \begin{minipage}{0.46\textwidth}
        \centering
        \includegraphics[width=\textwidth]{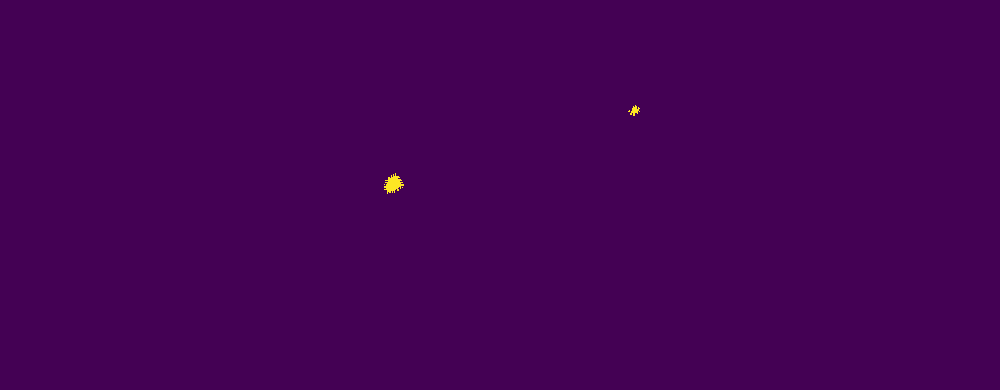}
        \caption{NC 2018 Precinct Level Set 25}
    \end{minipage}
\end{figure}

\begin{figure}[H]
    \centering
    \begin{minipage}{0.46\textwidth}
        \centering
        \includegraphics[width=\textwidth]{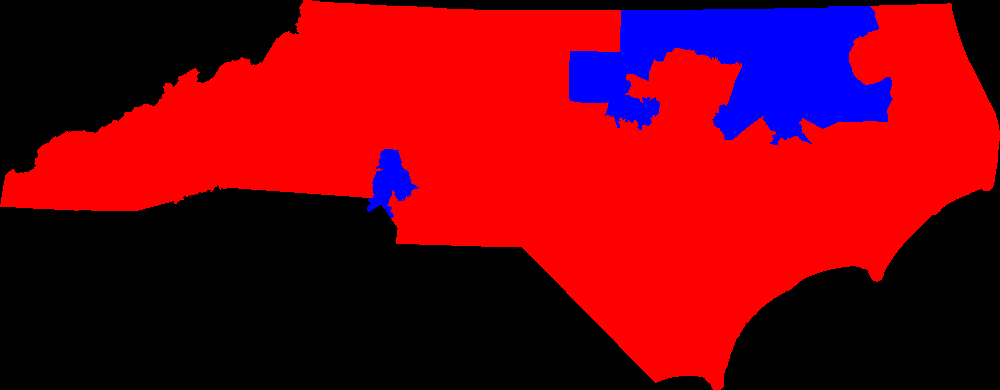}
        \caption{NC 2018 District Raster}
    \end{minipage}%
    \hfill
    \begin{minipage}{0.46\textwidth}
        \centering
        \includegraphics[width=\textwidth]{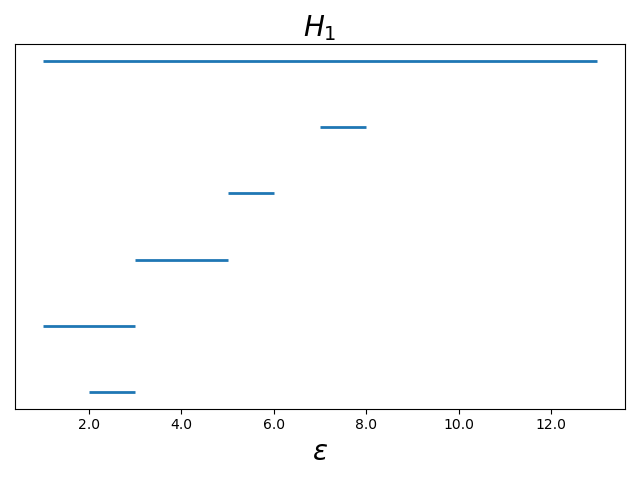}
        \caption{NC 2018 District Barcode}
    \end{minipage}
    \begin{minipage}{0.46\textwidth}
        \centering
        \includegraphics[width=\textwidth]{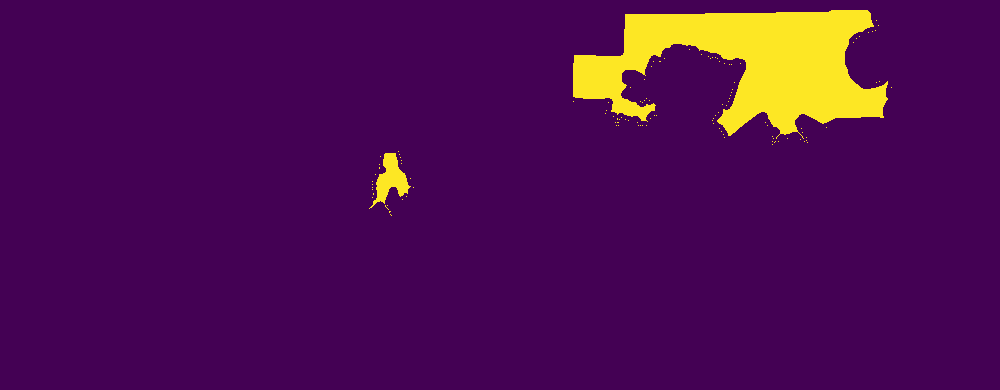}
        \caption{NC 2018 District Level Set 1}
    \end{minipage}%
    \hfill
    \begin{minipage}{0.46\textwidth}
        \centering
        \includegraphics[width=\textwidth]{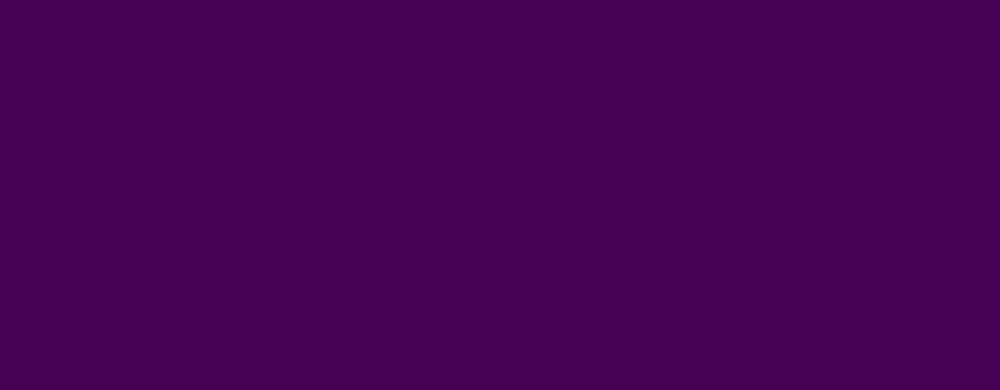}
        \caption{NC 2018 District Level Set 25}
    \end{minipage}
\end{figure}
\clearpage

\begin{figure}[H]
    \centering
    \begin{minipage}{0.46\textwidth}
        \centering
        \includegraphics[width=\textwidth]{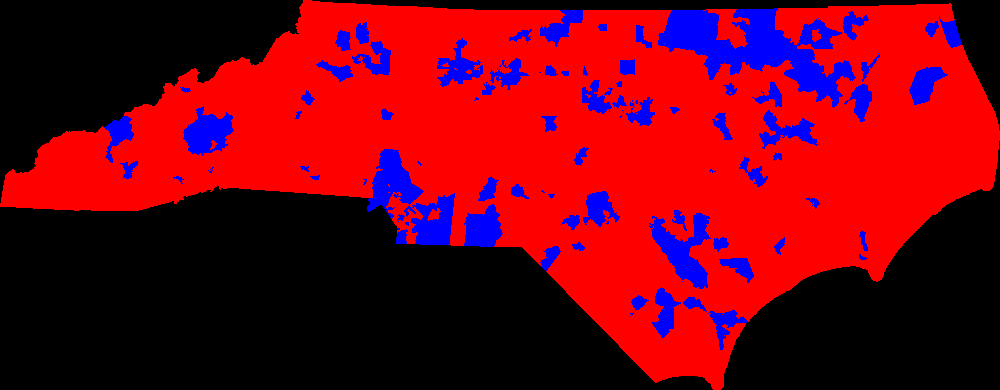}
\caption{NC 2020 Precinct Raster}
    \end{minipage}%
    \hfill
    \begin{minipage}{0.46\textwidth}
        \centering
        \includegraphics[width=\textwidth]{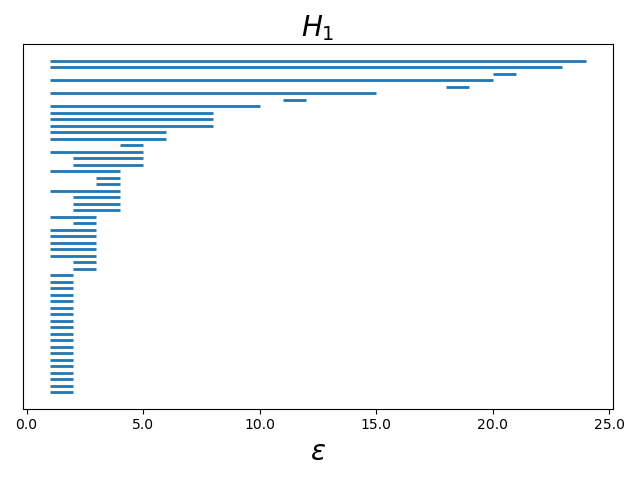}
        \caption{NC 2020 Precinct Barcode}
    \end{minipage}
    \begin{minipage}{0.46\textwidth}
        \centering
        \includegraphics[width=\textwidth]{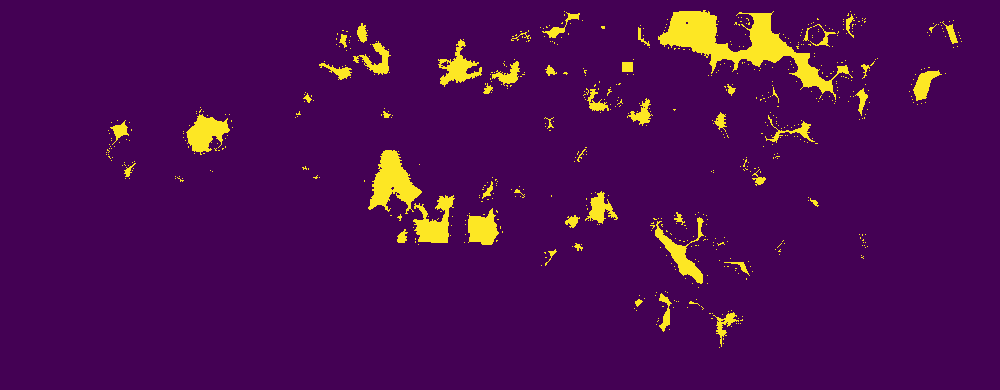}
        \caption{NC 2020 Precinct Level Set 1}
    \end{minipage}%
    \hfill
    \begin{minipage}{0.46\textwidth}
        \centering
        \includegraphics[width=\textwidth]{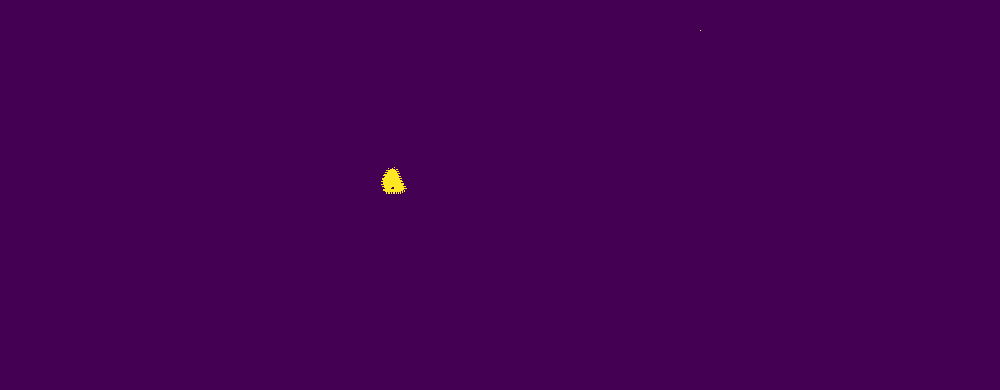}
        \caption{NC 2020 Precinct Level Set 25}
    \end{minipage}
\end{figure}

\begin{figure}[H]
    \centering
    \begin{minipage}{0.46\textwidth}
        \centering
        \includegraphics[width=\textwidth]{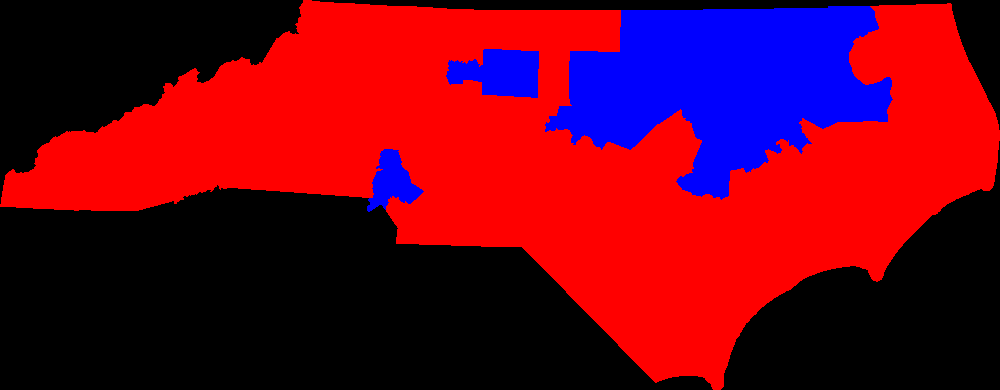}
        \caption{NC 2020 District Raster}
    \end{minipage}%
    \hfill
    \begin{minipage}{0.46\textwidth}
        \centering
        \includegraphics[width=\textwidth]{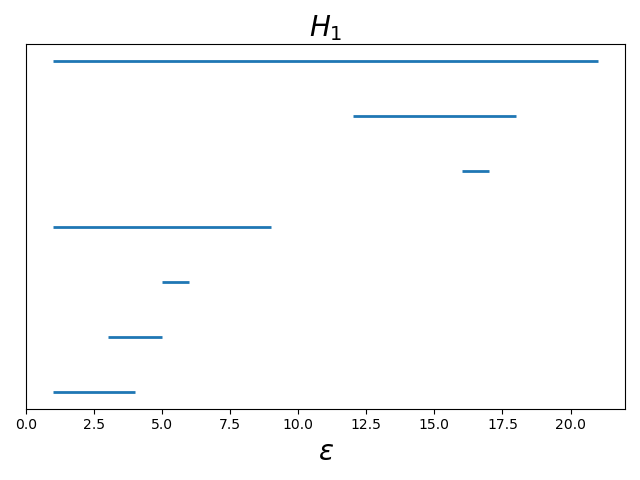}
        \caption{NC 2020 District Barcode}
    \end{minipage}
    \begin{minipage}{0.46\textwidth}
        \centering
        \includegraphics[width=\textwidth]{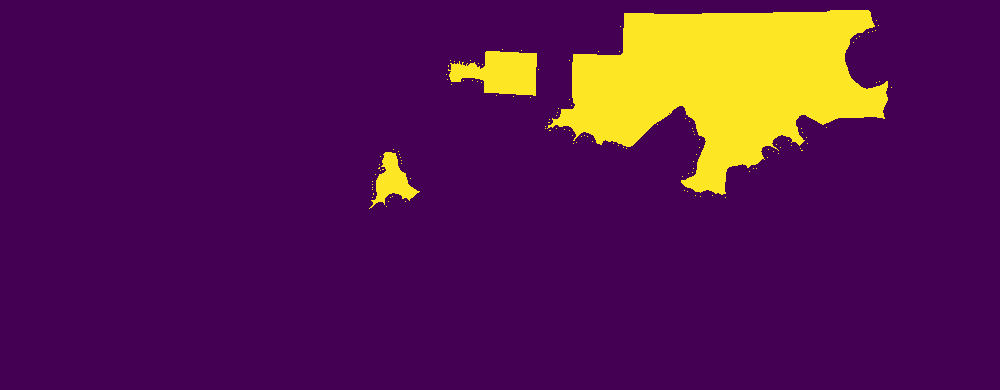}
        \caption{NC 2020 District Level Set 1}
    \end{minipage}%
    \hfill
    \begin{minipage}{0.46\textwidth}
        \centering
        \includegraphics[width=\textwidth]{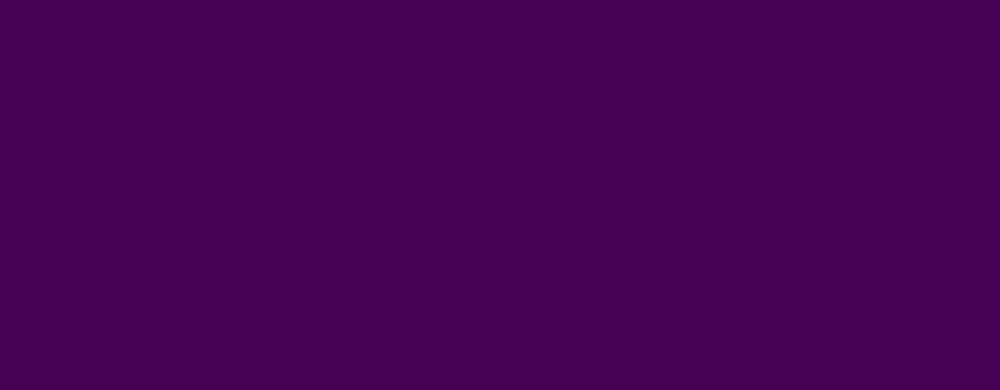}
        \caption{NC 2020 District Level Set 25}
    \end{minipage}
\end{figure}
\clearpage

\begin{figure}[H]
    \centering
    \begin{minipage}{0.46\textwidth}
        \centering
        \includegraphics[width=\textwidth]{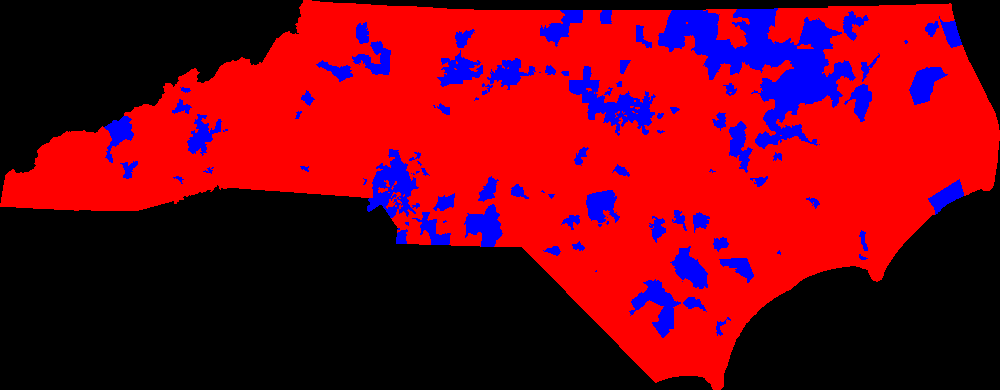}
\caption{NC 2022 Precinct Raster}
    \end{minipage}%
    \hfill
    \begin{minipage}{0.46\textwidth}
        \centering
        \includegraphics[width=\textwidth]{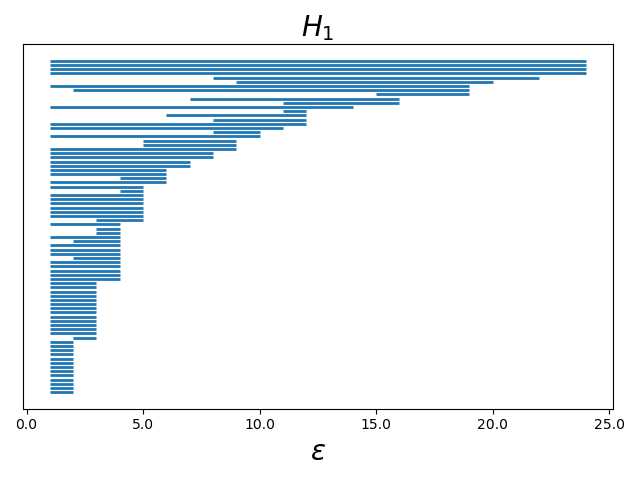}
        \caption{NC 2022 Precinct Barcode}
    \end{minipage}
    \begin{minipage}{0.46\textwidth}
        \centering
        \includegraphics[width=\textwidth]{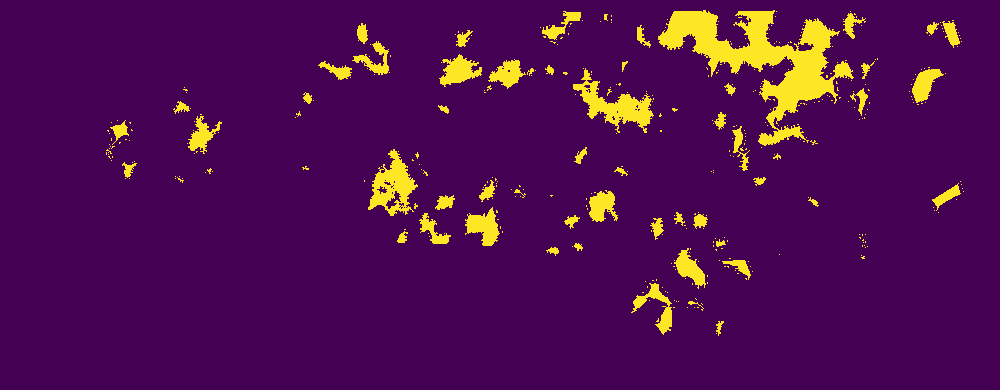}
        \caption{NC 2022 Precinct Level Set 1}
    \end{minipage}%
    \hfill
    \begin{minipage}{0.46\textwidth}
        \centering
        \includegraphics[width=\textwidth]{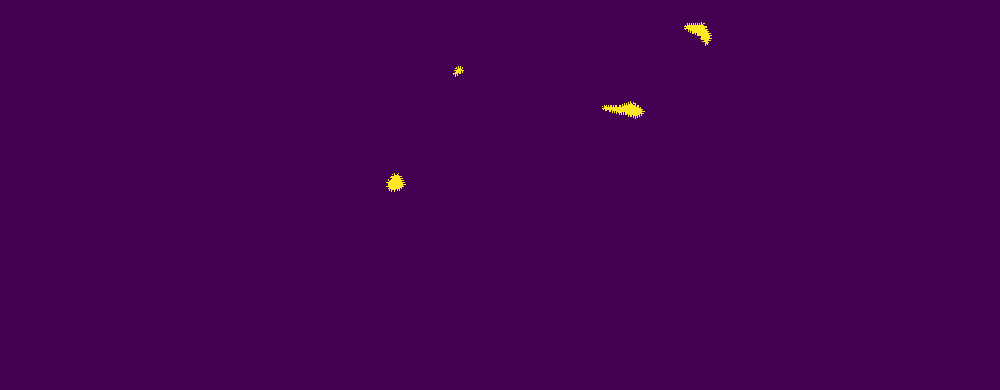}
        \caption{NC 2022 Precinct Level Set 25}
    \end{minipage}
\end{figure}

\begin{figure}[H]
    \centering
    \begin{minipage}{0.46\textwidth}
        \centering
        \includegraphics[width=\textwidth]{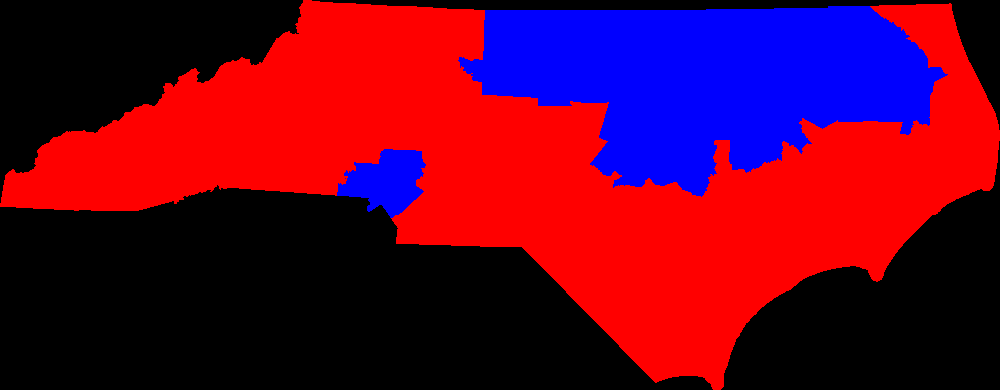}
        \caption{NC 2022 District Raster}
    \end{minipage}%
    \hfill
    \begin{minipage}{0.46\textwidth}
        \centering
        \includegraphics[width=\textwidth]{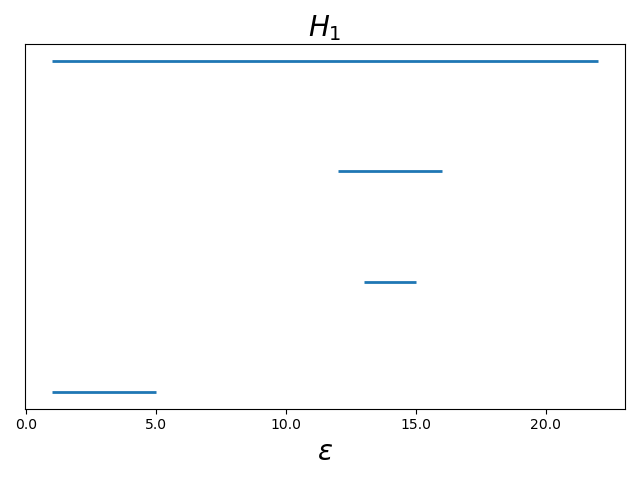}
        \caption{NC 2022 District Barcode}
    \end{minipage}
    \begin{minipage}{0.46\textwidth}
        \centering
        \includegraphics[width=\textwidth]{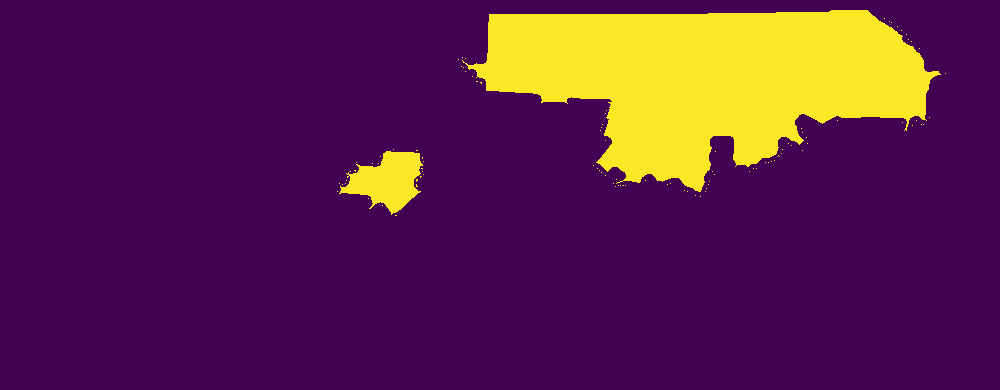}
        \caption{NC 2022 District Level Set 1}
    \end{minipage}%
    \hfill
    \begin{minipage}{0.46\textwidth}
        \centering
        \includegraphics[width=\textwidth]{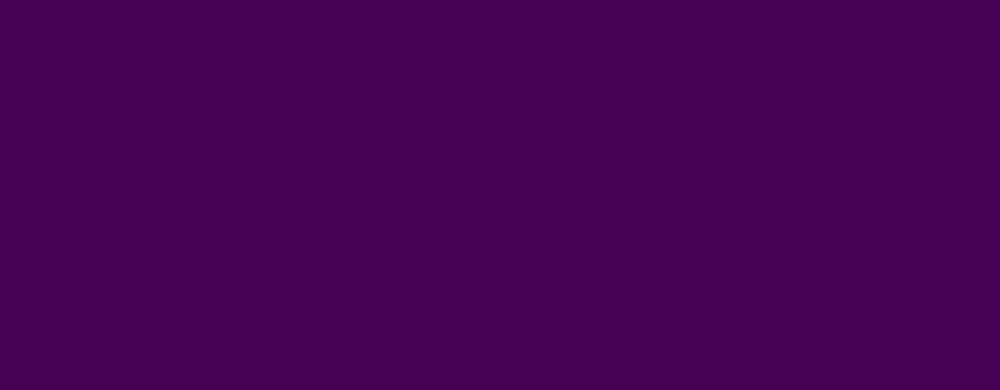}
        \caption{NC 2022 District Level Set 25}
    \end{minipage}
\end{figure}
\clearpage

\begin{figure}[H]
    \centering
    \begin{minipage}{0.46\textwidth}
        \centering
        \includegraphics[width=\textwidth]{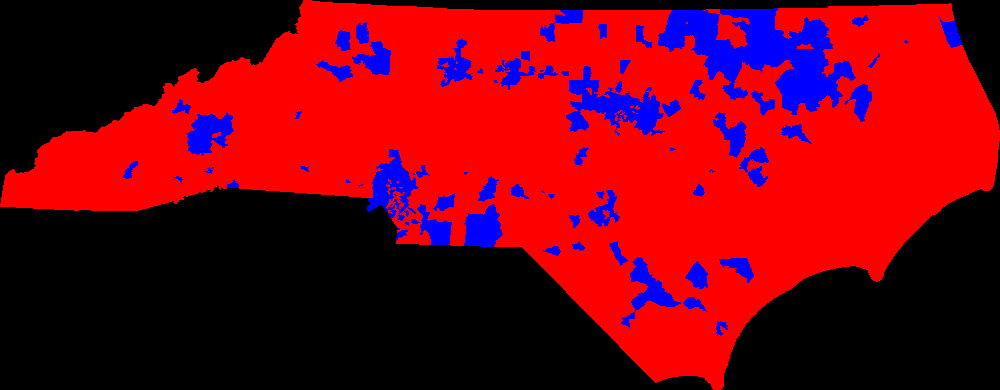}
\caption{NC 2024 Precinct Raster}
    \end{minipage}%
    \hfill
    \begin{minipage}{0.46\textwidth}
        \centering
        \includegraphics[width=\textwidth]{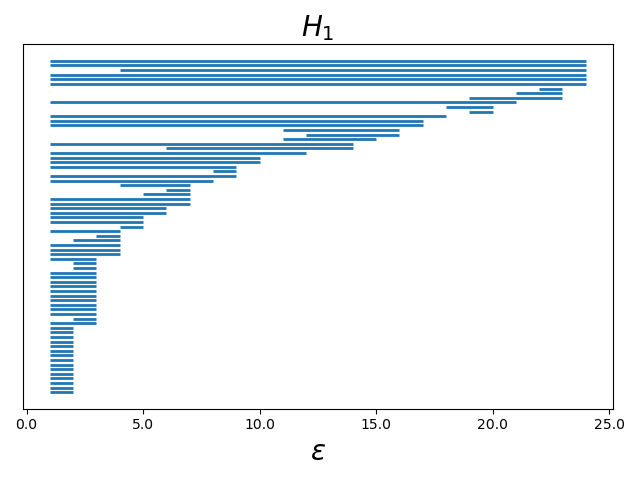}
        \caption{NC 2024 Precinct Barcode}
    \end{minipage}
    \begin{minipage}{0.46\textwidth}
        \centering
        \includegraphics[width=\textwidth]{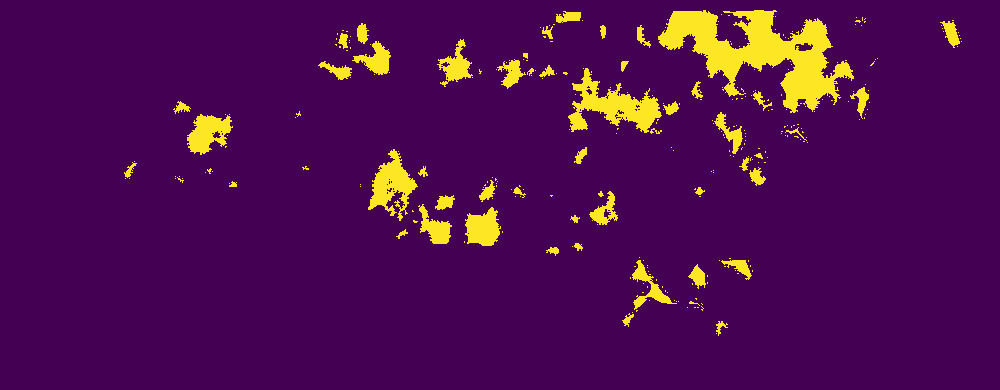}
        \caption{NC 2024 Precinct Level Set 1}
    \end{minipage}%
    \hfill
    \begin{minipage}{0.46\textwidth}
        \centering
        \includegraphics[width=\textwidth]{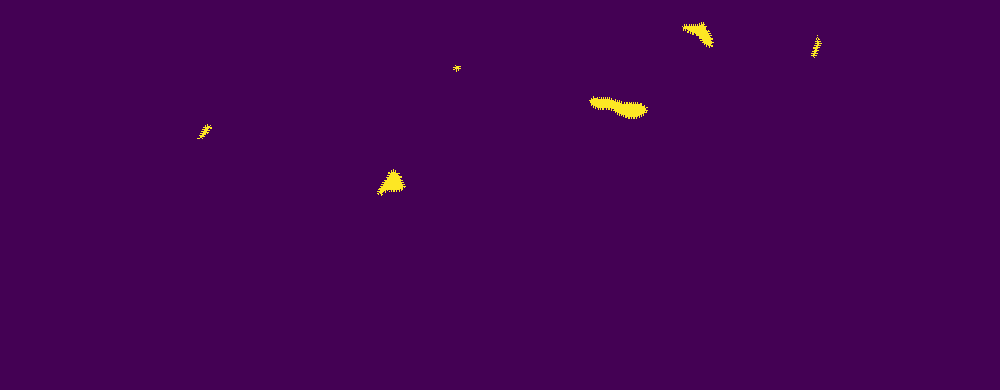}
        \caption{NC 2024 Precinct Level Set 25}
    \end{minipage}
\end{figure}

\begin{figure}[H]
    \centering
    \begin{minipage}{0.46\textwidth}
        \centering
        \includegraphics[width=\textwidth]{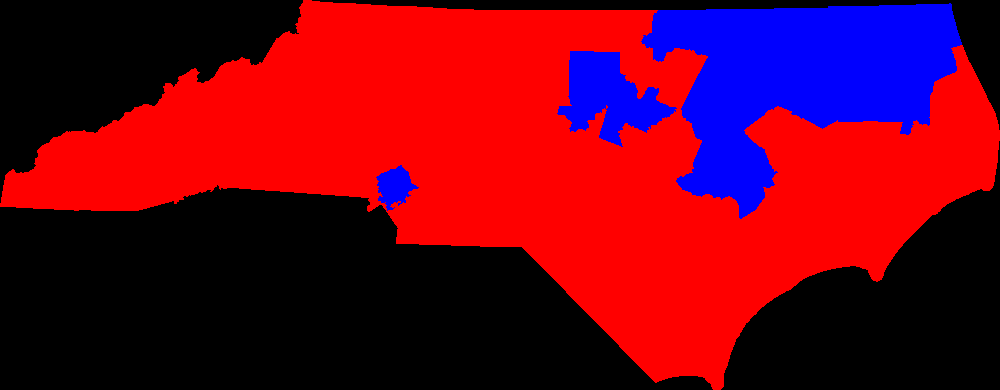}
        \caption{NC 2024 District Raster}
    \end{minipage}%
    \hfill
    \begin{minipage}{0.46\textwidth}
        \centering
        \includegraphics[width=\textwidth]{nc_2024_district_barcode.png}
        \caption{NC 2024 District Barcode}
    \end{minipage}
    \begin{minipage}{0.46\textwidth}
        \centering
        \includegraphics[width=\textwidth]{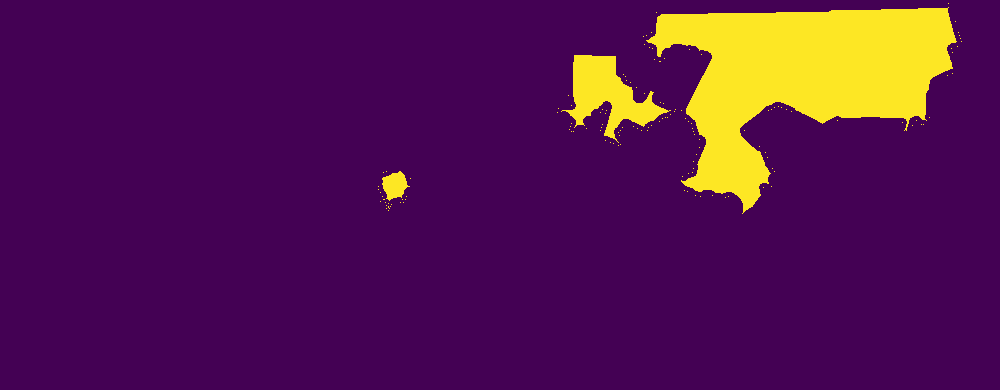}
        \caption{NC 2024 District Level Set 1}
    \end{minipage}%
    \hfill
    \begin{minipage}{0.46\textwidth}
        \centering
        \includegraphics[width=\textwidth]{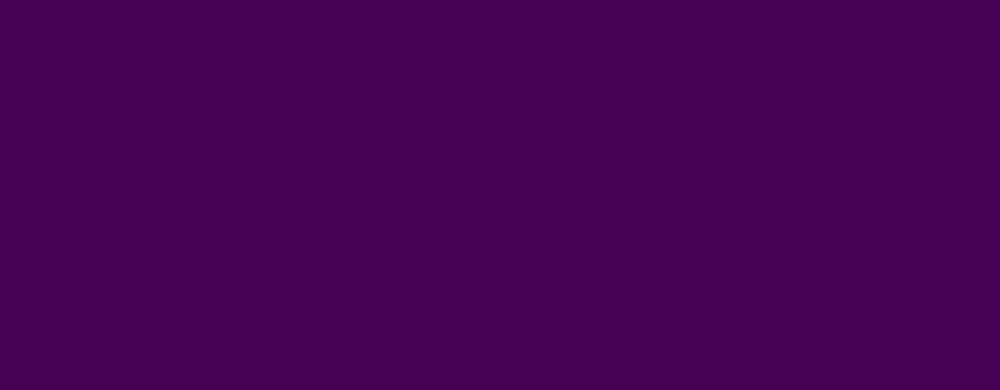}
        \caption{NC 2024 District Level Set 25}
    \end{minipage}
\end{figure}

\end{document}